\documentclass[sn-standardnature]{sn_jnl}

\jyear{2021}%

\theoremstyle{thmstyleone}%
\theoremstyle{thmstyletwo}%

\theoremstyle{thmstylethree}%

\usepackage{graphicx}
\usepackage{subfig}
\usepackage{float}

\begin{document}

\title[Article Title]{Increasing and Diverging  Greenhouse Gas Emissions of Urban Wastewater Treatment in China}


\author[1]{\fnm{Yujun} \sur{Huang}}

\author*[1]{\fnm{Fanlin} \sur{Meng}}\email{baicaimd@gmail.com}

\author*[1]{\fnm{Shuming} \sur{Liu}}\email{shumingliu@mail.tsinghua.edu.cn}

\author[2]{\fnm{Kate} \sur{Smith}}

\affil*[1]{\orgdiv{School of Environment}, \orgname{Tsinghua University}, \orgaddress{\city{Beijing}, \country{China}}}


\affil[2]{\orgname{Aurecon}, \orgaddress{\city{Neutral Bay}, \state{New South Wales}, \country{Australia}}}


\abstract{Upgrading effluent standards of wastewater treatment plants (WWTPs) and repairing sewerage systems leads to contradictions and synergies between water pollution control and climate change mitigation. This affects historical trajectories and characteristics of greenhouse gas (GHG) emissions from China’s WWTPs, which stay inadequately studied. Here we establish emissions inventories of China’s WWTPs using plant-level WWTP operational data. We find that removed amount of chemical oxygen demand and ammonia nitrogen increased 0.8 and 1.3 times during 2009-2019, while WWTP GHG emissions increased 1.8 times, being 6 times national GHG emissions growth rate. Increasing sludge yield and electricity intensity became primary driving factors in 2015 because of stricter effluent standards and lower influent contaminant concentration. We defined Functional Unit-Gini coefficient to quantify divergence of WWTP GHG emissions, which grew from 0.20 in 2009 to 0.29 in 2019. Diversified sludge disposal methods and energy structure increased the inequality, while upgrading effluent standards decreased it.}

\maketitle

\section{Introduction}\label{sec1}

In recent years, China's urban wastewater treatment rate increased and effluent quality requirements strengthened, driven by national policies such as the Action Plan for Prevention and Control of Water Pollution, released in 2015 \cite{state_council_of_china_water_2015}. The wastewater treatment rate of cities (including counties) in China increased from 57\% in 2007 to 94\% in 2017 \cite{mohurd_statistical_2018}, and the treated wastewater volume increased by 1.3 times during this period. The proportion of effluent reaching Class 1A (an effluent quality standard stricter than Class 1B implemented before, see Supplementary Table 1) increased from 6\% in 2007 to 56\% in 2017 \cite{smith_evaluation_2019}. However, problems of rainwater inflow, groundwater infiltration and wastewater exfiltration (IIE) in sewerage systems became increasingly serious in China \cite{cao_leakage_2019}, which cause both black and odorous water bodies and low wastewater treatment efficiency because of low influent chemical oxygen demand (COD) in WWTPs \cite{zhao_pin-pointing_2020}. A survey in 2021 for 467 WWTPs in China shows that actual average influent COD concentration of the WWTPs was only 0.7 times of designed average value \cite{zhang_current_2021}. 

Related studies found that stricter effluent standards lead to more greenhouse gas (GHG) emissions \cite{zhu_raising_2013,rahman_life-cycle_2016}. Average GHG intensity of eight investigated WWTPs reaching Class 1A in China was 17 kg CO${_2}$ eq/(kg PO$_{4}^{3-}$ eq removed), approximately 1.5 times that of four investigated WWTPs reaching Class 1B \cite{zhu_raising_2013}. Life-cycle analysis through theoretical models indicate that the GHG emissions in WWTPs with nutrient removal are more than three times those with organic matter removal only \cite{rahman_life-cycle_2016}, because much N$_2$O is generated during nitrogen removal \cite{kampschreur_nitrous_2009}, which contributes to 5.6\% of China’s national N$_2$O emissions in 2014 \cite{ncsc_second_2018}. Advanced technologies such as membrane filtration applied for stricter standard also consume much electricity, resulting in much higher levels of indirect GHG emissions \cite{rahman_life-cycle_2016}.

Water pollution control and climate change mitigation are both critical targets for sustainable development. Reducing GHG emissions has become a great challenge for water sector management \cite{rothausen_greenhouse-gas_2011,zhang_urbanization_2019}, and wastewater treatment plays an increasingly important role \cite{mo_can_2012,zhang_hidden_2017}. In 2021, the National working Conference on Ecological and Environmental Protection declared the importance of implementing both pollutant removal and GHG emissions reduction. For policymakers dealing with water, energy and climate change, it is challenging but necessary to characterize nationwide WWTP GHG emissions in China \cite{zhang_greenhouse_2021}. Moreover, differences of technology and environment between WWTPs results in GHG inequality (divergence of GHG intensity) of WWTPs. China seeks to consider equality when implementing climate change mitigation actions during its climate change actions \cite{mi_economic_2020}, and informed discussion about “fairness” in WWTP GHG control needs a quantitative understanding of historical variation in WWTP GHG inequality.

Some studies have explored relationships between WWTPs and GHG emissions at national level. Direct emissions during treatment and indirect emissions from electricity use were believed to be the most important sources of GHG emissions from WWTPs \cite{zhang_hidden_2017}. But they assume influent quality, sludge yield and electricity intensity of all WWTPs in China to be the same and static. Therefore, the growth rate of GHG emissions and the proportion of each GHG source are easily misestimated, and differences among WWTPs cannot be quantified. Contradictions and synergies exist between wastewater pollutant removal and electricity consumption. Regarding contradictions, previous research found that upgrading standard from Class 1B to Class 1A could result in 2\%-36\% greater electricity use for China’s WWTPs \cite{smith_evaluation_2019}. Regarding synergies, dealing with IIE problems prevents water pollution, and it raises electricity use efficiency because it increases influent COD concentration. Electricity consumption of WWTPs in China is estimated to be reduced by at least 20\% if influent COD increases to more than 500 mg/L \cite{niu_energy_2019}. However, the energy structure for electricity generation differs among regions, so the impact on GHG emissions is different from that of electricity. Moreover, few studies focus on all GHG emissions sources with nationwide plant-level data. Previous studies on carbon inequality focus on human society \cite{mi_economic_2020}, and few studies pay attention to carbon inequality in wastewater treatment.

This study establishes a GHG emissions inventory for China’s WWTPs from 2009 to 2019. To the best of our knowledge, this inventory, for the first time, covers real monthly plant-level operational data of China’s WWTPs and regional differences among WWTPs. We explore regional differences and affecting factors of GHG emissions at provincial, city and county levels. Then we examine socioeconomic drivers of spatiotemporal changes in GHG emissions using a Logarithmic Mean Divisia Index (LMDI) method. We define a functional unit Gini (FU-Gini) coefficient to quantify the inequality in GHG intensity between WWTPs. We find that GHG emissions from WWTPs have been growing much faster than national GHG emissions, and since 2015, the main factors driving this increase have shifted from wastewater pollutant quantity to sludge generation and electricity use intensities. GHG inequality of wastewater treatment has grown continuously. These findings have policy implications from a national perspective as they help understand the trade-offs and synergies between water pollution control and mitigating climate change, and the inequalities among WWTPs across space and time.

\section{Results}\label{sec2}

\subsection{Wastewater treatment GHG emissions and characteristics in China’s WWTPs}\label{subsec1}

China’s WWTPs generated 58.3 Mt CO${_2}$ eq in 2019, and it equals to 2.2 days of China’s CO${_2}$ emissions and 4.4 months of CO${_2}$ emissions in Beijing in the same year \cite{guan_assessment_2021}. The total emissions are 1.3 times those in the US (2019) \cite{us_environmental_protection_agency_inventory_2021}, 3.3 times those in the European Union (2019) \cite{water_information_system_for_europe_country_2021}, and 22.6 times those in the UK (2018) \cite{brown_uk_2021}. The average GHG intensity was 0.922 kg CO${_2}$ eq/m$^3$ in 2019, while in 2011 it was 0.807 kg CO${_2}$ eq/m$^3$, being 3.8 times those in China’s urban water supply \cite{smith_contribution_2016}. The breakdown of WWTP GHG emissions in 2019 shown by supplementary Fig. 1 reveals that the process of sludge treatment and disposal (STD) (47\%), electricity use (26\%) and collection and treatment (20\%) are the three main GHG sources.

\begin{figure}[h]%
	\centering
	\includegraphics[width=1.0\textwidth]{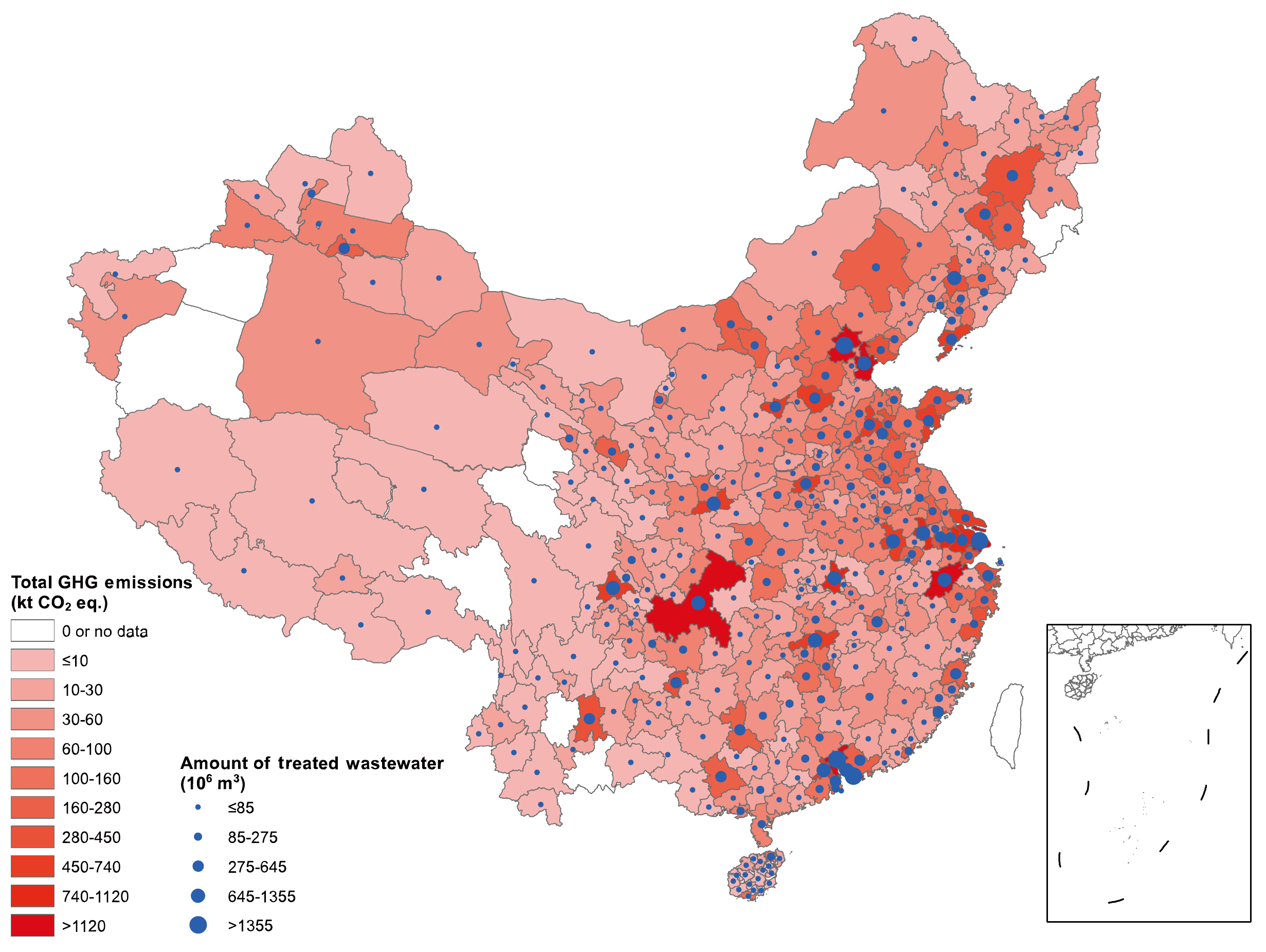}
	\caption{\textbf{GHG emissions associated with WWTP operation, 2019.} The colors of the cities represent GHG emissions. Darker red colors indicate higher GHG emissions, estimated by the life-cycle GHG emissions of five sources during WWTP operation. The blue points show the amount of wastewater treated by the cities. Larger points represent more treated wastewater. The Pearson correlation coefficients between GHG emissions and amount of treated wastewater in cities (0.937, P=0.000) reveal strong correlations between the variables at the 0.01 significance level (two-sided).}\label{fig:1}
\end{figure}

Fig. \ref{fig:1} shows China’s WWTP GHG emissions of all cities in 2019. Relatively few cities contribute disproportionately large ratio of the total emissions. Megacities were responsible for 13.6\% of total GHG emissions in 2019, though there are only 7 megacities among 1975 cities and counties, with 8.4\% of China’s population. Greater contaminant removal amount in bigger cities contribute to this imbalance. Beijing, Shenzhen and Shanghai are the top 3 cities in WWTP GHG emissions, treated wastewater volumes and contaminant removal based on six water quality indices (COD, biochemical oxygen demand (BOD), suspended solids (SS), ammonia nitrogen (NH$_3$-N), total nitrogen (TN) and total phosphorus (TP)) (Supplementary Table 2). Greater wastewater volumes in these more developed cities are due to larger populations and higher wastewater treatment rates (Supplementary Fig. 2) because of more developed sewerage networks. Lower outflow concentrations and higher influent concentrations of contaminants (Supplementary Fig. 3) also increase removed amount of the six indices in bigger cities. Lower average outflow concentrations of contaminants are due to stronger implementations of stricter standards in bigger cities (Supplementary Fig. 4). Higher influent concentrations in bigger cities are due to more implementation of separate but not combined sewerage systems (Supplementary Table 3) preventing rainwater inflow, and less pipes in disrepair in bigger cities because of better economy and technology, which reduce groundwater infiltration and wastewater exfiltration. 

\begin{figure*}[t!]
	\centering
	\subfloat[\label{fig:2a}]{
		\includegraphics[scale=0.25]{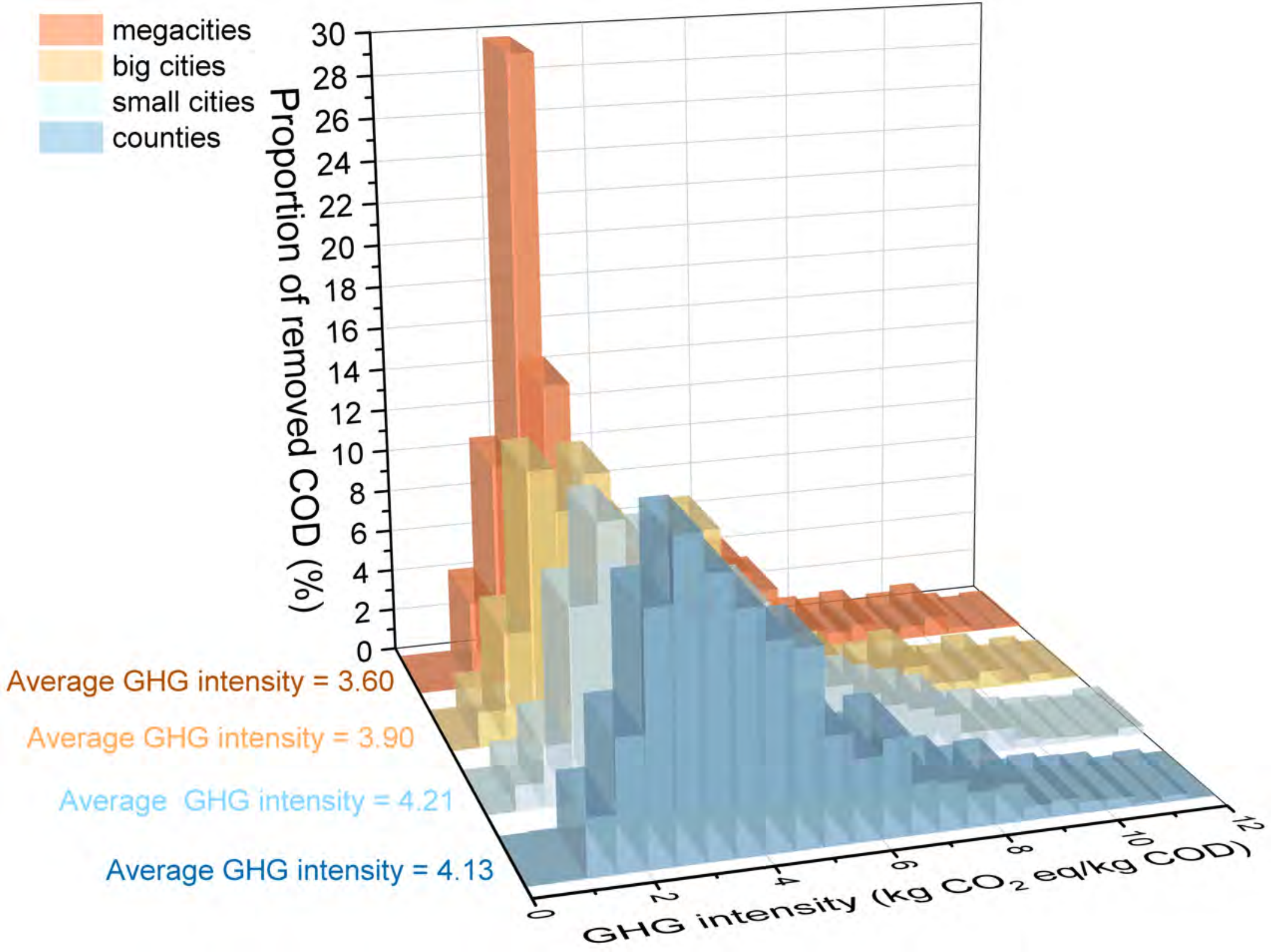}}
	\subfloat[\label{fig:2b}]{
		\includegraphics[scale=0.25]{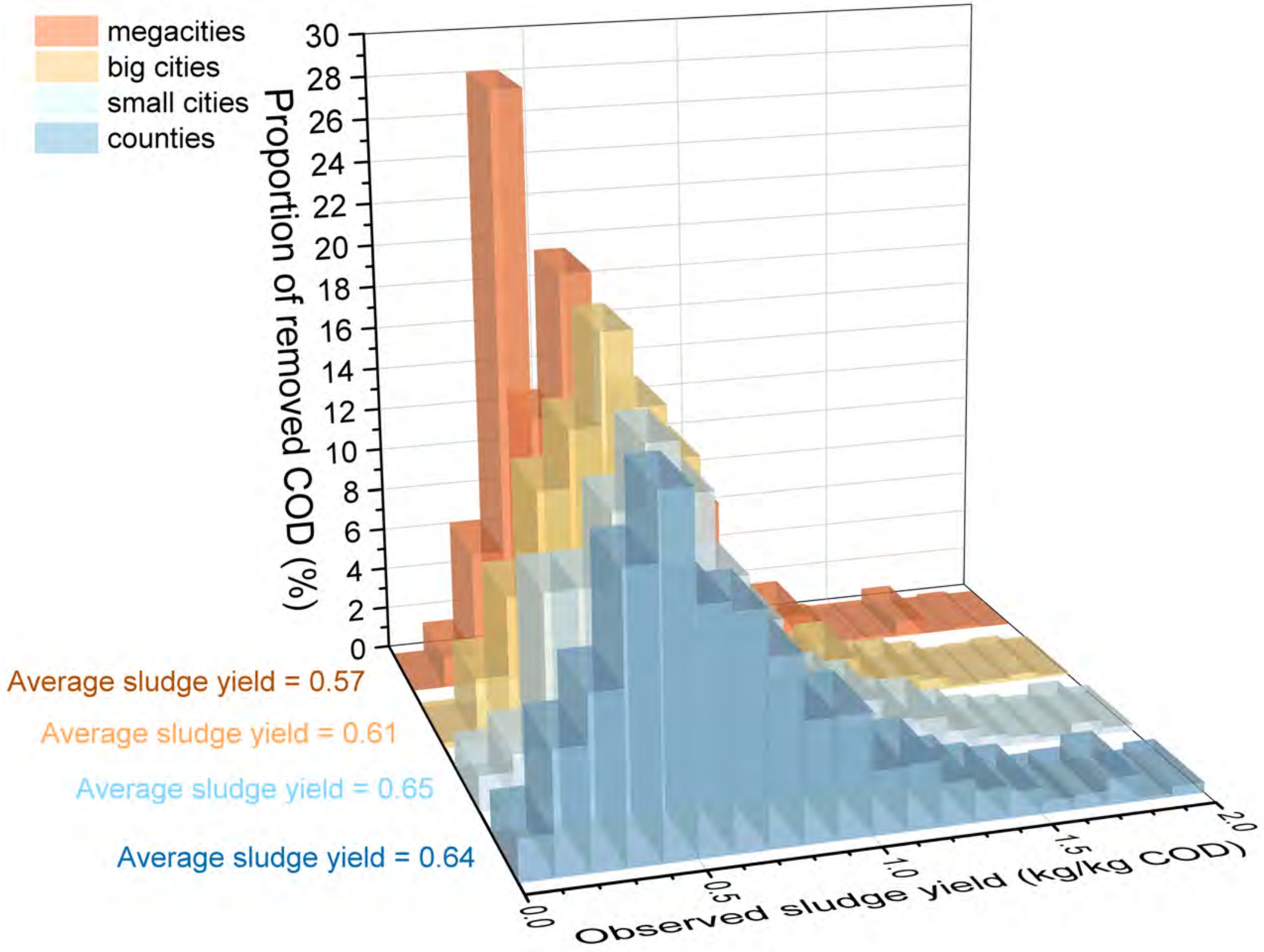}}
	
	\subfloat[\label{fig:2c}]{
		\includegraphics[scale=0.25]{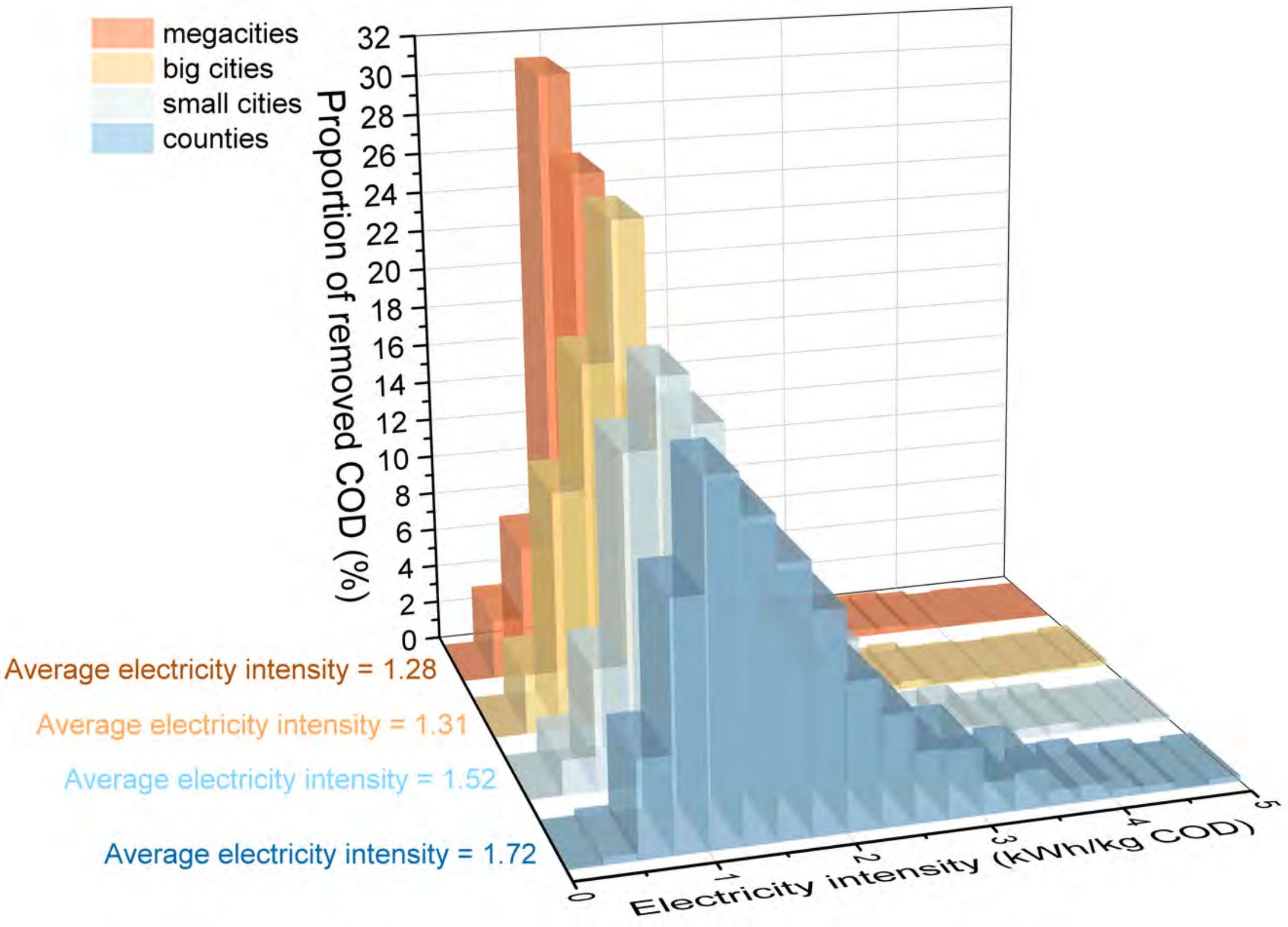}}
	\caption{\textbf{Distribution of WWTP operational data, 2019. a,} GHG intensity, \textbf{b,} sludge yield, \textbf{c,} electricity intensity of WWTPs in different scales of cities and counties. Supplementary Table 4 shows the criteria for city scale classification.}
	\label{fig:2}
	\centering
\end{figure*}

However, GHG intensity (GHG emissions per unit of COD removal) in bigger cities tends to be lower due to lower sludge yield and electricity intensity (see definitions in Methods, Fig. \ref{fig:2}), and relatively higher influent COD concentration. Sludge yield and electricity intensity are both inversely proportional to influent COD concentration (Supplementary Fig. 5). In 2019, average sludge yield and electricity intensity were 0.49 kg/(kg COD removed) and 0.98 kWh/(kg COD removed) in WWTPs with 400-450 mg/L influent COD concentration, and 0.74 kg/(kg COD removed) and 1.83 kWh/(kg COD removed) in WWTPs with 150-200 mg/L influent COD concentration.  

\begin{figure}[h]%
	\centering
	\includegraphics[width=1\textwidth]{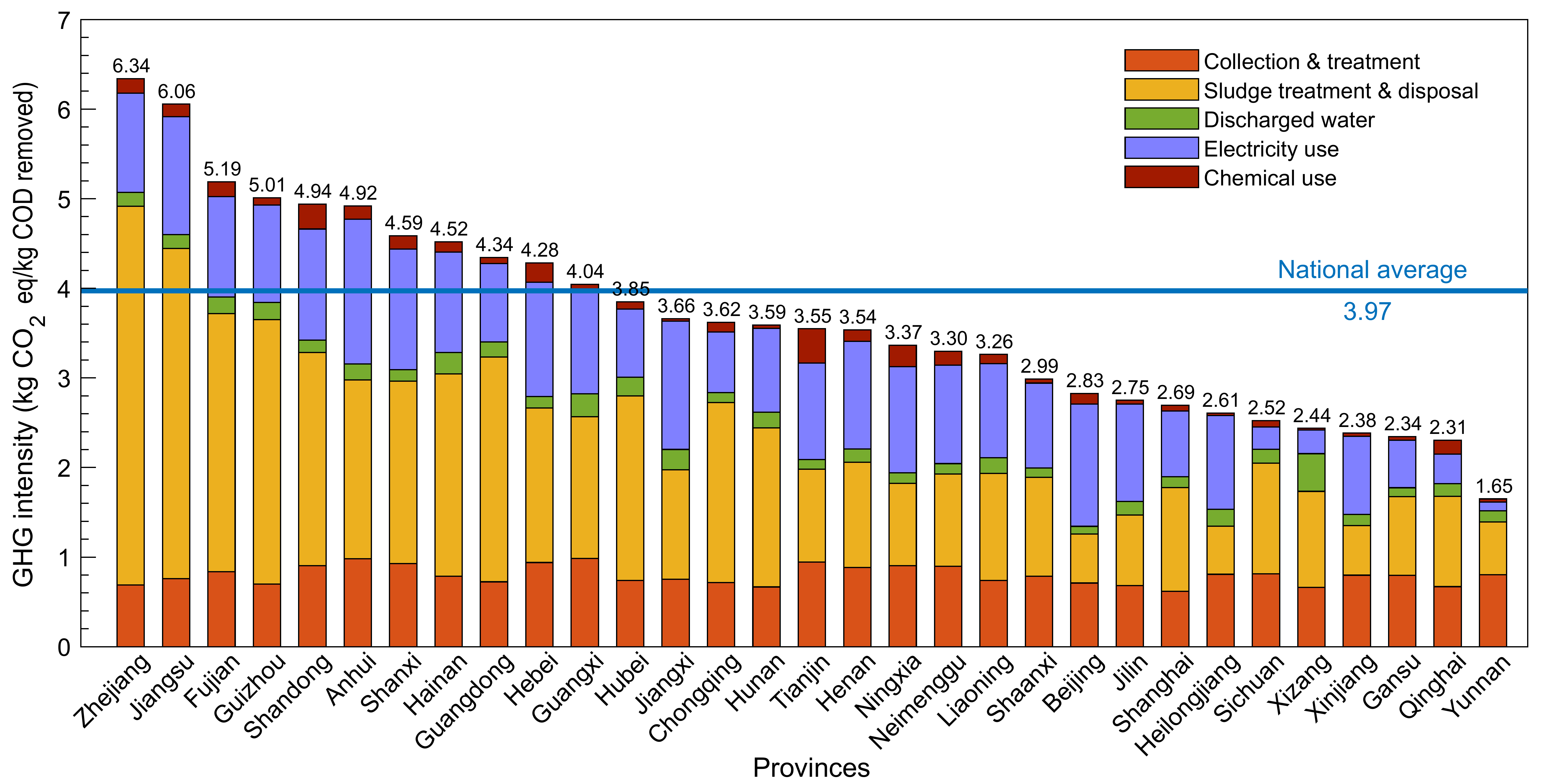}
	\caption{\textbf{GHG intensity of 31 provinces and municipalities in China, 2019.} The provinces are ranked by GHG intensity after dividing the GHG emissions from WWTP operation by the amount of removed COD. Contribution of the GHG sources is shown by different colors.}\label{fig:3}
\end{figure}

GHG intensity also varies across provinces (Fig. \ref{fig:3}). Zhejiang has the highest GHG intensity (6.34 kg CO${_2}$ eq /(kg COD removed)). It is 3.8 times that in Yunnan and 1.6 times the national average. Jiangsu emits the most GHG (6.75 Mt) and ranks the second in GHG intensity (6.06 kg CO${_2}$ eq/(kg COD removed)). STD and electricity use are the main sources for almost all provinces and municipalities. For instance, GHG emissions from STD make up 67\% of total GHG emissions in Zhejiang, and GHG emissions from electricity use contribute the most in Beijing (48\%). 
Intensity factors and emissions factors (see definitions in Methods) jointly affect the GHG intensity of STD and electricity use in different provinces. Regarding intensity factors, upgrading the effluent standard from Class 1B to Class 1A increased sludge yield and electricity intensity by 17\% and 12\% on average (Supplementary Fig. 6). GDP in the provinces is positively related with the proportion of wastewater treated to 1A or a stricter standard (Supplementary Table 5). Therefore, total GHG intensity tends to be higher in provinces with higher GDP. Regarding emissions factors, taking Beijing and Shanghai as an example, GHG intensities for the two municipalities are almost equal, but the contribution from STD in Shanghai is greater than in Beijing, although there is no significant difference in sludge yield (Supplementary Table 6). In Beijing, there is a higher rate of land application of sludge than in Shanghai, where sanitary landfill and incineration are the most common methods of STD and have higher emissions factors (Supplementary Fig. 7). In contrast, GHG intensity from electricity use in Beijing is greater than in Shanghai, with slightly higher electricity intensity (Supplementary Table. 5). Electricity used in Beijing is mainly from thermal power. In comparison, most electricity consumed in Shanghai is from Sichuan (21\%) and Hubei (16\%) (Supplementary Fig. 8), where electricity is mainly from hydropower. 

\subsection{Trends and driving factors of skyrocketing wastewater treatment GHG emissions}\label{subsec2}

\begin{figure*}[t!]
	\centering
	\subfloat[\label{fig:4a}]{
		\includegraphics[scale=0.17]{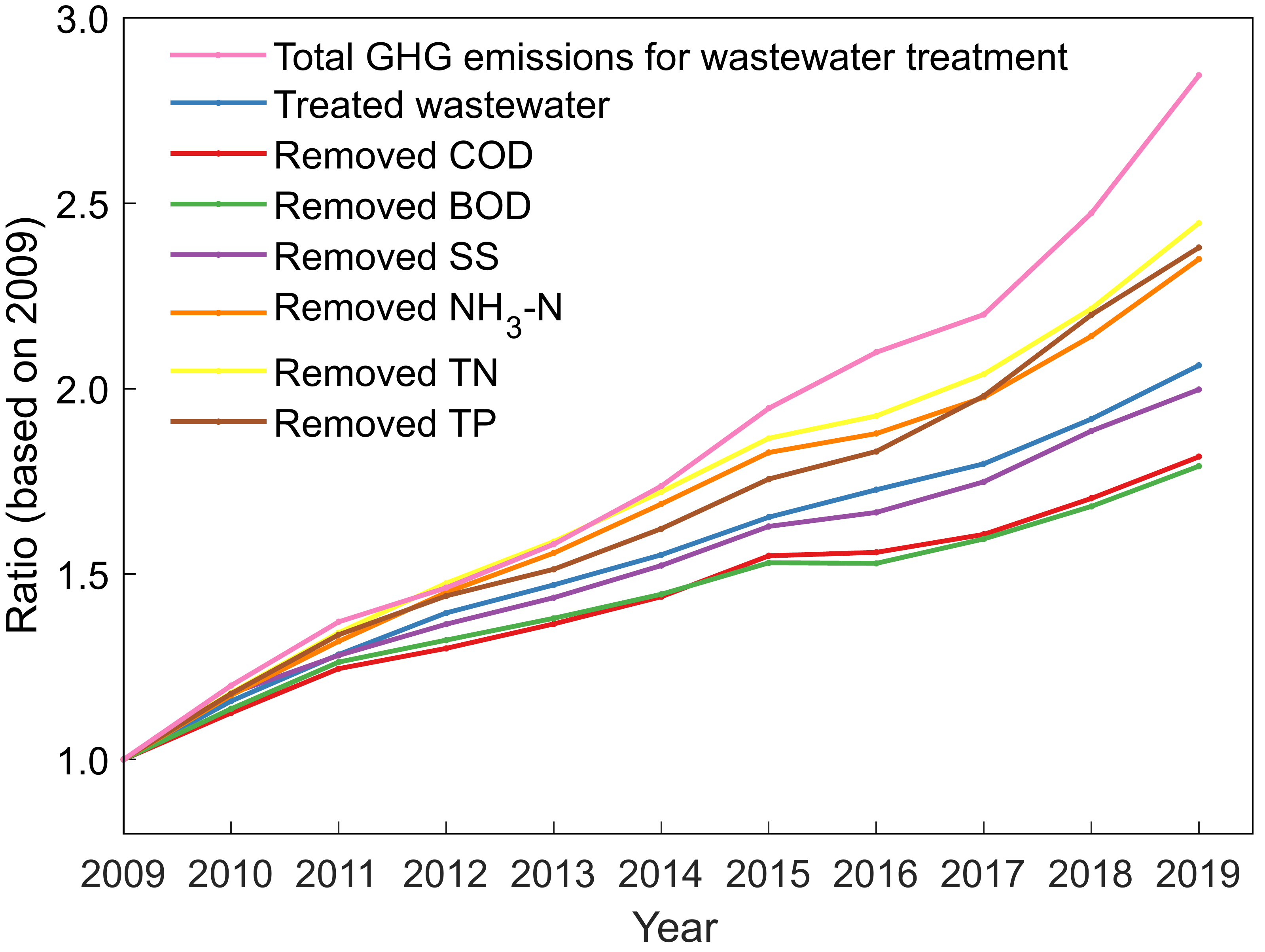}}
	\subfloat[\label{fig:4b}]{
		\includegraphics[scale=0.16]{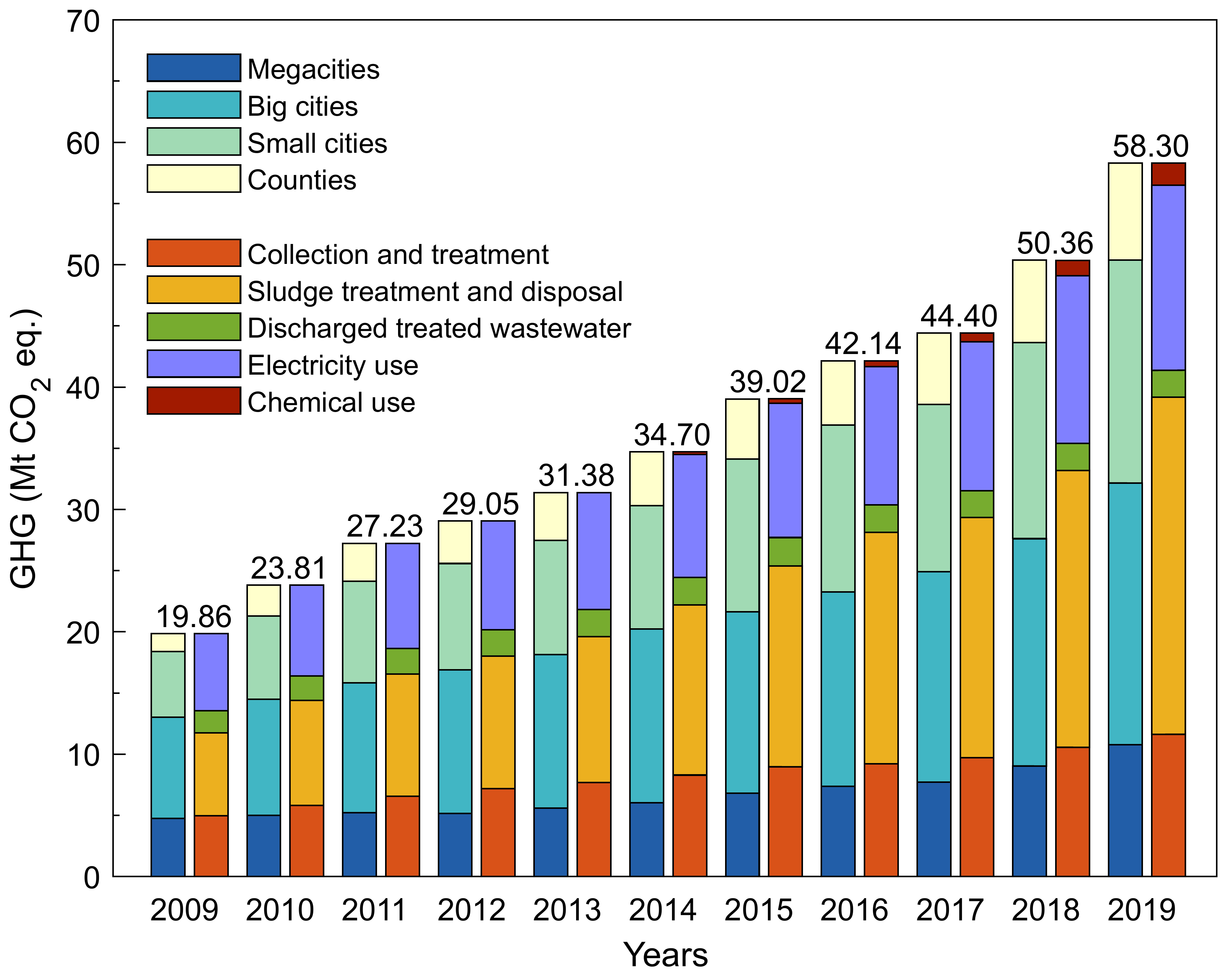}}
	\caption{\textbf{Increasing GHG emissions of China’s wastewater treatment, 2009-2019. a,} Increase in wastewater treatment indices since 2009. Considering the consistency of system boundary and low proportion of emissions from discharged water and chemical use, GHG emissions from discharged water and chemical use is not included in all the years. \textbf{b,} Bar charts showing GHG emissions from four types of cities (left corresponding to every year), and the colors from blue to yellow correspond to the scale of the cities, including 7 megacities, 64 big cities, 572 small cities and 1526 counties. The right bar charts corresponding to every year represent GHG emissions from five GHG sources. For chemical use, only data since 2014 are available.}
	\label{fig:4}
	\centering
\end{figure*}

By 2019, total GHG emissions from all sectors in China increased by 0.3 times compared to emissions in 2009 \cite{unep_emissions_2020}. By comparison, during the same period, WWTP GHG emissions increased by 1.8 times (Fig. \ref{fig:4a}), being significantly quicker than the growth in treated wastewater volume and the reduction in COD, BOD, SS, NH$_3$-N, TN and TP. Therefore, the growth in wastewater treatment GHG emissions is driven by both increased pollutant removal and changes in GHG intensity. Growing pollutant removal can be attributed to increasing wastewater treatment rate and volume.
Increased GHG intensity can be attributed to stricter standards and fewer influent organics. The removed amount of NH$_3$-N, TN and TP increased faster than the volume of treated wastewater because of lower corresponding effluent concentrations due to stricter standards (Supplementary Fig. 9, 10). Thus, GHG intensity of all GHG sources increases \cite{foley_comprehensive_2010}. Lower influent COD and BOD concentration raises the GHG intensity of WWTPs, because it reduced influent COD/TN (and BOD/TN) ratio (Supplementary Fig. 5) and limited the efficiencies of denitrification and biological phosphorus removal \cite{cao_leakage_2019}, or increased the likelihood WWTPs need to add external carbon for nutrient removal. Fig. \ref{fig:4b} shows the GHG emissions composition of cities of different population. From 2009 to 2019, GHG emissions by WWTPs in counties have grown significantly (increasing by 4.4 times), which reflects the development of wastewater treatment facilities in counties. STD and electricity use act as the two fastest increasing GHG emissions sources (growing by 3.1 and 1.4 times). 

\begin{figure}[h]%
	\centering
	\includegraphics[width=1\textwidth]{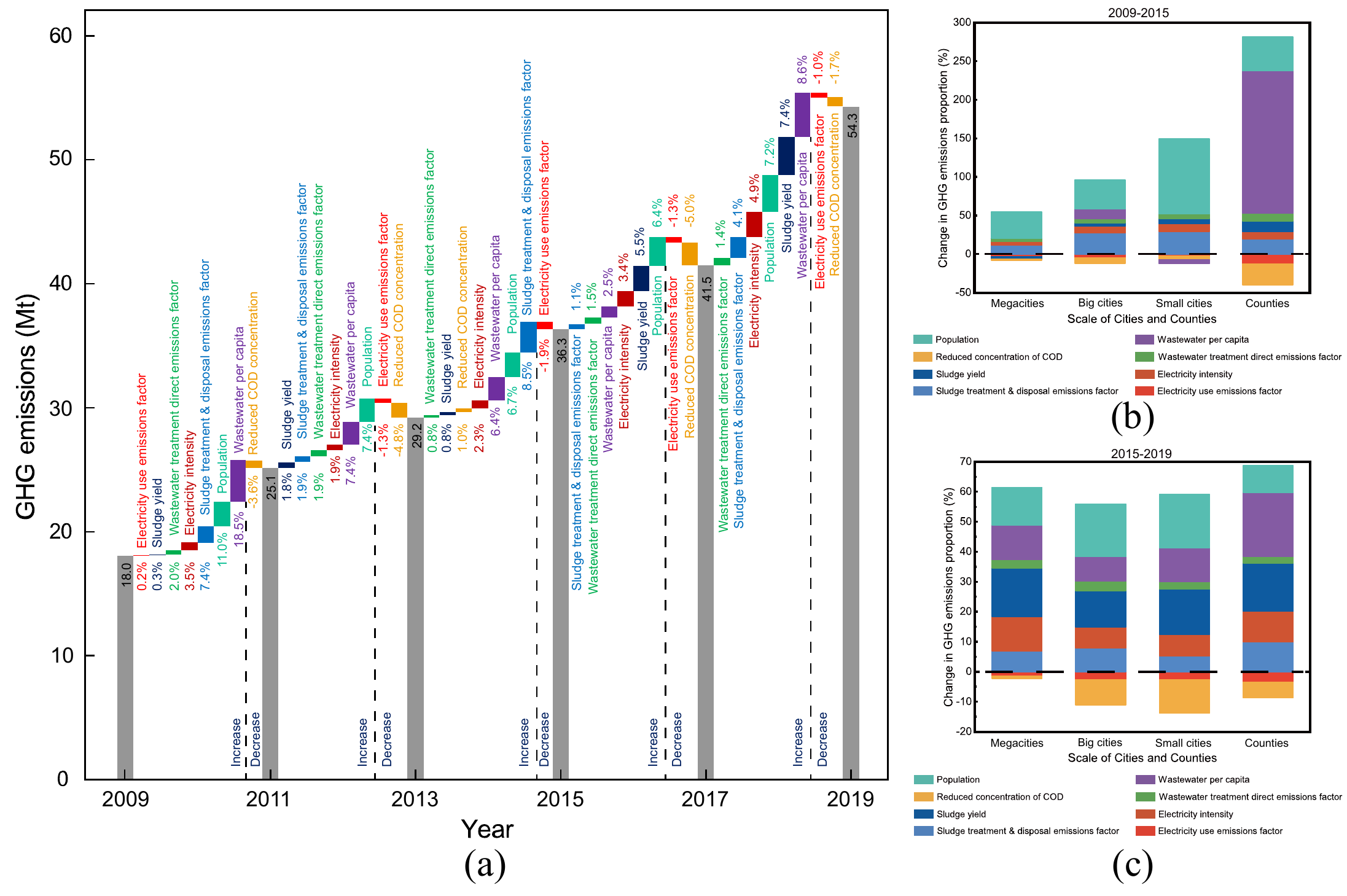}
	\caption{\textbf{Contribution of eight drivers (see definitions in Methods) to the increase in wastewater treatment GHG emissions. a,} During the five periods of 2009-2011, 2011-2013, 2013-2015, 2015-2017 and 2017-2019 in China. \textbf{b,} During 2009-2015, \textbf{c,} During 2015-2019, in cities and counties of different scales. LMDI was applied as an index decomposition analysis to quantify the relative contribution of the variation of the eight factors. The length of each bar reflects the contribution of each driver to total GHG emissions in \textbf{a} and corresponding scale of cities or counties in \textbf{b} and \textbf{c}.}\label{fig:5}
\end{figure}

To further quantify the driving forces of wastewater treatment GHG emissions growth, LMDI analysis is applied, considering eight factors (Fig. \ref{fig:5}). Wastewater per capita has always pushed total GHG emissions upward, increasing from 64.7 m$^3$/capita in 2009 to 94.6 m$^3$/capita in 2019. It was also the most important driver (8.6\%) in the latest period (2017-2019). It can be attributed to an increase in the number and size of wastewater collection and treatment facilities and the rate of wastewater collection and treatment.
The total period 2009-2019 was affected by events in 2015, when the Action Plan for Prevention and Control of Water Pollution was promulgated and implemented. The main driving forces changed significantly in that year. Between 2009 and 2015, population and GHG intensity of STD were the main forces driving the increase in total GHG emissions due to urbanization and popularization of sludge incineration and building materials utilization (Supplementary Fig. 11). Between 2015 and 2019, electricity intensity and sludge yield became important driving factors due to stricter standards, catching up with the pace of the former two main driving forces.
In contrast, driving factors of reduced COD concentration (COD removed per unit water volume) and electricity use emissions factor appeared acting to decrease the total emissions almost all the time. Reduced COD concentration continues to decrease because the influent concentration of COD keeps dropping (Supplementary Fig. 12), indicating growing IIE problems in China’s sewerage \cite{cao_leakage_2019}. However, the decrease of total emissions contributed by decrease of reduced COD concentration cannot offset the increase contributed by electricity intensity and sludge yield. This confirms that although reductions in the COD removal would seem to reduce treatment burden, reductions in COD influent concentration actually leads to decreased operational efficiency of WWTPs. With the popularization of low-carbon energy, the emissions factor of electricity use has been reduced, but to a limited extent. At present, it is difficult to neutralize increasing total GHG emissions caused by various factors.
Further LMDI decomposition is applied for cities and counties of different scales for the period 2009-2015 (Fig. \ref{fig:5}b) and the period 2015-2019 (Fig. \ref{fig:5}c). Between 2009 and 2015, growth in GHG emissions in smaller cities and counties was quicker than in bigger cities. Wastewater per capita acted as the main driving factor in counties (181\%) due to development of wastewater collection and treatment facilities. The driving factor of population contributes greater in cities (73\% of total increase in megacities) than in counties (18\% of total increase), reflecting the effect of population concentrating in bigger cities through urbanization. STD emissions factor was also an important driving factor of GHG emissions in bigger cities, due to rapid transformation in STD methods (Supplementary Fig. 7). Between 2015 and 2019, the difference in the growth rate in emissions for cities and counties of different scales lessened. Sludge yield and electricity intensity became important driving factors in cities and counties of all scales. In counties, wastewater per capita also contributed less in the period 2015-2019 than 2009-2015, indicating that wastewater collection and treatment in counties had undergone most development prior to 2015. Reduced COD concentration dramatically dropped, suppressing GHG emissions growth in big and small cities. This suggests that IIE problems had greater impact in cities of those scales (Supplementary Fig. 12). 

\subsection{Increasing GHG inequality in wastewater treatment}\label{subsec3}

The difference in GHG intensity between WWTPs is a type of inequality and Functional Unit Gini coefficient (FU-Gini) is defined to quantify this inequality. The FU-Gini coefficient of total WWTP GHG inequality in China has increased from 0.20 in 2009 to 0.29 in 2019 (Fig. \ref{fig:5}a). At the provincial level, FU-Gini coefficient increased in 22 and 24 of 31 provinces (including municipalities) in the periods 2009-2015 and 2015-2019 (Supplementary Table 7, 8 and 9). The FU-Gini coefficient of cities and counties of different scales also increased (Supplementary Table 10). Increased inequality in GHG emissions from STD and electricity use led to increased total GHG inequality as they are the main GHG sources. Diversifying STD methods (Supplementary Fig. 11) enhanced the inequalities, as FU-Gini coefficients of GHG emissions from STD increased 0.11 in the period 2009-2019, being 0.24 higher than FU-Gini coefficients of sludge yield in 2019. Difference of emissions factors for electricity use in provinces also enhanced the inequalities. In the period 2013-2015, progress of upgrading effluent standard and increase of electricity intensity was faster in provinces with higher emissions factors of electricity use (Supplementary Table 11). Therefore, FU-Gini coefficients of GHG emissions from electricity use increased 0.03, being 0.05 higher than that of electricity intensity in 2019. Statistical analysis revealed that the inequality of sludge yield and electricity intensity declines with stricter standard in China across space and time (see Supplementary Table 12-15 for details).
FU-Gini coefficients of wastewater collection and treatment are decreasing with growing equal distribution of reduced NH$_3$-N, TN and TP concentration, while average reduced contaminant concentration increased over the years (Supplementary Fig. 9). The inequality in GHG emissions associated with discharged treated wastewater increased before 2017 and then started to decrease. This is because this part of GHG emissions is directly related to effluent quality, and implementation of stricter standards varies between WWTPs due to differences in influent wastewater, technology and economic conditions \cite{zhang_current_2016}. After 2017, effluent quality converged with progress on upgrading to meet stricter standards. The FU-Gini coefficient of GHG emissions from chemical use is relatively big but decreases with time. Chemical use varies among WWTPs, but a growing number of WWTPs have begun to apply chemicals to remove phosphorus or add carbon sources to meet increasingly stringent standards. The effect of chemical use on the inequality in total GHG emissions is limited, because its contribution to total GHG emissions is relatively small. 

\begin{figure}[h]%
	\centering
	\includegraphics[width=1\textwidth]{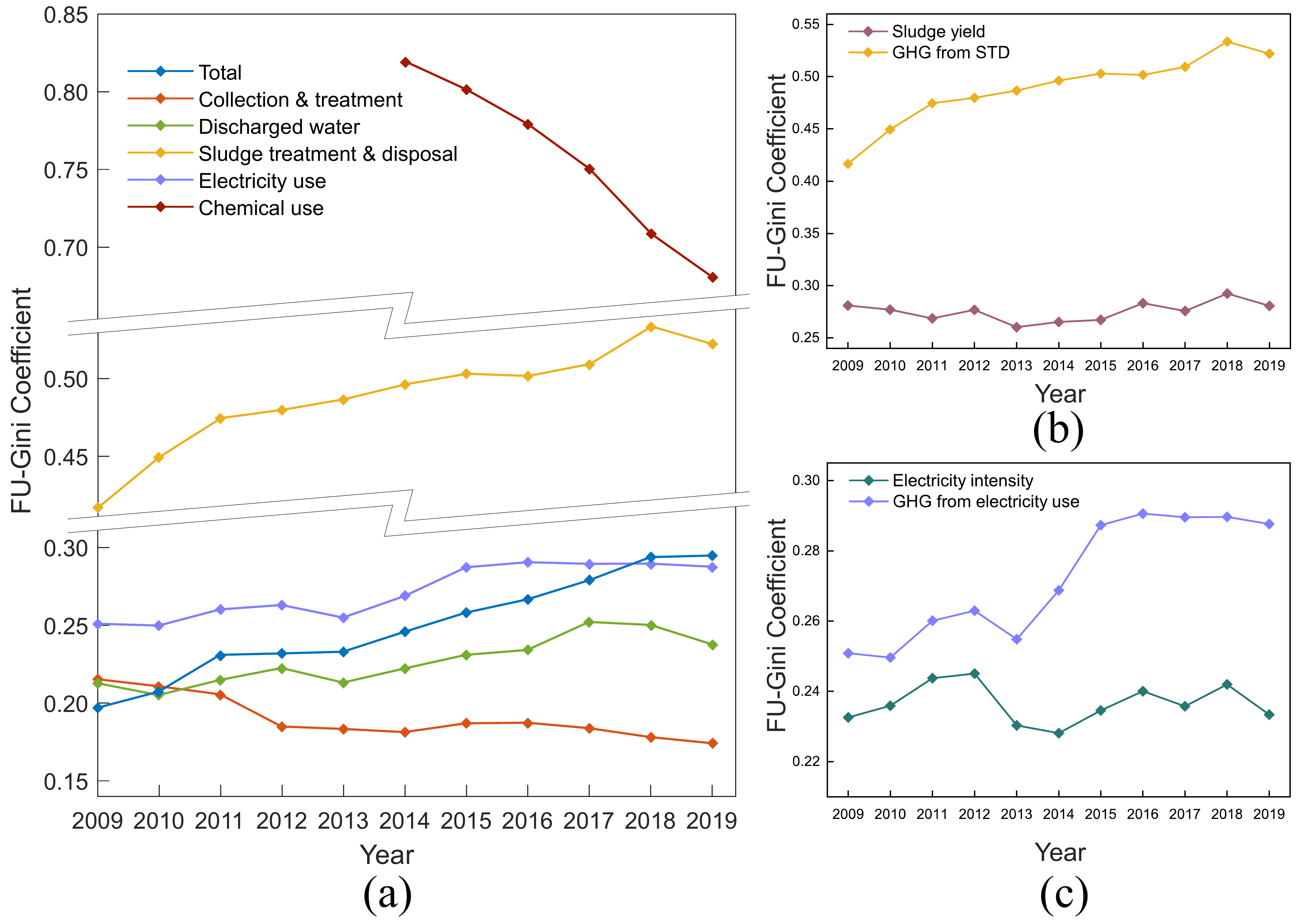}
	\caption{\textbf{FU-Gini coefficients of WWTP GHG intensity in China, 2009-2019. a,} FU-Gini coefficients of five emissions sources. For chemical use, only data since 2014 are available. \textbf{b,} FU-Gini coefficients of sludge yield and GHG emissions from STD. \textbf{c,} FU-Gini coefficients of electricity intensity and GHG emissions from electricity use.}\label{fig:6}
\end{figure}

\section{Discussion}\label{sec3}

To the best of our knowledge, our study is the first to provide a nationwide plant-level insight into WWTP GHG emissions in China. Previous research indicated that the main source of WWTP operating GHG emissions was direct emissions from wastewater treatment (a part of scope 1 emissions) and scope 2 emissions. Using real WWTP operational data, we find that emissions from STD have become an important part of total GHG emissions. STD methods and energy structure for electricity generation are important factors affecting GHG intensity. For provinces mainly using thermal power, electricity consumption should be more strictly controlled. For WWTPs applying sludge incineration or using treated sludge for building material, more attention should be paid to controlling sludge amount.
Stricter effluent standards lead to contradictions between water pollution control and climate change mitigation, but repairing sewerage systems to reduce IIE problems leads to synergies between them. In general, GHG intensity in wealthier provinces tends to be higher, because more WWTPs have been upgraded to advanced processes. Higher sludge yield and electricity intensity is facilitated by the better economic situation and capability found in wealthier areas. Also, water capacity in economically developed provinces tends to be worse \cite{jia_regionalization_2018}, leading to more stringent wastewater treatment requirements. GHG intensity in bigger cities tends to be lower despite higher total GHG emissions, which is mainly due to higher influent contaminant concentration. As IIE problems in China \cite{cao_leakage_2019} caused decreasing influent concentration of WWTPs, solving IIE problems would realize collaborative water pollution prevention and control, and wastewater treatment GHG emissions control. Higher influent contaminant concentration can increase WWTP operation efficiency \cite{niu_energy_2019}. Higher influent COD/TN ratio combined with higher influent COD raises TN removal efficiency \cite{ge_enhanced_2010}, and sludge yield may become lower \cite{liu_kinetic_1999}, decreasing GHG emissions from STD. Wastewater dilution by abnormal influent and infiltration increases hydraulic load, reducing wastewater treatment efficiency \cite{dirckx_groundwater_2016} and increasing energy use for pumping.
In the period 2009-2019, WWTPs in China removed more pollutants, while WWTP GHG emissions increased significantly quicker than total GHG emissions in China. With the continuous upgrading wastewater effluent standards, the intensity factors (sludge yield and electricity intensity) have become more important drivers leading to the growth of total GHG emissions. Stricter standards lead biological sludge to experience more nitrification and denitrification to remove more NH$_3$-N and TN, and adsorb more phosphorus compounds to reduce TP. If WWTPs apply dephosphorization reagents to reduce TP, more chemical sludge would be generated. Reaching stricter standards also leads to stricter technology control such as aeration, and more implementation of advanced technologies, increasing electricity use. Moreover, although China's power sector was undergoing low carbon transformation, its restraining effect for GHG emissions growth was relatively insignificant. Therefore, policies toward technologies and effluent standards of WWTPs are of great importance for GHG emissions control. 
Governments need to pay more attention to GHG inequality of WWTPs when considering climate change mitigation actions, as it benefits for effective and fair management for WWTP GHG emissions in China. Wastewater treatment GHG inequality has increased with GHG emissions growth in China’s provinces and cities. The diversification of STD methods and the low-carbon transition of energy structure led to higher GHG inequality (i.e., diverging GHG intensity). Upgrading process led WWTPs to apply more similar technologies and more standardized management, thus the distribution of sludge yield and electricity intensity becomes more equal, which benefits decreasing GHG inequality.
As China is promoting realizing its carbon peak, the proportion of wastewater treatment GHG emissions would increase rapidly if not controlled. The Central Economic Work Conference in December 2020 underlined coordinated pollution and GHG reduction. We suggest (1) a more complete trade-offs mechanism between stricter effluent standards and GHG emissions growth, (2) taking increasing influent contaminant concentration as an important way for collaboratively reducing water pollution and GHG emissions, constructing more separated sewerage system, and rehabilitating sewerage systems with serious IIE problems, (3) Responsibility of WWTPs GHG emissions control be considered during WWTPs upgrading according to local conditions including economy and technology level, STD methods, energy structure for electricity use, influent wastewater quality, local water environment carrying capacity, need for water reuse, etc., (4) the gap in WWTP GHG intensity can be narrowed by simultaneously upgrading the effluent standard in WWTPs with poorer effluent quality and reduce the GHG intensity of WWTPs with high GHG intensity. (5) promote energy recovery technologies in suitable condition \cite{smith_can_2018}, such as thermal and chemical energy recovery \cite{hao_energy_2019}, anaerobic sludge digestion \cite{yan_net-zero-energy_2017}, and STD method of land application \cite{yang_current_2015}.

\section{Methods}\label{sec4}

\subsection{Data Sources}\label{subsec1}
Operational data for WWTPs is obtained from the National Urban Wastewater treatment management information portal. This portal contains detailed monthly operational data for WWTPs in 27 provinces and 4 municipalities (i.e., 31 areas) of China, except Hong Kong, Macao and Taiwan. Monthly data is available for the period from 2009 to 2019. The database included 5458 WWTPs in cities and counties in 2019. Cities are further classified into 3 scales according to their population, and the thresholds are in Supplementary Table 4.
Six water quality indices are used for identifying the standard of WWTP effluent (Supplementary Table 1). These are chemical oxygen demand (COD), biochemical oxygen demand (BOD), suspended solids (SS), ammonia nitrogen (NH$_3$-N), total nitrogen (TN), and total phosphorus (TP). The standards stipulated in China also include other indices, such as animal and vegetable oils, petroleum, etc. These additional indices are not recorded in the database, and the six known indexes are the most representative indices for wastewater. Therefore, other indices are not considered in this study. Effluent wastewater for a particular WWTP is considered to reach a standard when all six indices meet the standard.

\subsection{GHG emissions inventories}\label{subsec2}
This research applies bottom-up procedures to estimate GHG emissions of WWTP, i.e., calculate the operating GHG emissions every month, and then sum up. The system boundary for wastewater treatment (Supplementary Fig. 13) shows the processes that produce GHG emissions. There are three scopes for GHG accounting \cite{ranganathan_greenhouse_2015}. Scope 1 refers to direct GHG emissions from owned or controlled sources. Scope 2 refers to indirect GHG emissions from the generation of consumed electricity. Scope 3 refers to indirect GHG emissions from sources not owned or controlled. Five GHG sources are considered from the three scopes, they are:
(1) Wastewater collection and treatment (scope 1),
(2) Discharged treated wastewater (scope 1), 
(3) Sludge treatment and disposal (STD) (most are direct emissions belonging to scope 1 \cite{liu_life_2013}, electricity consumption excluded),
(4) Electricity use (scope 2, including electricity consumption of STD),
(5) Chemical use (scope 3). 
GHG emissions from the construction phase of WWTPs takes up about 10-35\% of life-cycle GHG emissions of WWTPs \cite{mo_can_2012,tangsubkul_life_2005}, which are often neglected \cite{maktabifard_energy_2020}. GHG emissions from the demolishing and disposal phase are estimated to be negligible, and total GHG emissions were found to be mainly related with GHG emissions from operation phase \cite{lundie_life_2004}. Therefore, only the operational phase is included in the system boundary. The interprovincial electricity transmission loss is not accounted for because the data is not available, and the loss is not more than 3\% \cite{zhang_modeling_2020} so the effect is limited. As electricity consumption takes up more than 80\% of energy use in approximately 86\% of WWTPs in the US \cite{carlson_energy_2007}, energy use other than electricity consumption is not taken into account. Reclaimed water reuse is not accounted for because only 16\% of treated wastewater was reused in 2017 \cite{china_urban_water_association_urban_2016}, and it also need to be pumped with electricity consumption for water supply, which is also the most important GHG emissions source for clean-water supply \cite{smith_contribution_2016}.
The GHG emissions inventories are constructed virtually following the life-cycle thinking method and ISO 14040/44 methodological guidelines \cite{iso_iso_2006}, which uses the corresponding emissions factor for bottom-up calculation (equation (\ref{equation:1})).

\begin{equation}
	\text{EF}_{k} = \sum_{i}^{I}{\sum_{j}^{J}{{(AD}_{i,k} \times \text{CF}_{i,j,k} \times \text{GWP}_{j})}}\label{equation:1}
\end{equation}

Where \emph{EF\textsubscript{k}} denotes the GHG emissions of process
\emph{k} (kg CO\textsubscript{2} eq.); \emph{AD\textsubscript{i,k,l}}
represents the activity amount of type \emph{i} in process \emph{k} (MWh
for electricity use and kg AD for other emissions);
\emph{CF\textsubscript{i,j,k}} indicates the type \emph{j} GHG
(CH\textsubscript{4}, N\textsubscript{2}O or CO\textsubscript{2} eq.)
life-cycle emission factor of activity of type \emph{i} in process
\emph{k} (kg GHG/(MWh or kg AD)); \emph{GWP}\textsubscript{j} is the
global warming potential of type \emph{j} GHG (kg CO\textsubscript{2}
eq./kg GHG), and Supplementary Table 16 shows it.

The specific GHG sources decide specific meaning of
\emph{AD\textsubscript{i,k}} and different
\emph{CF\textsubscript{i,j,k}} in each process: for GWP source (1) and
(3), WWTPs use different technologies (Supplementary Table 17); for GHG
source (2), destination of WWTPs' discharge varies (Supplementary Table
18); for GWP source (4), the energy structure differs between
provinces \cite{china_electric_power_press_china_2018}, and \emph{CF\textsubscript{i,j,k}} of
power generation technologies varies (Supplementary Table 19); for GWP
source (5), every WWTP uses a specific type and amount of
chemical.

CH\textsubscript{4} and N\textsubscript{2}O emissions are the source of
GHG emissions in wastewater collection and treatment, and in discharge
of treated wastewater. Carbon dioxide (CO\textsubscript{2}) emissions
are mostly generated from biogenic organic matter in human excreta or
food waste, so they are not considered \cite{ipcc_2019_2019}. During
wastewater collection, CH\textsubscript{4} is generated in sewerage
systems and emitted from structures of WWTPs, thus it is included in
emission factors of wastewater treatment process.

For electricity use in China, interprovincial electricity transmission
takes up 17.6\% of total generated electricity in 2017. For provinces,
the rate of imported electricity from other provinces can be from 0.2\%
(Tibet) to 62.3\% (Beijing) \cite{china_electricity_council_annual_2018}. Therefore, the
electricity used by the WWTP comes from both the province it locates and
other provinces. The electricity imported from other provinces is
assumed to be generated by corresponding province (i.e., no secondary
electricity transmission). The Annual Compilation of Statistics of Power
Industry provides the data of electricity generation
and use in China, 2017 \cite{china_electricity_council_annual_2018}. It also reveals the interprovincial electricity
transmission in 2017. Nevertheless, some transmission data are not
between provinces, but between regional grids or between a regional grid
and a province, and this kind of data is disaggregated. Supplementary
Table 20 shows the name of the regional grids, the provinces they
include and the abbreviations of the provinces. For a regional grid
importing electricity, the disaggregation is based on the share of power
consumption of the provinces served by this regional grid; in the case
of exporting electricity, the electricity is disaggregated by referring
to the proportion of power generation of the corresponding provinces.
Supplementary Fig. 8 shows the interprovincial electricity transmission
in 2017. Based on weighting method, the final emission factor
\emph{CF\textsubscript{k}} of electricity use in a province is
calculated by equation (\ref{equation:2}) and (\ref{equation:3}). As GHG type of electricity use in the
data are all CO\textsubscript{2} eq., \emph{GWP\textsubscript{i}} is 1
and not accounted.

\begin{equation}
	\text{CF}_{k,self} = \sum_{i}^{I}{(\text{CF}_{i} \times R_{k,i})}\label{equation:2}
\end{equation}

\begin{equation}
	\(\text{CF}_{k} = \frac{\text{CF}_{k,self} \times \left( E_{k,generated} - E_{k,exported} \right) + \sum_{j}^{S\backslash k}{{(CF}_{j,self}*E_{k,j})}}{\left( E_{k,generated} - E_{k,exported} \right) + \sum_{j}^{S\backslash k}E_{k,j}}\)\label{equation:3}
\end{equation}

Where \emph{CF\textsubscript{k,self}} denotes the emission factor of
generating electricity in province \emph{k} (kg CO\textsubscript{2}
eq./MWh); \emph{CF\textsubscript{i}} represents the emission factor of
power generation technology \emph{i} (kg CO\textsubscript{2} eq./MWh);
\emph{R\textsubscript{k,i}} is the proportion of electricity generated
by technology \emph{i} in province \emph{k}. \emph{CF\textsubscript{k}}
indicates the final emission factor of electricity use of province
\emph{k} (kg CO\textsubscript{2} eq./MWh);
\emph{E\textsubscript{k,generated}} is the generated electricity by
province \emph{k} (MWh); \emph{E\textsubscript{k,exported}} is the
exported electricity generated by province \emph{k} (MWh);
\emph{E\textsubscript{k,j}} is the amount of electricity transmitted
from province \emph{j} to province \emph{k}; \emph{S} refers to the set
of China's provinces.
Shannon’s diversity index \cite{shannon_mathematical_1948} is applied to quantify the diversity of STD methods based on equation (\ref{equation:4}).
\begin{equation}
	H = -\sum_{i=1}^{I}{p_{i}\log_{I}{p_{i}}}\label{equation:4}
\end{equation}
where \emph{H} is the Shannon’s diversity index, \emph{p$_{i}$} denotes the proportion of sludge (in dry mass) treated and disposed with the \emph{i} th method (see all methods in Supplementary Table 22). The higher the \emph{H}, the more diversified the STD methods are.

\subsection{LMDI}\label{subsec3}
Both the increase in volume of treated wastewater
(quantity indicator) and GHG intensity (intensity indicator) promote GHG
emissions to grow. To quantify and compare different driving forces,
LMDI is applied to conduct index decomposition analysis. LMDI can
decompose the change of a variable into quantity indicators and
intensity indicators related with the variable \cite{ang_lmdi_2015}. This
method was commonly used for studies on GHG emissions to assist policy
decision-making \cite{guan_structural_2018}. In this study, wastewater treatment
GHG emissions of China are decomposed as in equation (4). As GHG
emission sources of low proportion, discharged treated wastewater and
chemical use are not included.

\begin{equation}
\centering
\begin{aligned}
	\text{GHG}_{\text{total}} &= \text{GHG}_{\text{treat}} + \text{GHG}_{\text{sludge}} + \text{GHG}_{\text{elec}}\\
	&=P \times \frac{V}{P} \times \frac{M}{V} \times \left( \frac{\text{GHG}_{\text{treat}}}{M} + \frac{S}{M} \times \frac{\text{GHG}_{\text{sludge}}}{S} + \frac{E}{M} \times \frac{\text{GHG}_{\text{elec}}}{E} \right)\\
	&=P \times Q \times C \times \left( F_{\text{treat}} + I_{s\text{ludge}} \times F_{\text{sludge}} + I_{\text{elec}} \times F_{\text{elec}} \right)\
\end{aligned}
\centering
\end{equation}

Where \emph{GHG}\textsubscript{total} is the total wastewater treatment
GHG emissions in China; \emph{GHG}\textsubscript{treat},
\emph{GHG}\textsubscript{sludge} and \emph{GHG}\textsubscript{elec} are
GHG emissions from wastewater collection and treatment (in direct ways),
STD, and electricity use; \emph{V} is the amount of treated wastewater;
\emph{M} is the removed amount of COD; \emph{S} is the amount of treated
and disposed sludge; \emph{E} is electricity consumption amount. Thus,
\emph{GHG}\textsubscript{total} is further represented by three quantity
factors and six intensity indicators.

Three quantity factors are related with the amount of contaminants:

\emph{P} is population;

\emph{Q} = \emph{V}/\emph{P} is amount of wastewater per capita,
reflecting combined effect by habit of people, wastewater collection and
treatment rate, ratio of mixed industrial wastewater, and inflow /
infiltration / exfiltration condition of sewerage
systems \cite{cao_leakage_2019,mattsson_normalization_2016}. It determines the amount of wastewater to
treat together with population;

\emph{C} = \emph{M}/\emph{V} is the average reduced concentration of
COD, representing the condition of removing contaminant in wastewater;

Two Intensity factors are related with amount of electricity use and
sludge to treat and dispose per unit of COD:

\emph{I}\textsubscript{sludge} = \emph{S}/\emph{M} is treated and
disposed sludge per unit of removed COD, measuring the observed sludge
yield during wastewater treatment.

\emph{I}\textsubscript{elec} = \emph{E}/\emph{M} is electricity use per
unit of removed COD, i.e., electricity intensity. It indicates the
energy intensity.

Three emissions factors:

\emph{F}\textsubscript{treat} = \emph{GHG}\textsubscript{treat}/\emph{M}
is wastewater treatment direct emissions factor, from wastewater
collection and treatment. It mainly reflects the ratio of COD and TN
removal, as most of the emission factors of CH\textsubscript{4} and
N\textsubscript{2}O are the same.

\emph{F}\textsubscript{sludge} =
\emph{GHG}\textsubscript{sludge}/\emph{S} is STD emissions factor, from
wastewater collection and treatment, reflecting the variation of ratio
of different STD approaches.

\emph{F}\textsubscript{elec} = \emph{GHG}\textsubscript{elec}/\emph{E}
is electricity use emissions factor, reflecting the changes of energy
structure in the provinces.

Therefore, the change of the total wastewater treatment GHG emissions in
China in year \emph{t} compared with that of year \emph{t}-1 can be
represented by equation (5).

\begin{equation}
	\centering
	\begin{aligned}
		\mathrm{\Delta}\text{GHG}_{\text{total}} &= \sum_{i}^{}{L\left( \text{GHG}_{i}^{t},\text{GHG}_{i}^{t - 1} \right)\ln\left( \frac{P^{t}}{P^{t - 1}} \right)}\\
		&+ \sum_{i}^{}{L\left( \text{GHG}_{i}^{t},\text{GHG}_{i}^{t - 1} \right)\ln\left( \frac{Q^{t}}{Q^{t - 1}} \right)}\\
		&+ \sum_{i}^{}{L\left( \text{GHG}_{i}^{t},\text{GHG}_{i}^{t - 1} \right)\ln\left( \frac{C^{t}}{C^{t - 1}} \right)}\\
		&+ L\left( \text{GHG}_{\text{treat}}^{t},\text{GHG}_{\text{treat}}^{t - 1} \right)\ln\left( \frac{F_{\text{treat}}^{t}}{F_{\text{treat}}^{t - 1}} \right)\\
		&+ L\left( \text{GHG}_{\text{sludge}}^{t},\text{GHG}_{\text{sludge}}^{t - 1} \right)\left\lbrack \ln\left( \frac{I_{\text{sludge}}^{t}}{I_{\text{sludge}}^{t - 1}} \right) + \ln\left( \frac{F_{\text{sludge}}^{t}}{F_{\text{sludge}}^{t - 1}} \right) \right\rbrack\\
		&+ L\left( \text{GHG}_{\text{elec}}^{t},\text{GHG}_{\text{elec}}^{t - 1} \right)\left\lbrack \ln\left( \frac{I_{\text{elec}}^{t}}{I_{\text{elec}}^{t - 1}} \right) + \ln\left( \frac{F_{\text{elec}}^{t}}{F_{\text{elec}}^{t - 1}} \right) \right\rbrack\\
		&= \mathrm{\Delta}\text{GHG}_{P} + \mathrm{\Delta}\text{GHG}_{Q} + \mathrm{\Delta}\text{GHG}_{C} + \mathrm{\Delta}\text{GHG}_{F_{\text{treat}}}\\
		&+ \mathrm{\Delta}\text{GHG}_{I_{s\text{ludge}}} + \mathrm{\Delta}\text{GHG}_{F_{\text{sludge}}} + \mathrm{\Delta}\text{GHG}_{I_{\text{elec}}} + \mathrm{\Delta}\text{GHG}_{F_{\text{elec}}}
	\end{aligned}
	\centering
\end{equation}

Here, \emph{i} represents each GHG source, including wastewater
collection and treatment (direct emission), STD, and electricity use.
\(L\left( \text{GHG}_{i}^{t},\text{GHG}_{i}^{t - 1} \right) = {(GHG}_{i}^{t} - \text{GHG}_{i}^{t - 1})/\left\lbrack \ln\left( \text{GHG}_{i}^{t} \right),\ln\left( \text{GHG}_{i}^{t - 1} \right) \right\rbrack\)
is the logarithmic mean weight. \(\mathrm{\Delta}\text{GHG}_{P}\),
\(\mathrm{\Delta}\text{GHG}_{Q}\), \(\mathrm{\Delta}\text{GHG}_{C}\),
\(\mathrm{\Delta}\text{GHG}_{F_{\text{treat}}}\),
\(\mathrm{\Delta}\text{GHG}_{I_{s\text{ludge}}}\),
\(\mathrm{\Delta}\text{GHG}_{F_{\text{sludge}}}\),
\(\mathrm{\Delta}\text{GHG}_{I_{\text{elec}}}\),
\(\mathrm{\Delta}\text{GHG}_{F_{\text{elec}}}\)changes because of
population growth, more wastewater per capita, variation of reduced COD
concentration, the ratio of COD and TN removal, generating intensity of
sludge, variation of ratio of different STD approaches, electricity
intensity effect, and changes of energy structure in the provinces,
respectively.

\subsection{Functional unit based Gini coefficient (FU-Gini)}\label{subsec4}
Gini coefficient \cite{gini_measurement_1921} is a common indicator quantifying the
difference in the income distribution. The basic income Gini coefficient
is represented by equation (6).

\begin{equation}
	G = \sum_{k = 1}^{K}{D_{k}Y_{k}} + 2\sum_{k = 1}^{K}{D_{k}{(1 - P}_{k})} - 1
\end{equation}

Where \emph{G} is the Gini Coefficient, and \emph{T} is the Theil index.
\emph{K} denotes the number of groups, and \emph{n} denotes the number
of individuals. \emph{D\textsubscript{k}}, \emph{Y\textsubscript{k}} and
\emph{P\textsubscript{k}} respectively represent the proportion of the
population, the proportion of income, and the cumulative income
proportion of group \emph{k} (\emph{k}=1, 2, 3, \ldots, \emph{K}), and
\emph{Y\textsubscript{i}} is the income proportion of individual
\emph{i} (\emph{i}=1, 2, 3, \ldots, \emph{n}). Gini Coefficient range
from zero to one. Zero means absolutely equal while one means completely
unequal, and larger value means more inequality.

\emph{FU-Gini and WWTP-Gini.} Some researches replace the income with
carbon emission to figure out inequality of it, but most of them focus
on the inequality among people. To quantify the GHG emissions inequality
of WWTPs, a straightforward idea is to regard a WWTP as an individual
(WWTP-Gini). But this method ignores the scale difference of WWTPs,
making the result deviate from the actual inequality. To make an extreme
assumption, if there are only 2 WWTPs in a city, one removes
1×10\textsuperscript{5} t COD a year, and the other removes 2000 t COD a
year, and the GHG intensity of the latter is 4 times that of the former.
Taking each WWTP as an individual, the Gini coefficient is 0.3. However,
as the former WWTP contributes 92.6\% of total GHG emissions, the latter
WWTP's contribution to inequality is quite limited. As the scale
corresponding to the functional unit matters more, FU-Gini is defined by
regarding one functional unit as an abstract individual. The variables
related with population in the equations (6) - (7) are replaced with the
corresponding variable related with the amount of functional unit (In
this paper's case, population is replaced with removed COD amount). In
the 2 WWTPs-city assumption, the FU-Gini is 0.05, much lower than the
Gini coefficient of 0.3 regarding each WWTP as an individual. The
FU-Gini is more reliable for policy design and feasibility of agreement.

\begin{appendices}

\section{Section title of first appendix}\label{secA1}

An appendix contains supplementary information that is not an essential part of the text itself but which may be helpful in providing a more comprehensive understanding of the research problem or it is information that is too cumbersome to be included in the body of the paper.




\end{appendices}


\bibliography{my_bibliography}


\end{document}


\includegraphics[width=4.11506in,height=2.49261in]{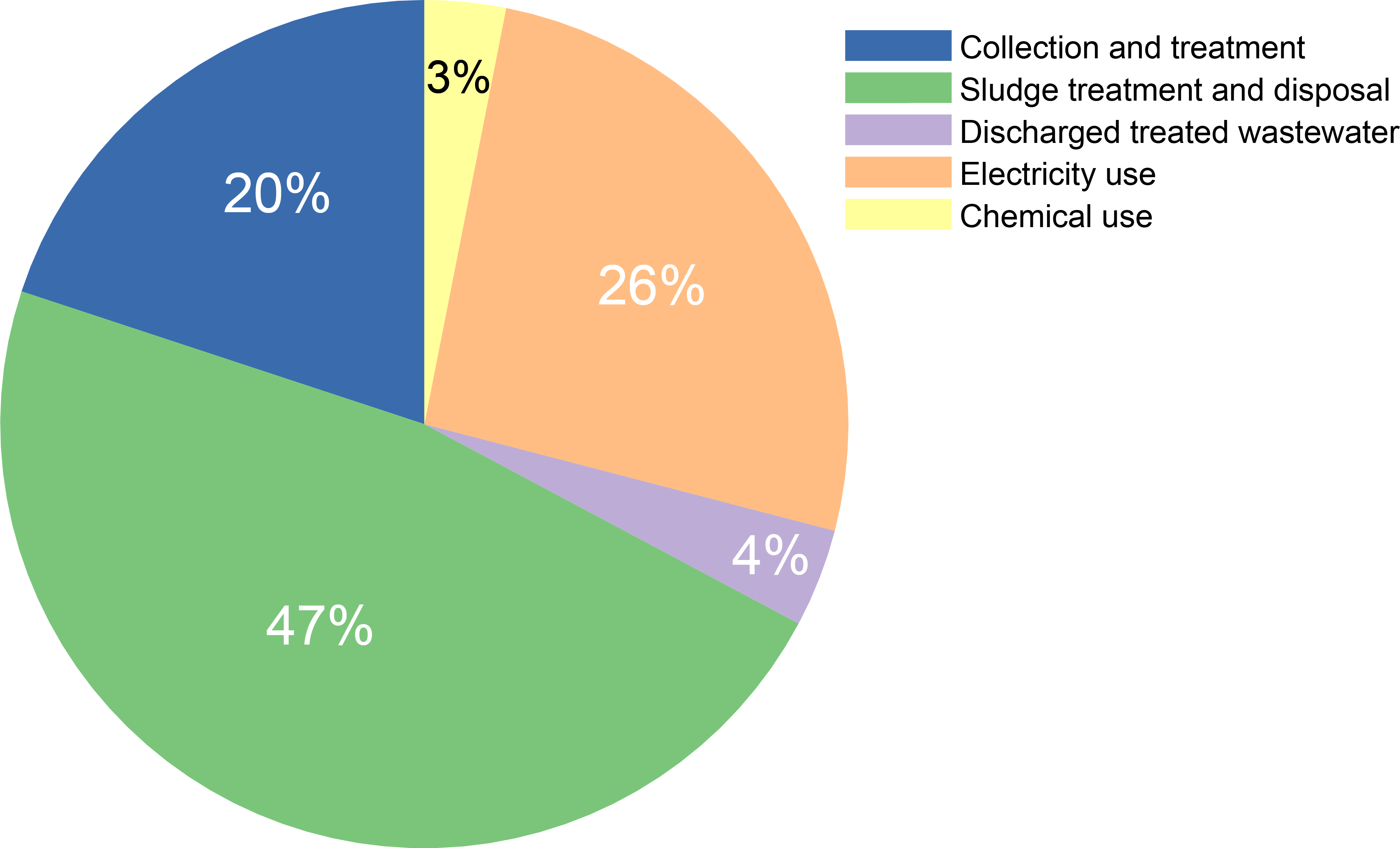}

\textbf{Supplementary Fig.} \textbf{1} Breakdown of China's WWTP GHG
emissions.

\includegraphics[width=3.36905in,height=2.49981in]{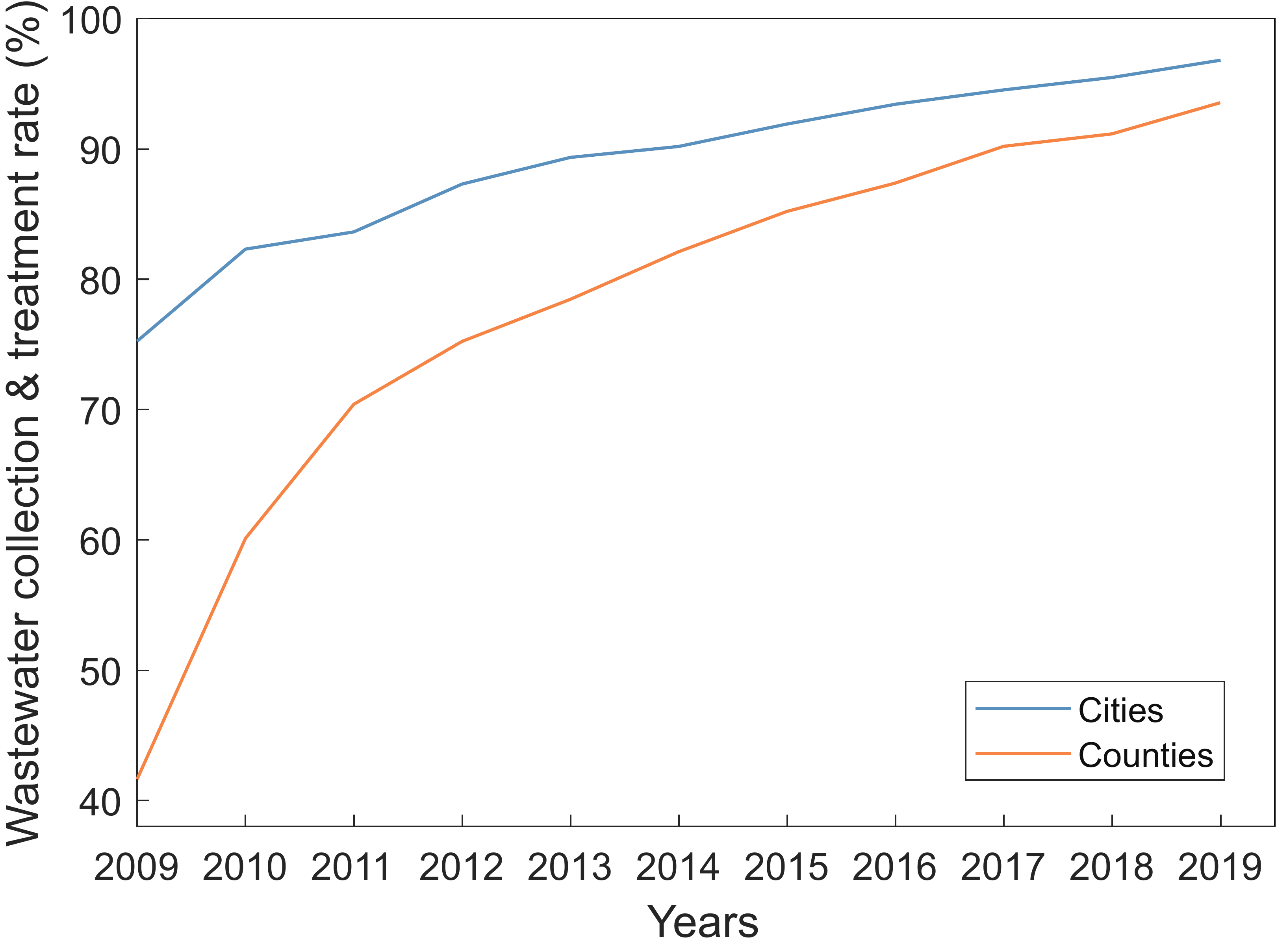}

\textbf{Supplementary Fig.} \textbf{2} Growth of wastewater treatment
rate in 2009-2019\textsuperscript{1}.

\begin{longtable}[]{@{}
  >{\raggedright\arraybackslash}p{(\columnwidth - 4\tabcolsep) * \real{0.1191}}
  >{\raggedright\arraybackslash}p{(\columnwidth - 4\tabcolsep) * \real{0.4405}}
  >{\raggedright\arraybackslash}p{(\columnwidth - 4\tabcolsep) * \real{0.4405}}@{}}
\toprule
\begin{minipage}[b]{\linewidth}\raggedright
\end{minipage} & \begin{minipage}[b]{\linewidth}\raggedright
Inflow
\end{minipage} & \begin{minipage}[b]{\linewidth}\raggedright
Outflow
\end{minipage} \\
\midrule
\endhead
COD & \includegraphics[width=2.41667in,height=2.019in]{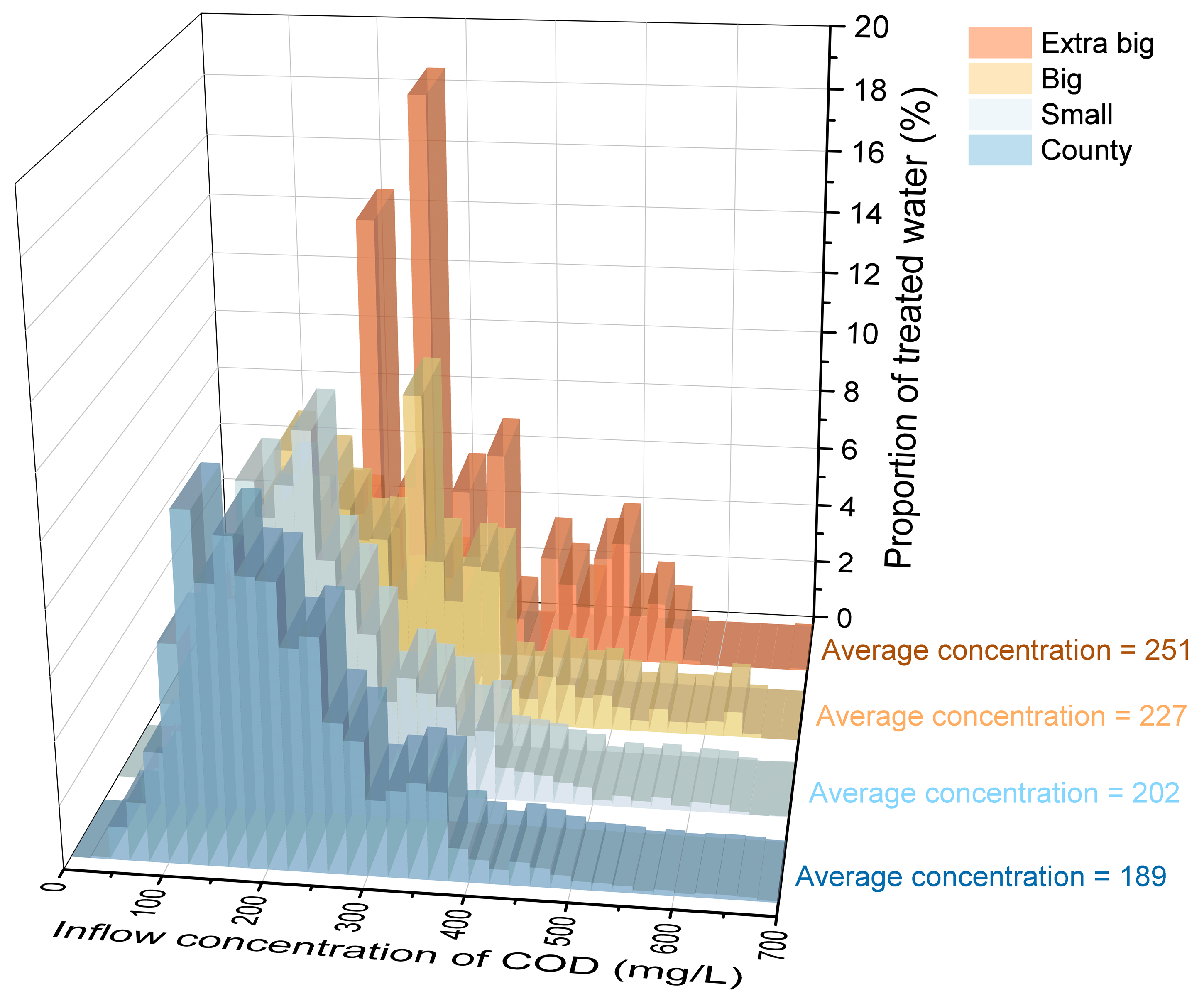}
&
\includegraphics[width=2.52639in,height=1.91667in]{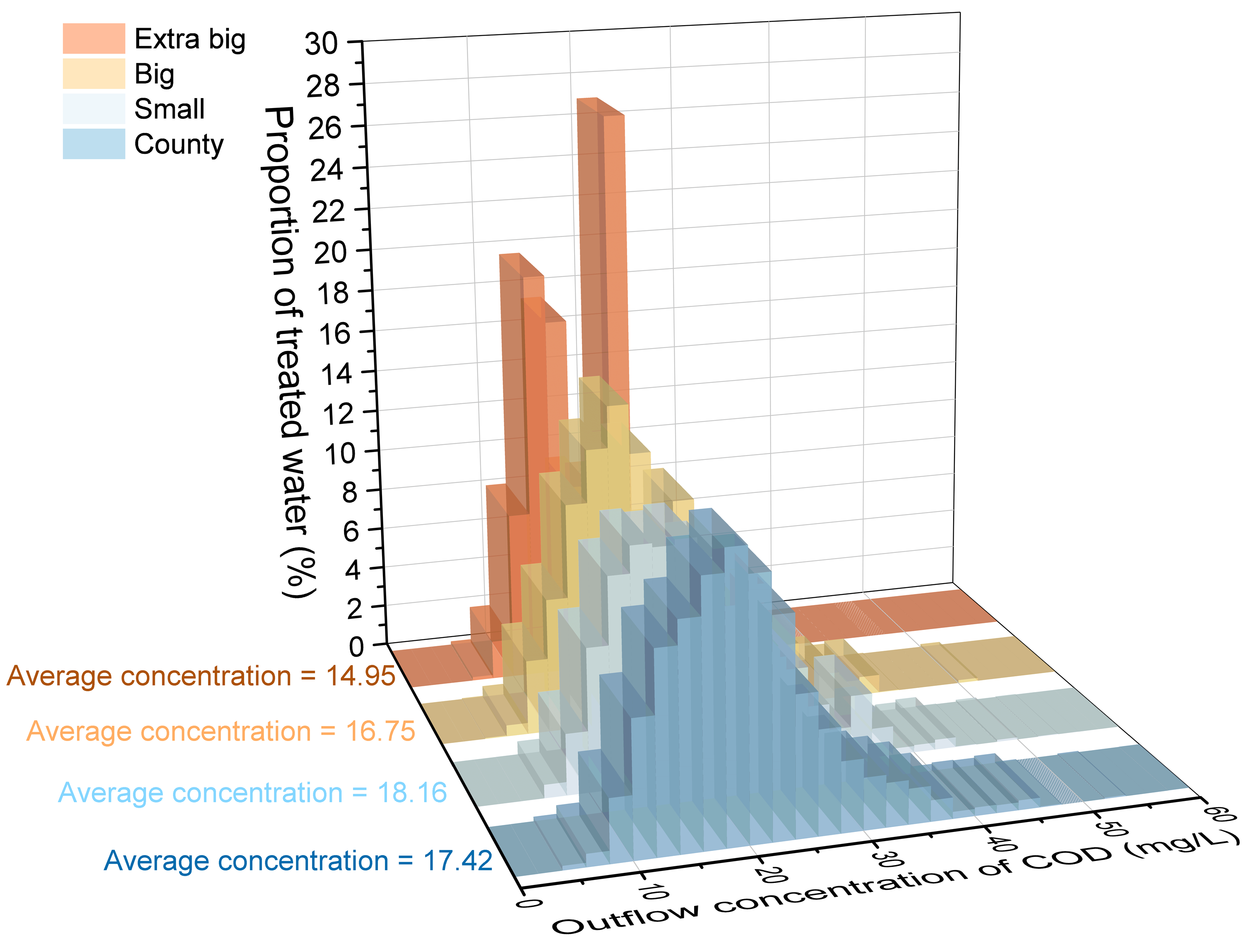} \\
BOD &
\includegraphics[width=2.43452in,height=1.99311in]{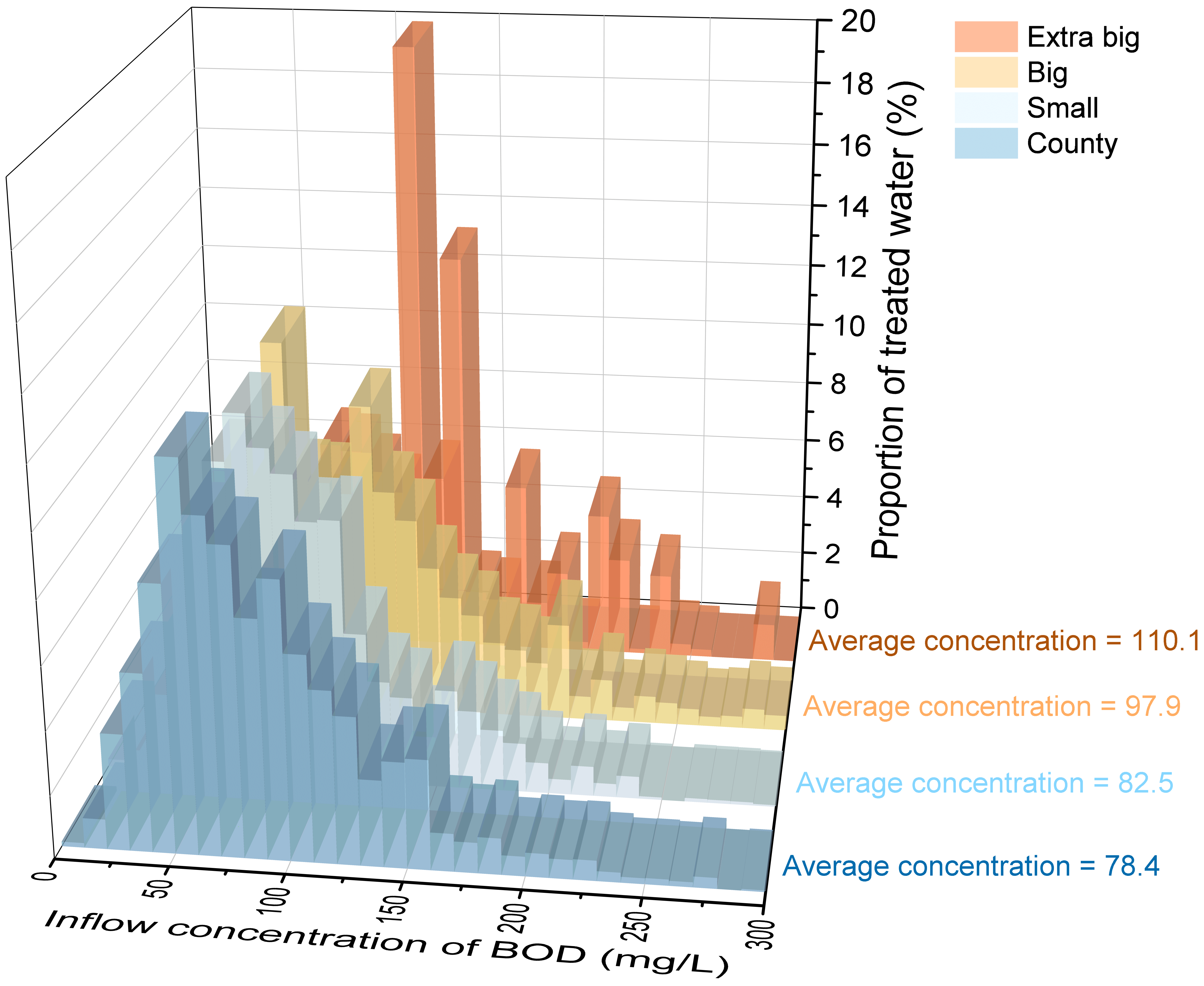} &
\includegraphics[width=2.49583in,height=1.89236in]{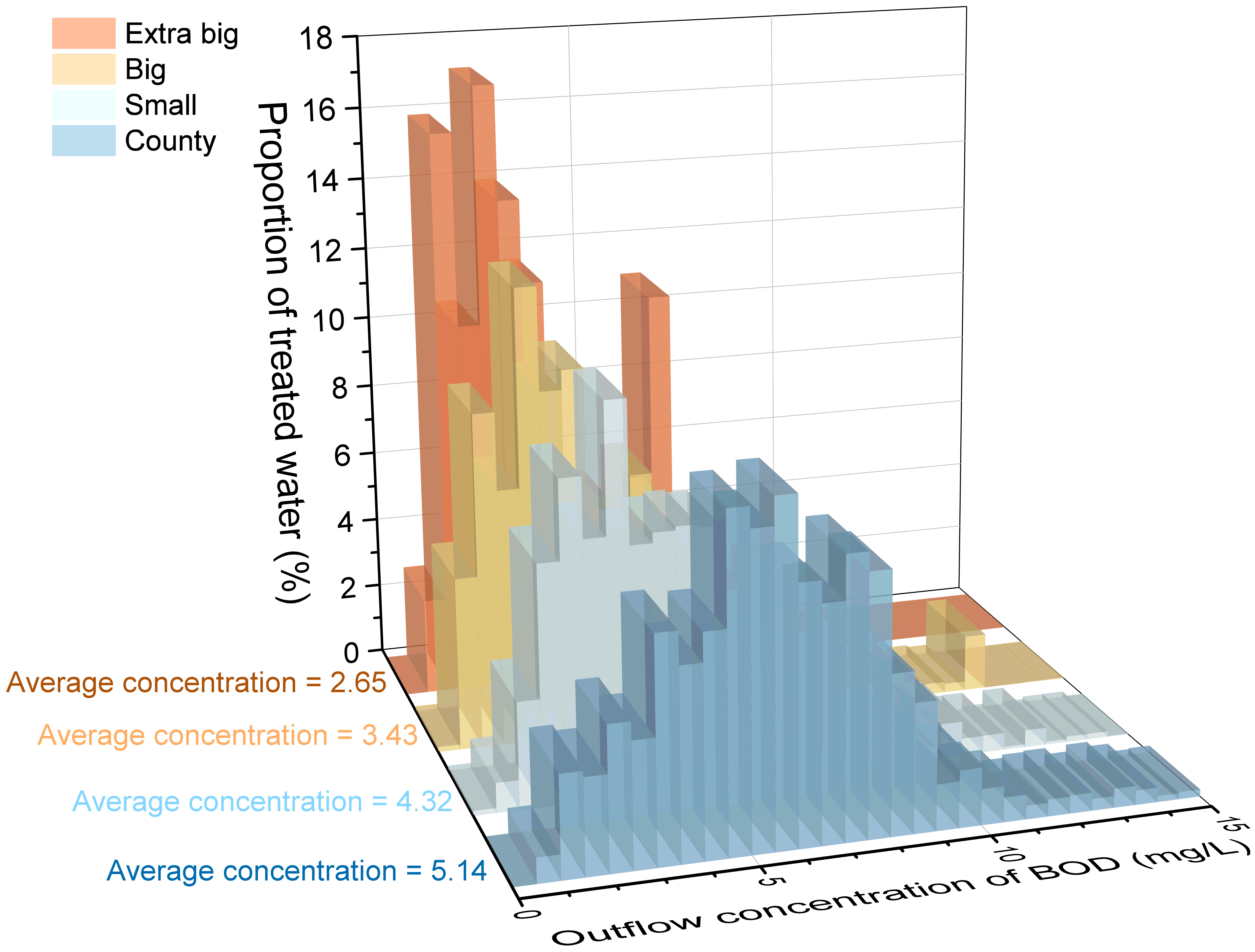} \\
SS &
\includegraphics[width=2.43403in,height=2.02377in]{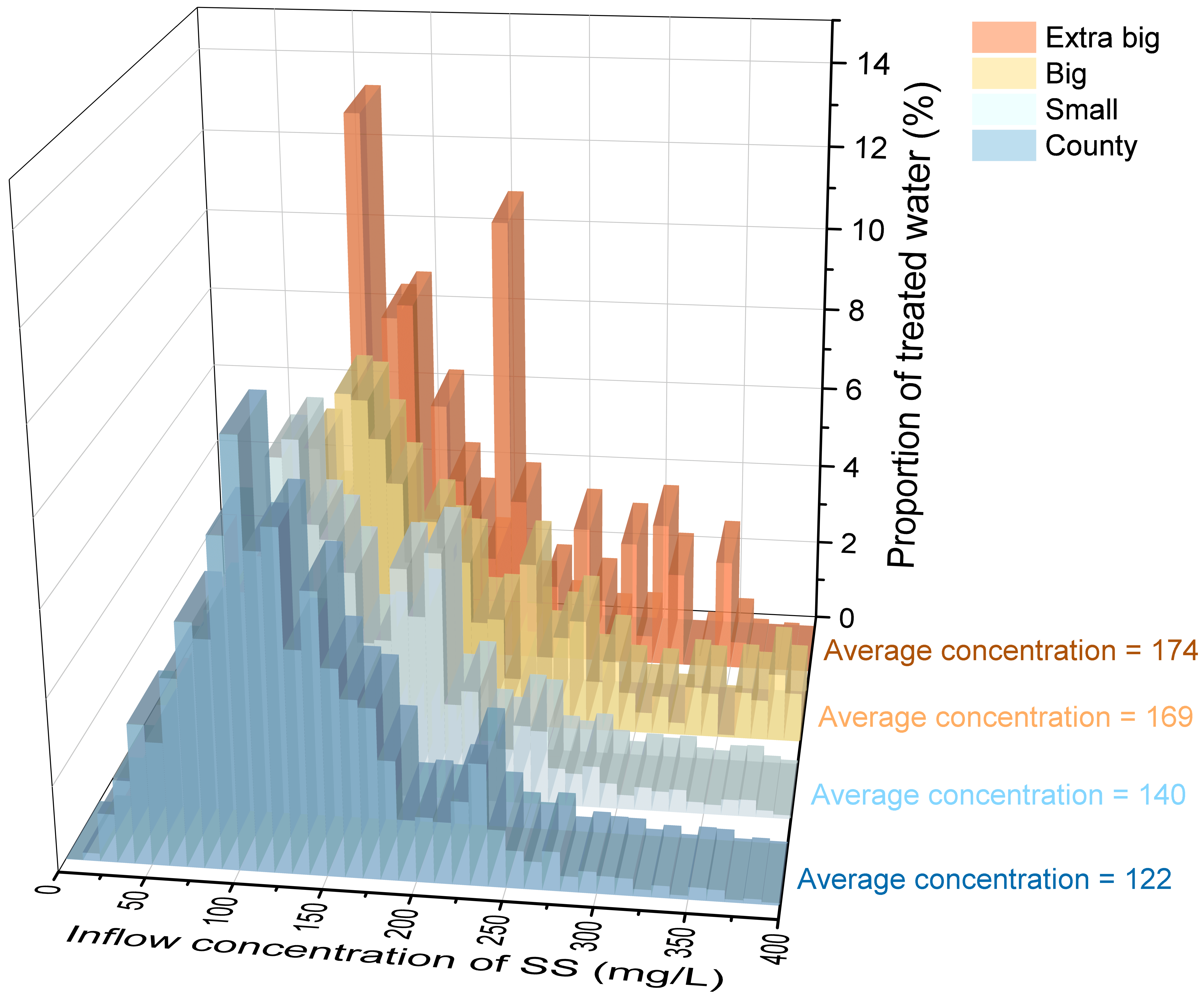} &
\includegraphics[width=2.49182in,height=1.89482in]{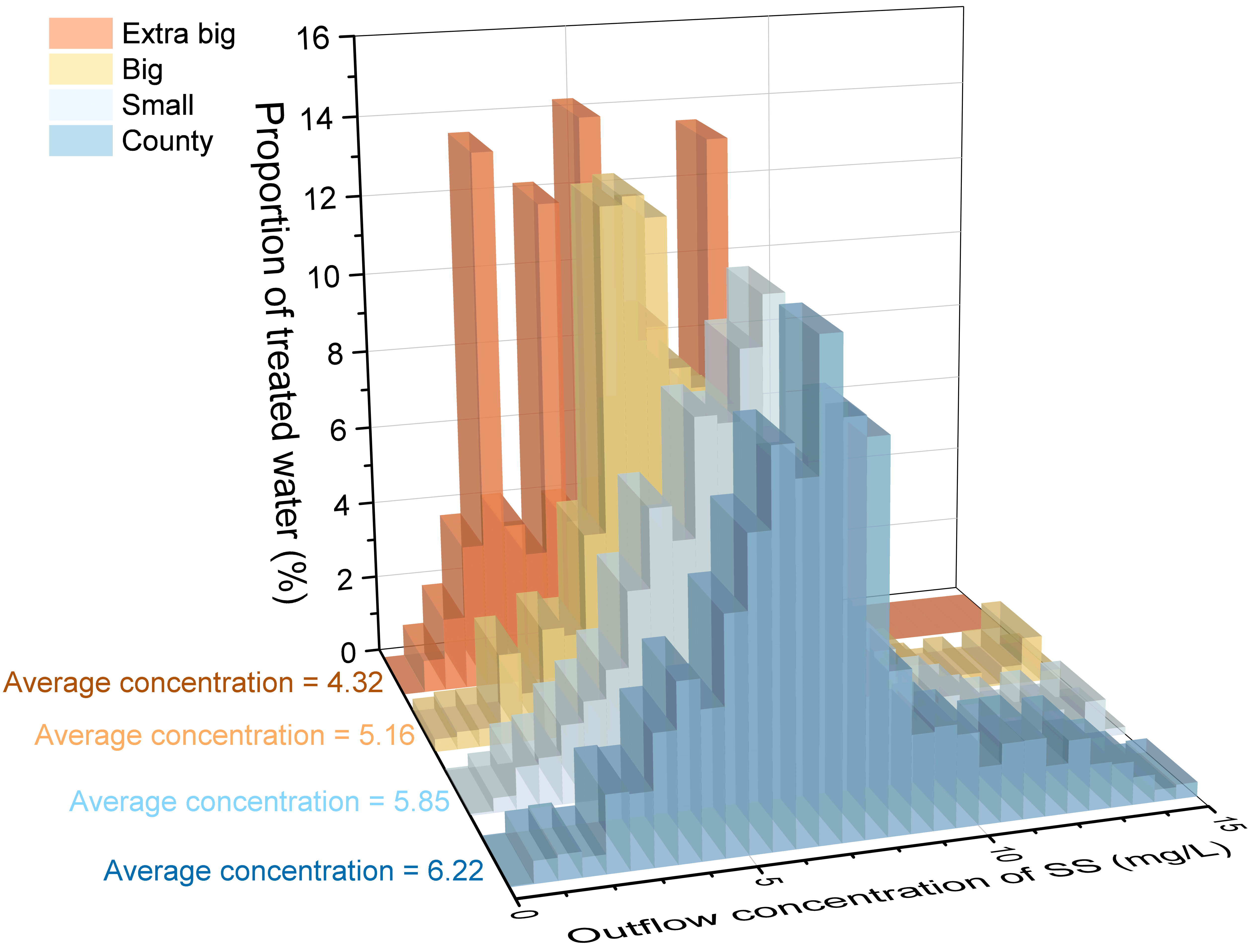} \\
NH\textsubscript{3} &
\includegraphics[width=2.44048in,height=1.9589in]{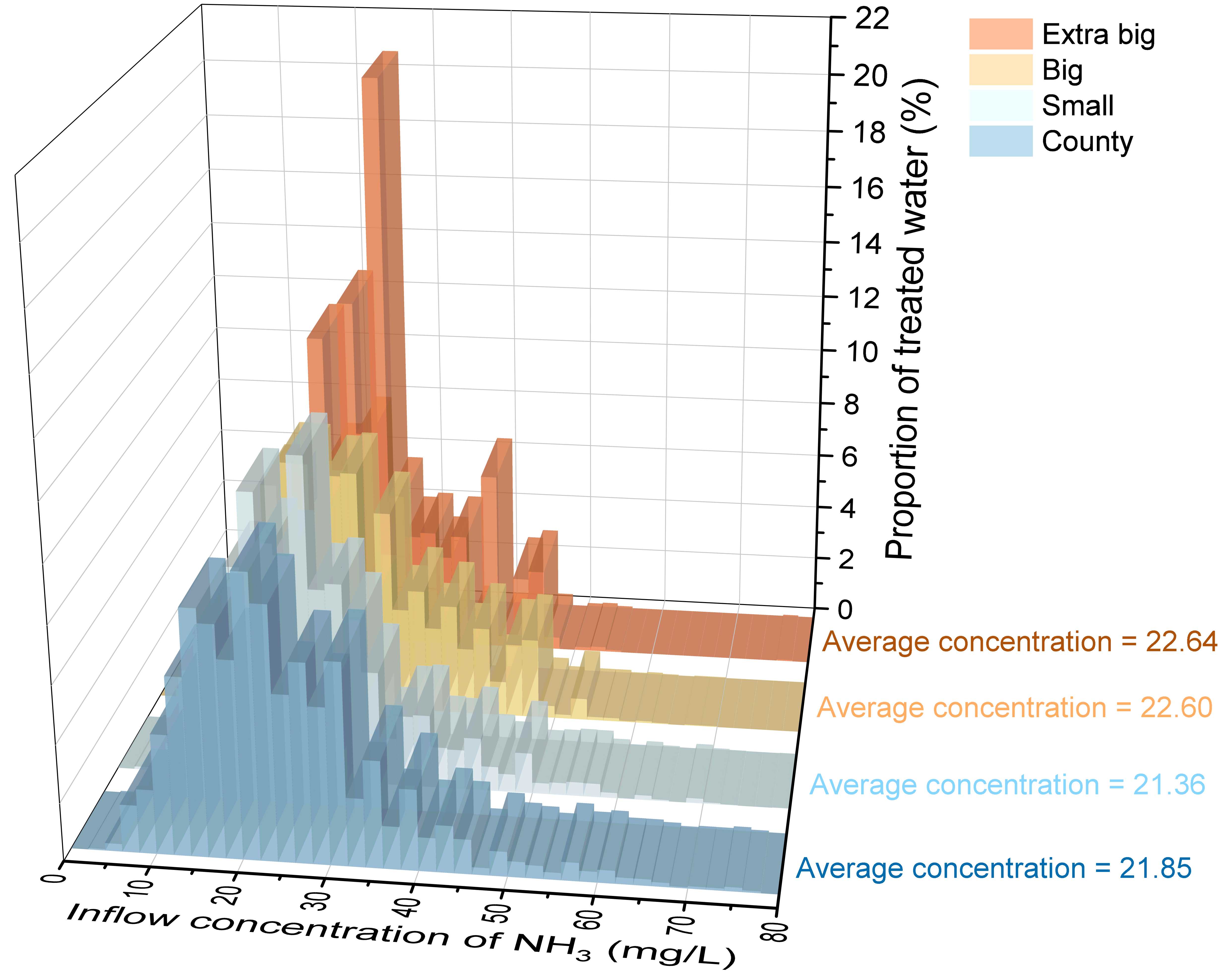} &
\includegraphics[width=2.4994in,height=1.875in]{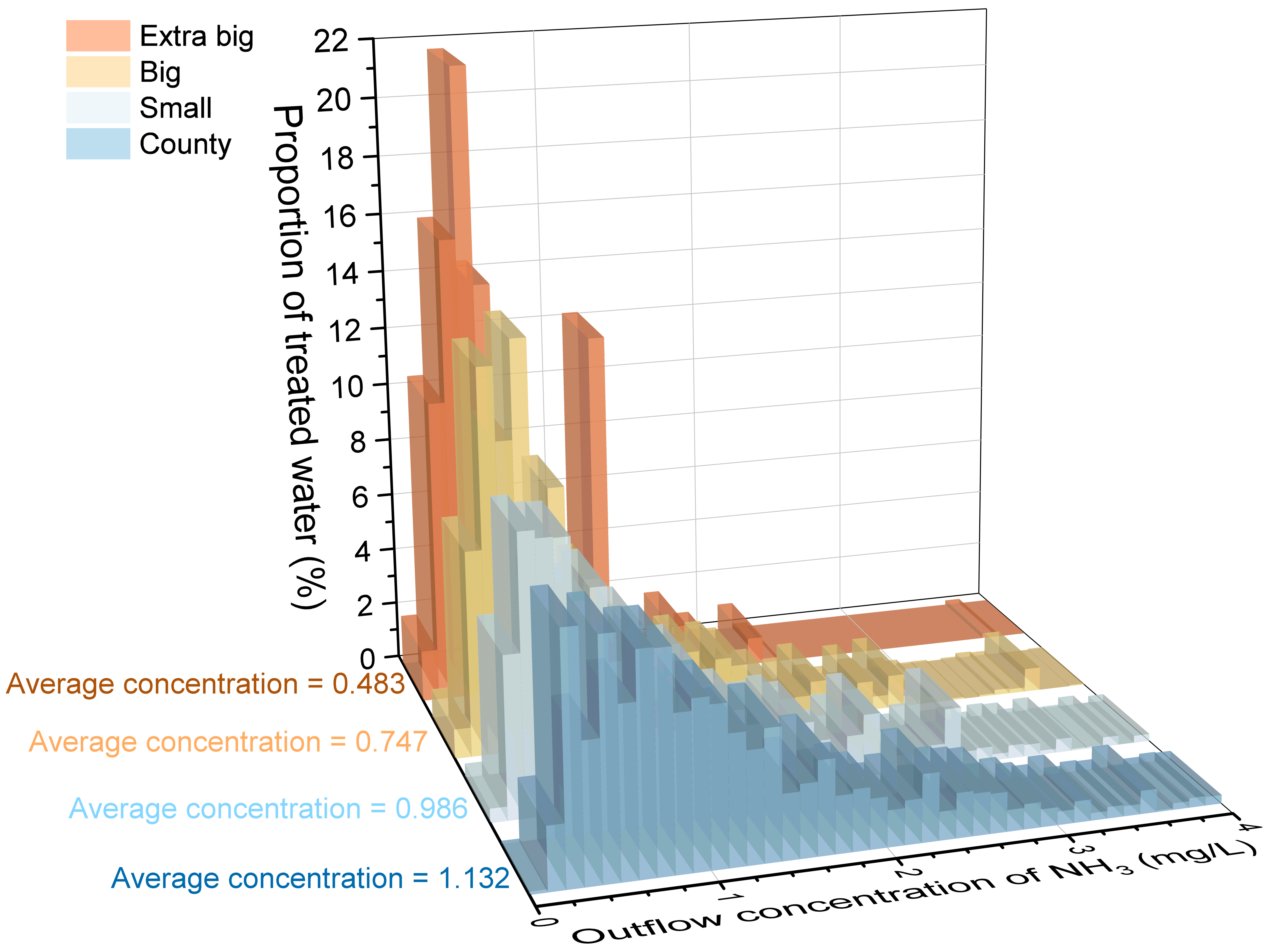} \\
TN &
\includegraphics[width=2.44643in,height=1.98872in]{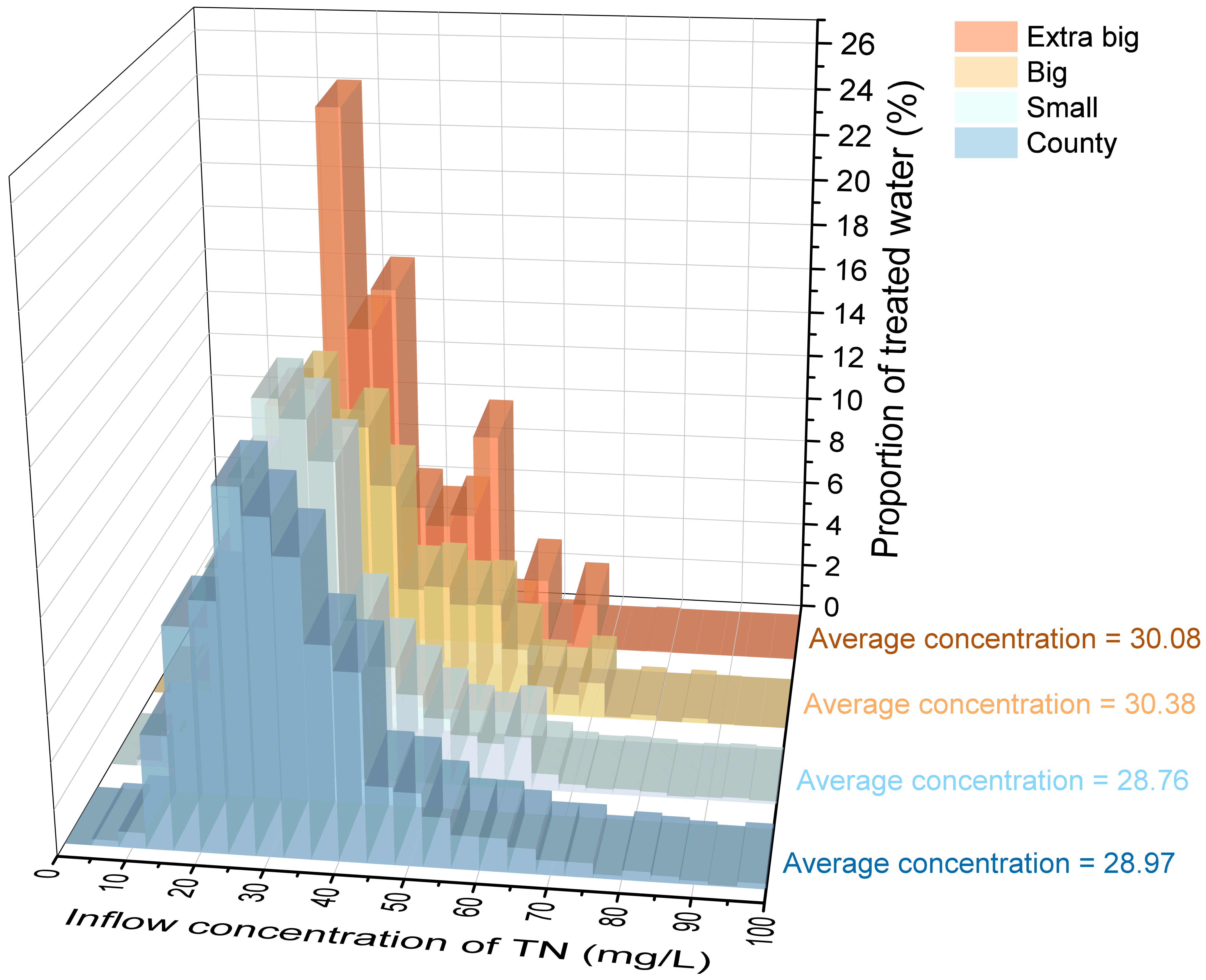} &
\includegraphics[width=2.48718in,height=1.90476in]{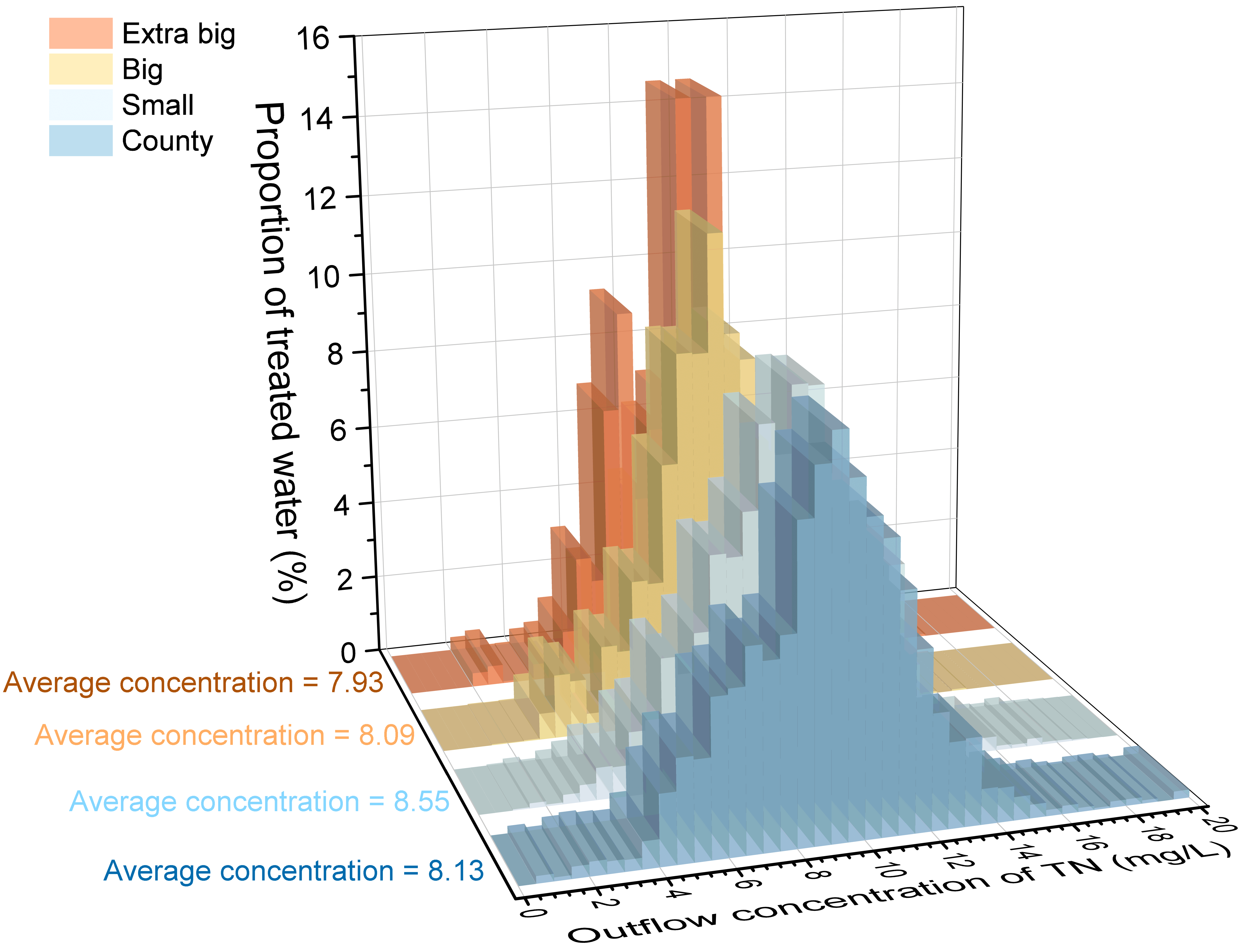} \\
TP &
\includegraphics[width=2.44583in,height=1.99236in]{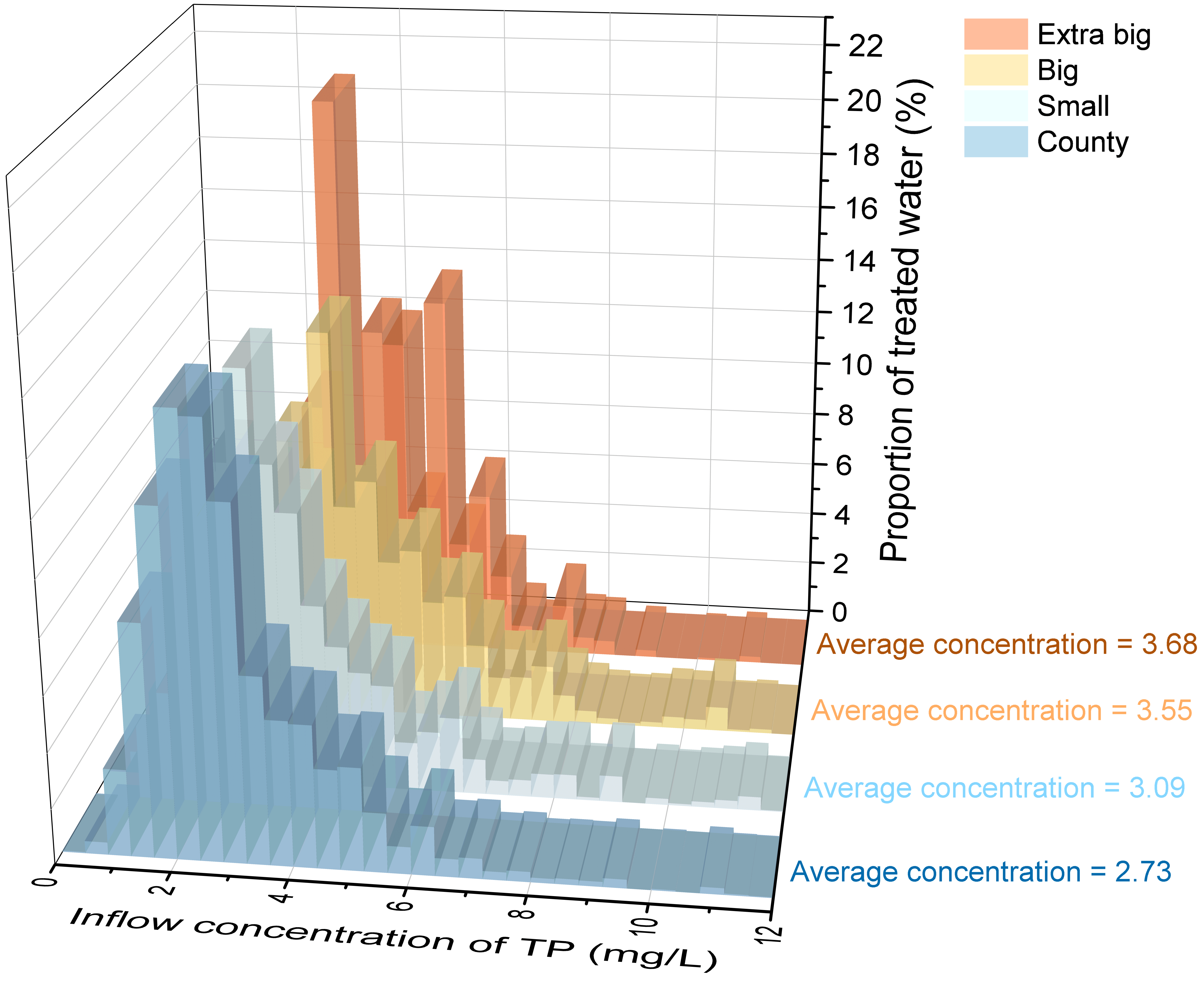} &
\includegraphics[width=2.47679in,height=1.85714in]{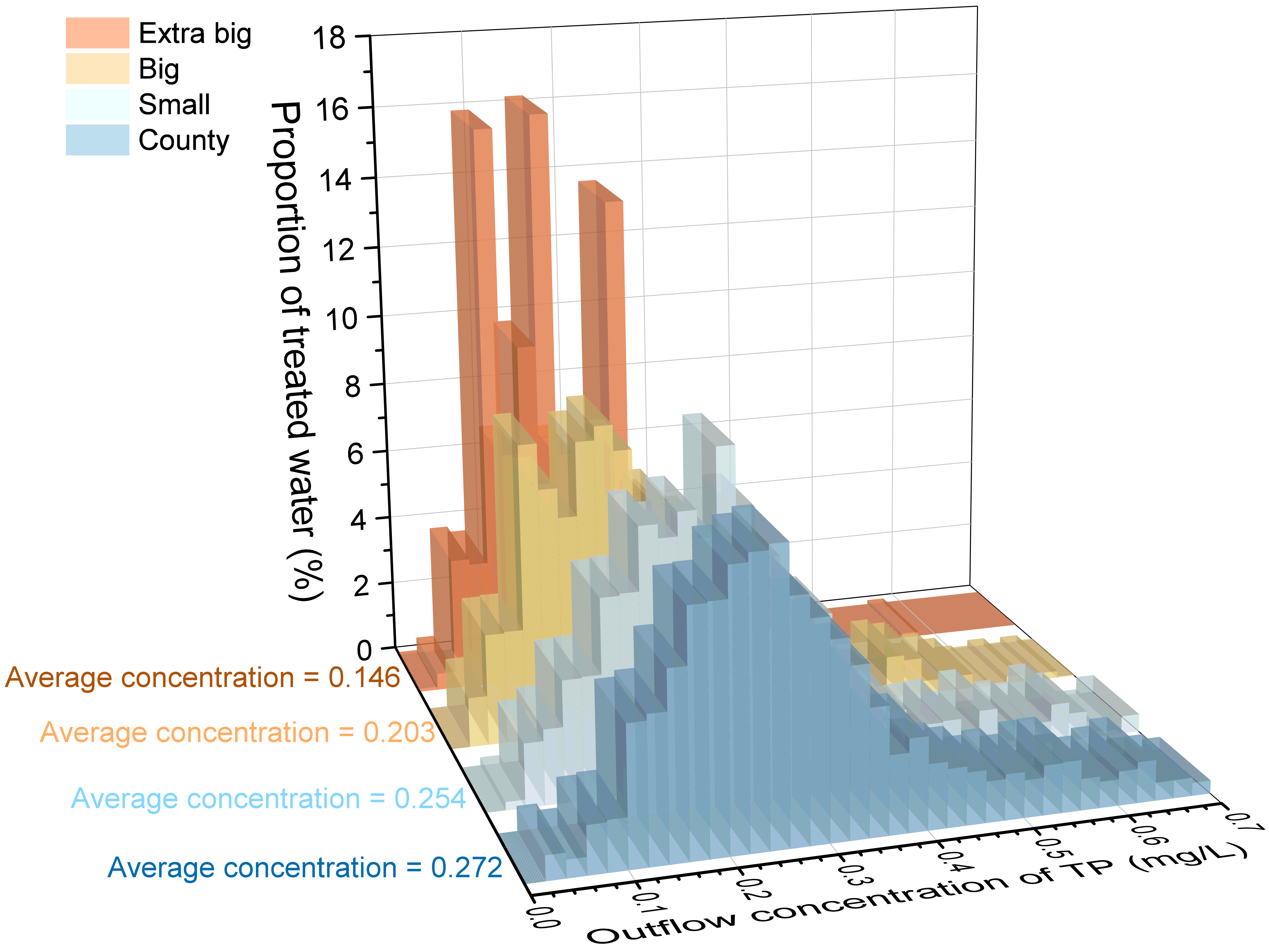} \\
COD/TN &
\multicolumn{2}{l}{\includegraphics[width=2.3742in,height=1.97917in]{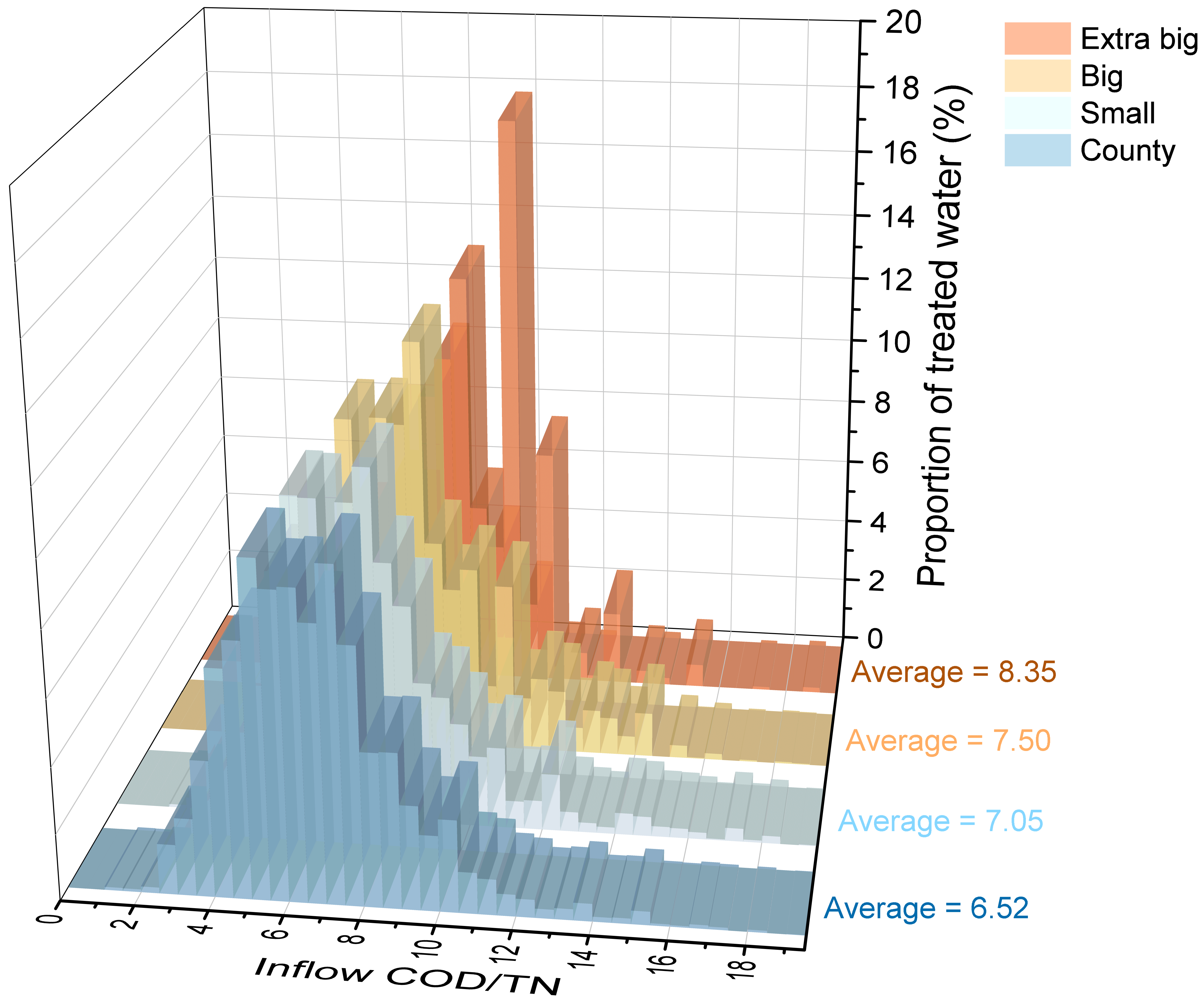}} \\
\bottomrule
\end{longtable}

\textbf{Supplementary Fig.} \textbf{3} The distribution of treated
wastewater with different \textbf{a,} COD inflow, \textbf{b,} COD
outflow, \textbf{c,} BOD inflow, \textbf{d,} BOD outflow, \textbf{e,} SS
inflow, \textbf{f,} SS outflow, \textbf{g,} NH3 inflow, \textbf{h,} NH3
outflow, \textbf{i,} TN inflow, \textbf{j,} TN outflow, \textbf{k,} TP
inflow, \textbf{l,} TP outflow, m, inflow COD/TN ratio, in different
scales of cities and counties.

\includegraphics[width=2in,height=1.8in]{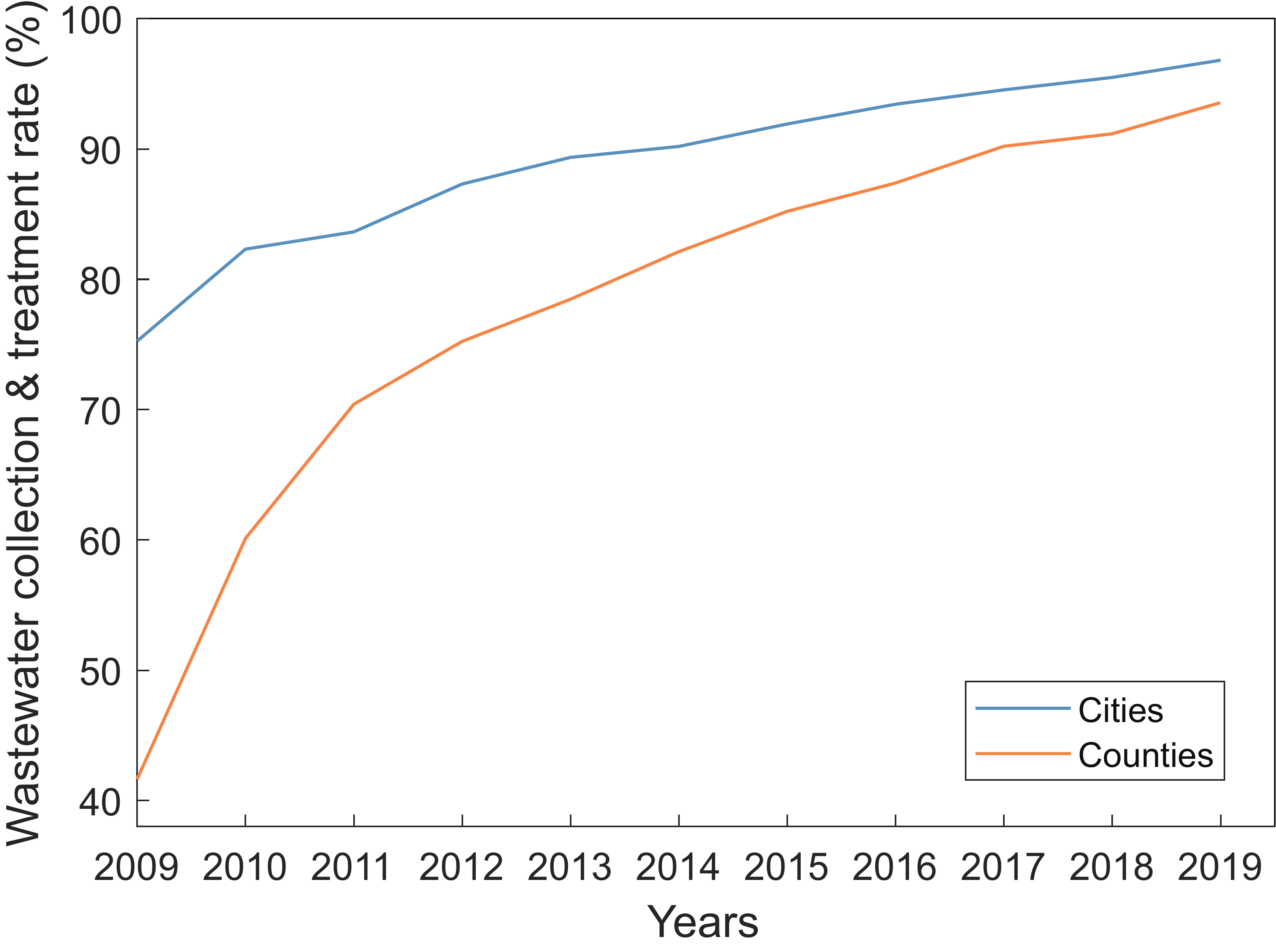}\includegraphics[width=2in,height=1.8in]{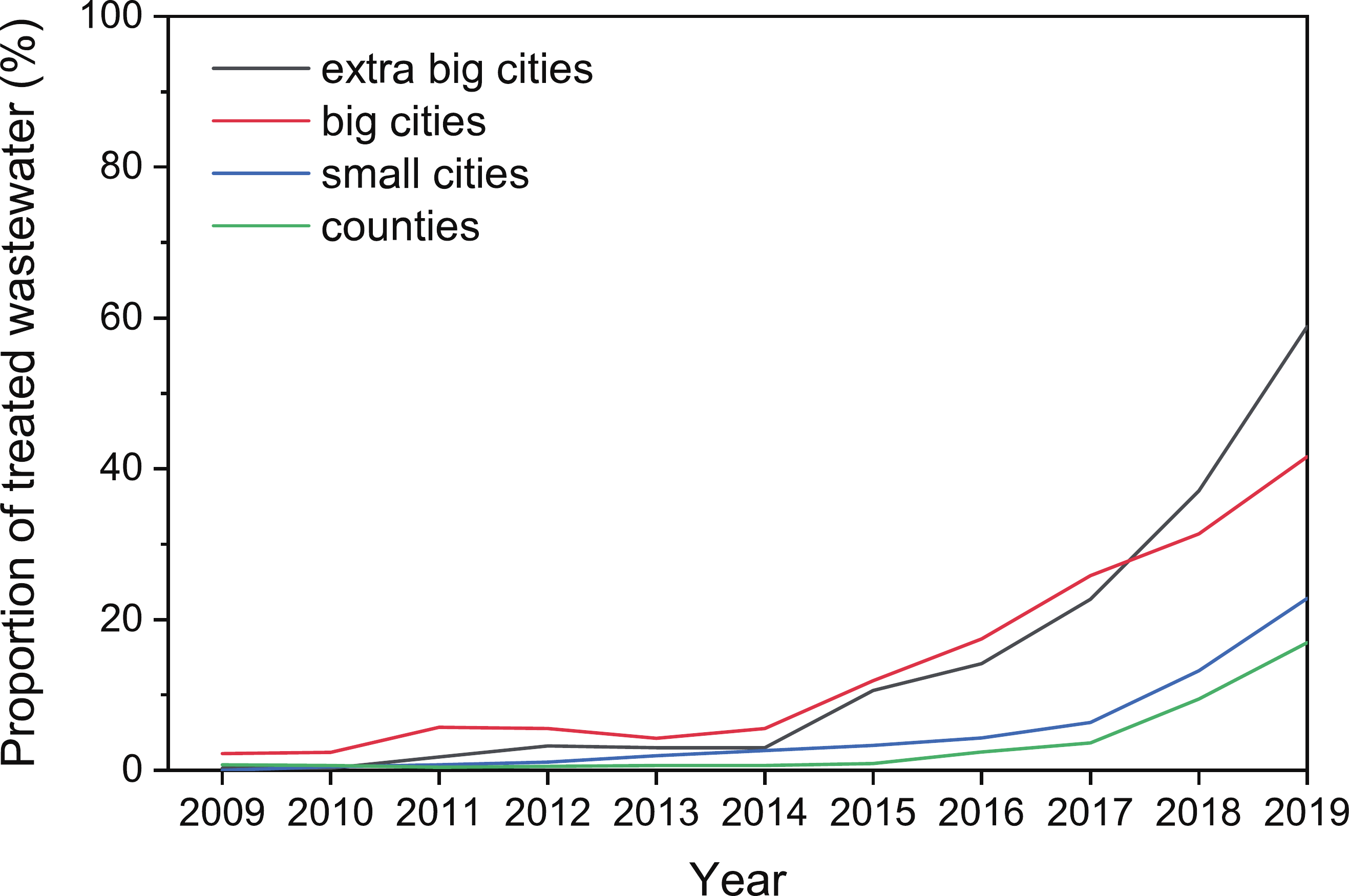}

\textbf{Supplementary Fig.} \textbf{4} Proportion of \textbf{a,}
wastewater treated to different standards in different scales of cities
and counties in 2019, \textbf{b,} special limitation standard, in
different scales of cities and counties in 2009-2019.

\includegraphics[width=2.83333in,height=2.31859in]{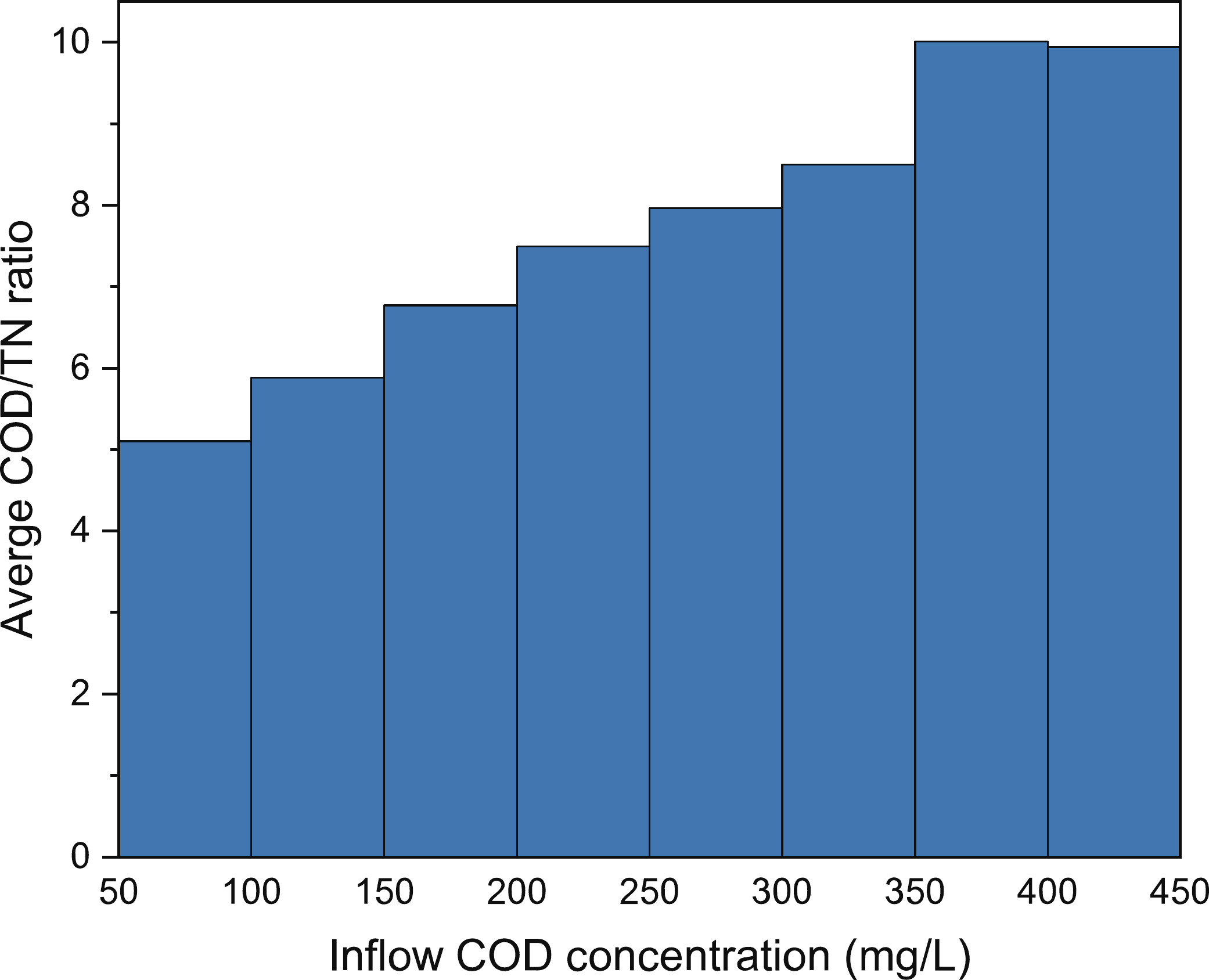}
\includegraphics[width=2.85208in,height=2.31161in]{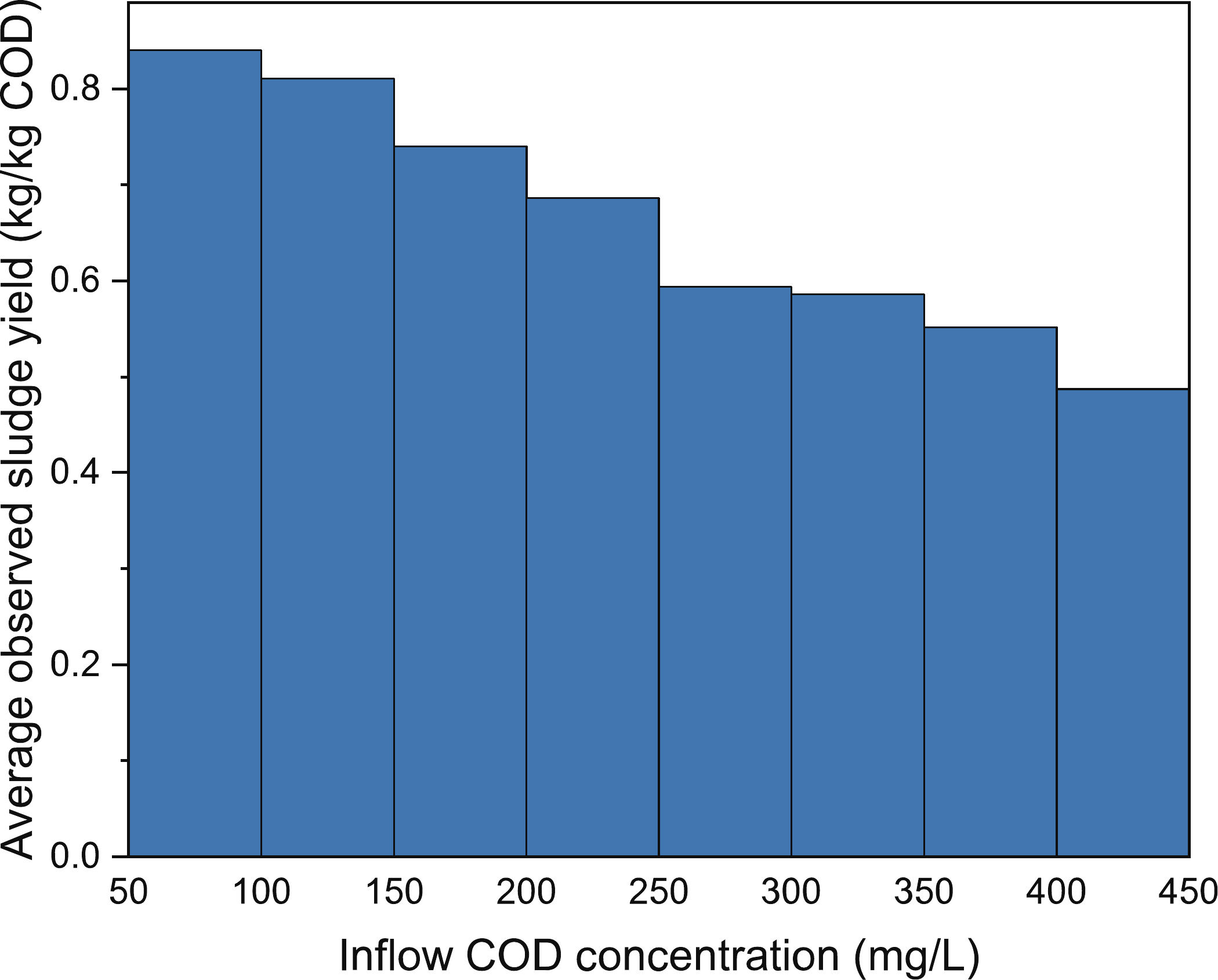}

\includegraphics[width=2.86739in,height=2.33333in]{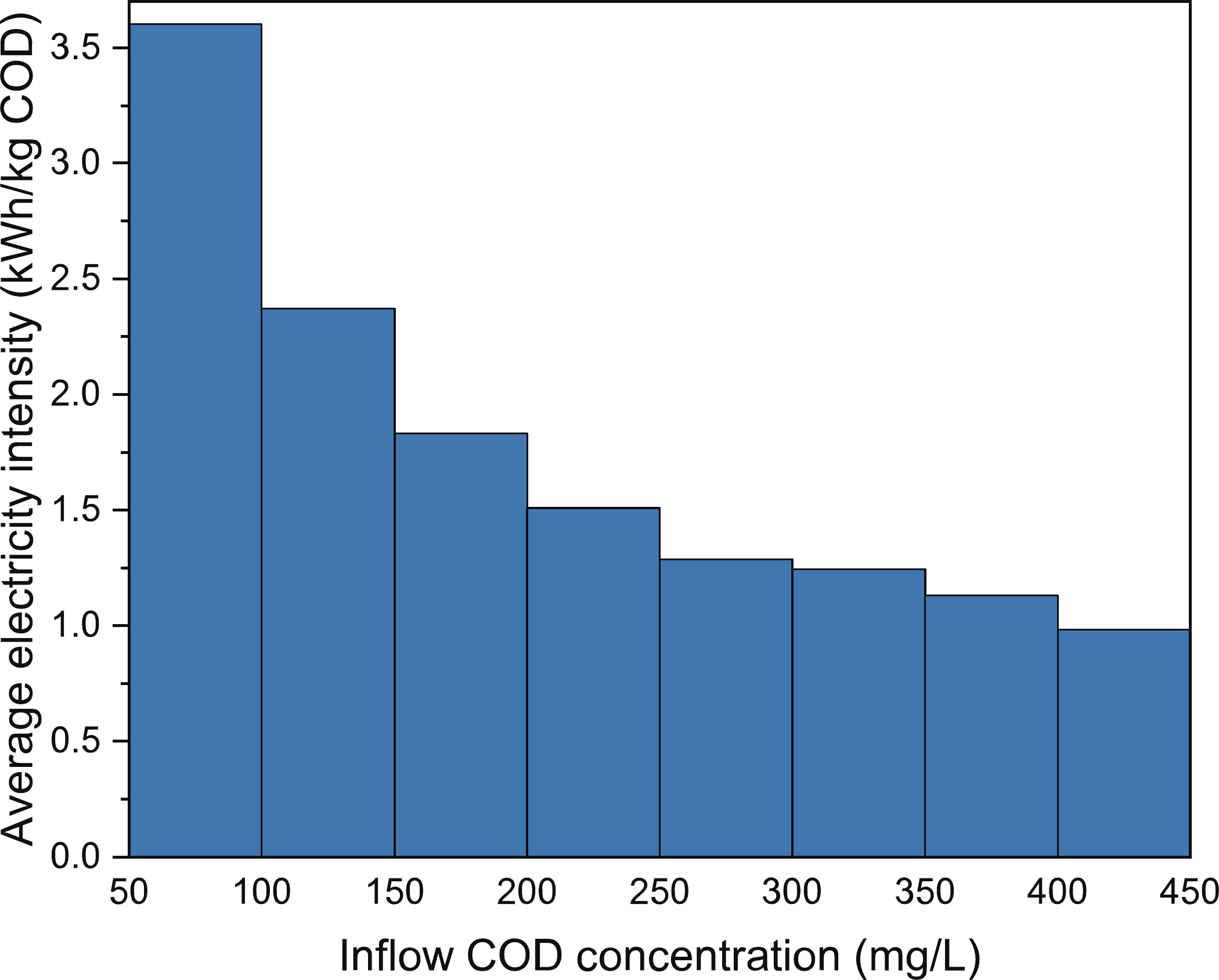}

\textbf{Supplementary Fig.} \textbf{5} Average \textbf{a,} COD/TN ratio,
\textbf{b,} observed sludge yield, \textbf{c,} electricity intensity in
each interval of inflow COD concentration range in China, 2019. The
average value is wastewater amount-weighted.

\includegraphics[width=5.7679in,height=2.01389in]{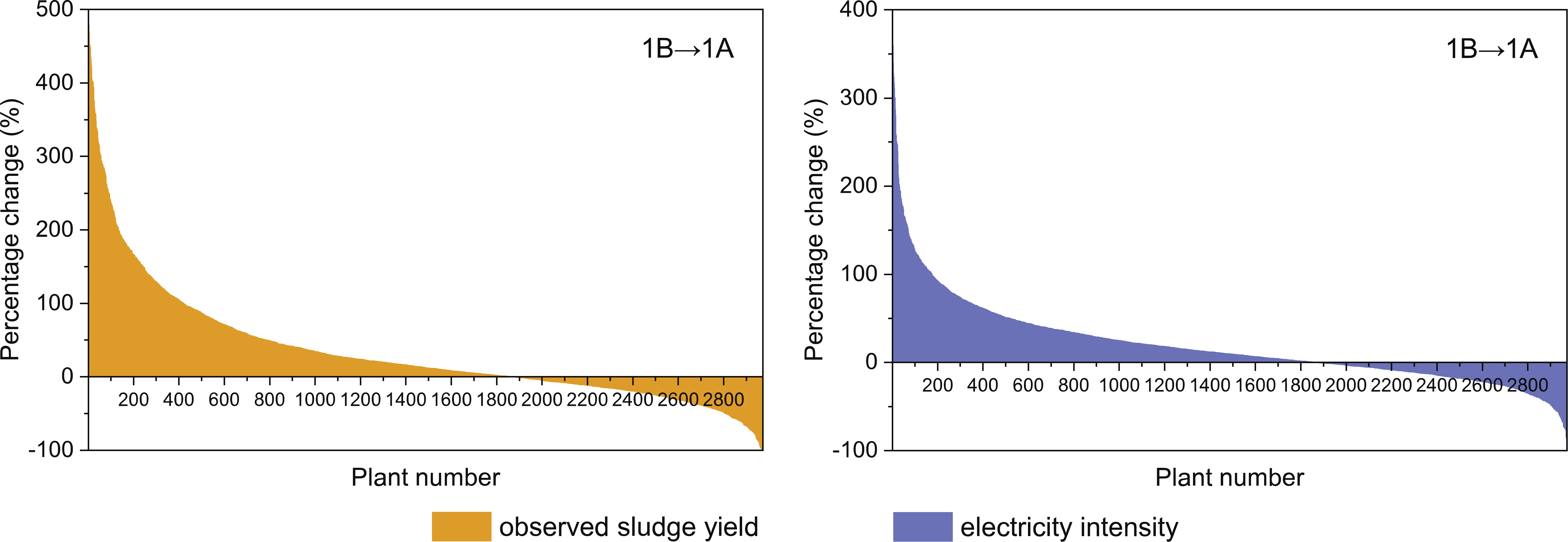}

\includegraphics[width=5.76806in,height=1.79653in]{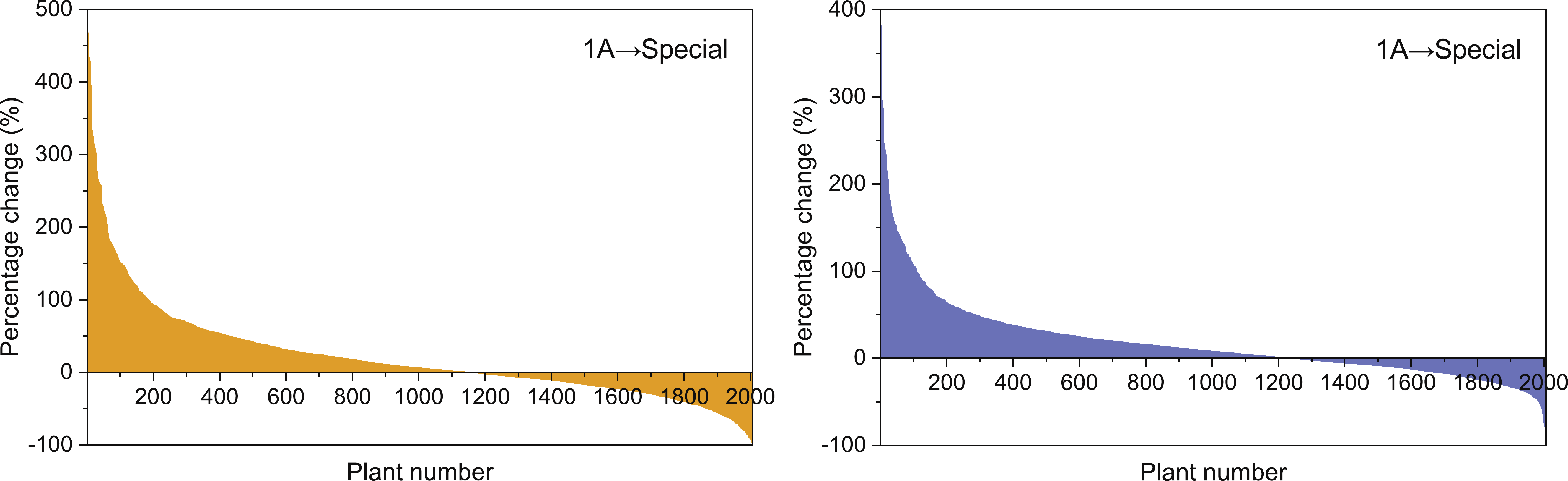}

\textbf{Supplementary Fig.} \textbf{6} Percentage change of \textbf{a,}
observed sludge yield, \textbf{b,} electricity intensity, from Class 1B
to Class 1A for 2972 plants across China, and \textbf{c,} observed
sludge yield, \textbf{d,} electricity intensity, from Class 1A to
special for 2006 plants across China. Each plant is represented by a
line, and upgrades were associated with an increase in sludge generation
and electricity use for most plants. Upgrades from Class 1B to Class 1A
resulted in average increase in observed sludge yield of 17.4\% and
electricity intensity of 11.8\%, and upgrades from Class 1A to special
resulted in average increase in observed sludge yield of 7.0\% and
electricity intensity of 9.0\%.

\includegraphics[width=3.44866in,height=2.16029in]{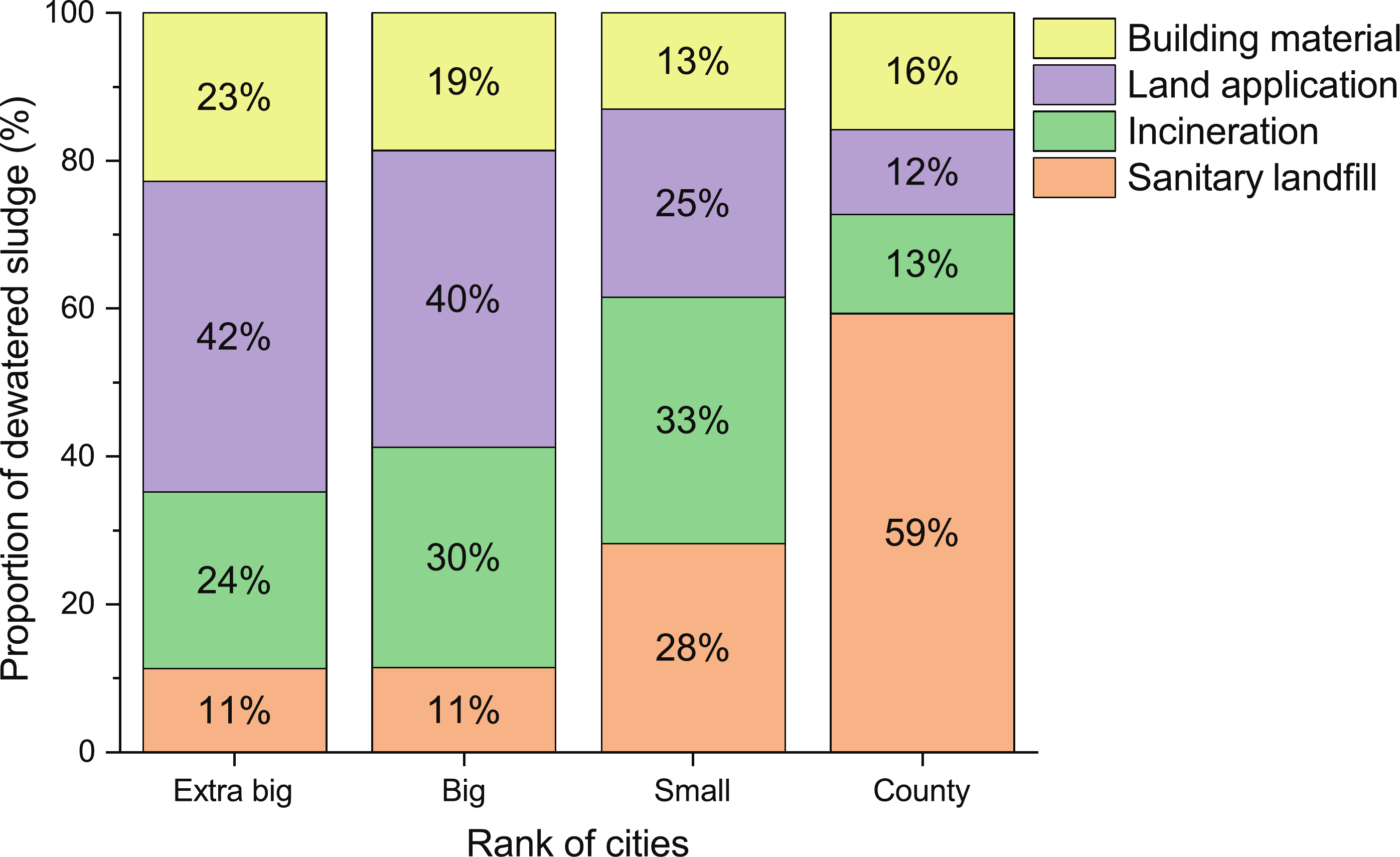}\includegraphics[width=2.30014in,height=2.16607in]{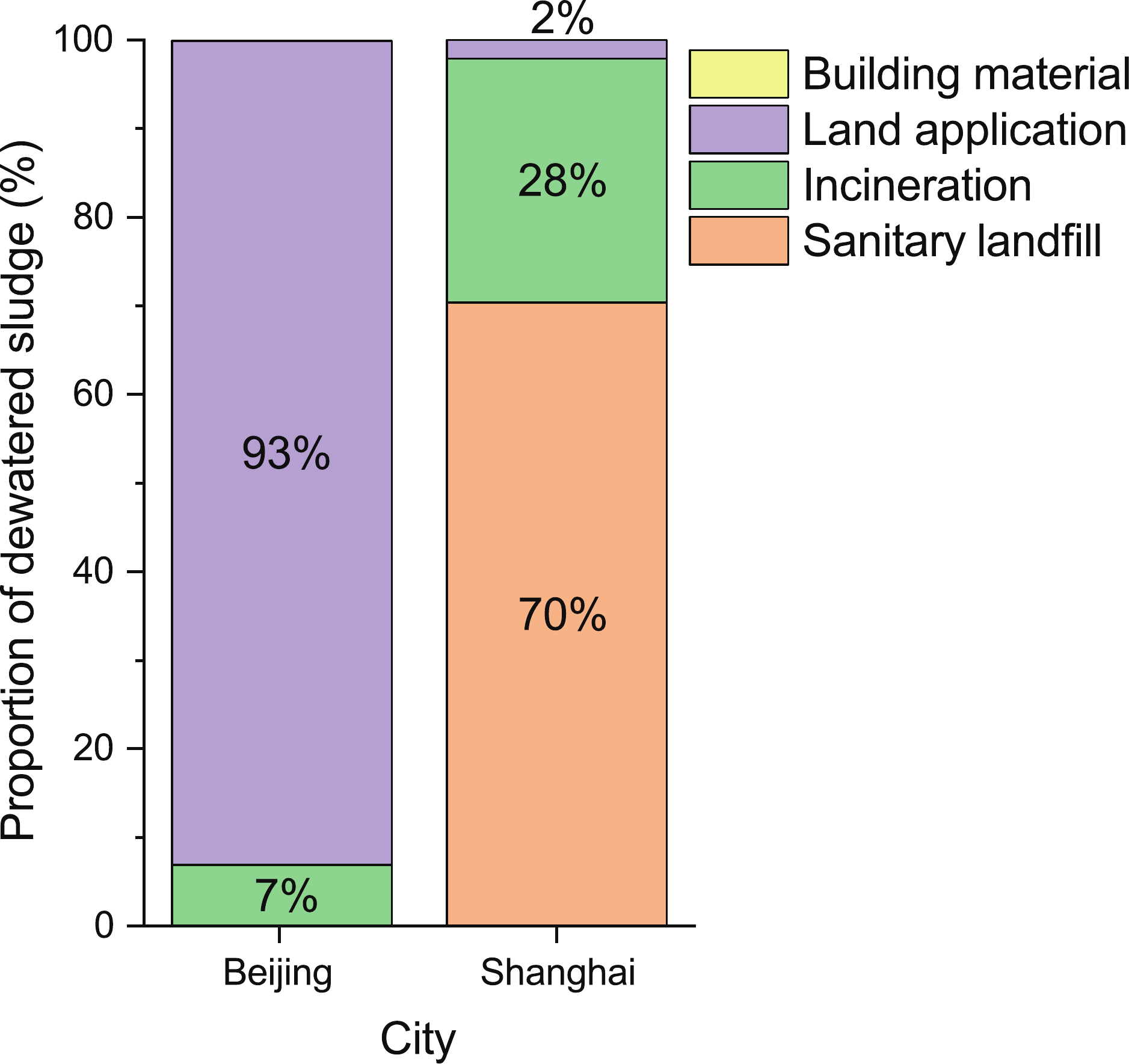}

\includegraphics[width=5.74376in,height=4.10417in]{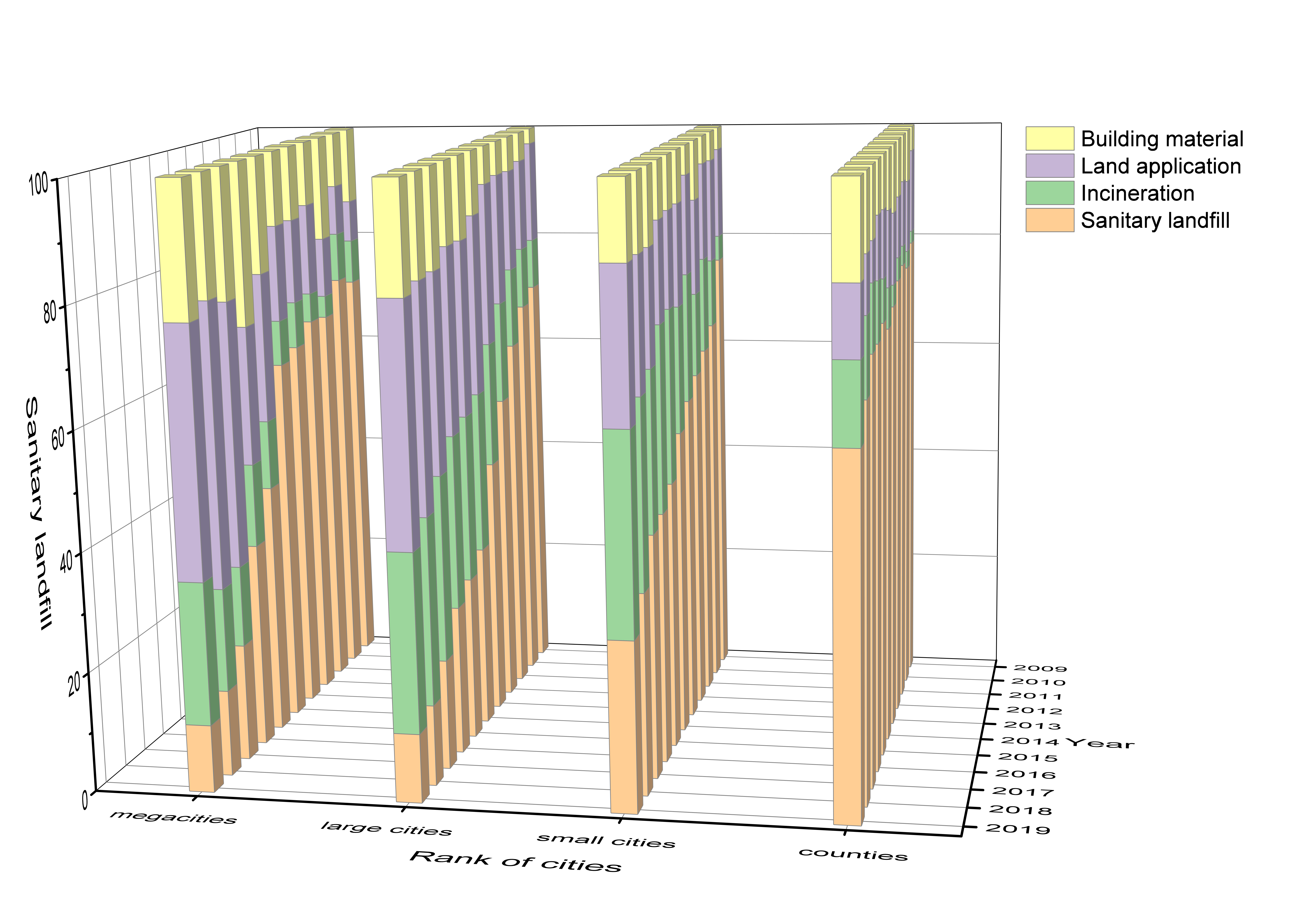}

\textbf{Supplementary Fig.} \textbf{7} Proportion of dewatered sludge
for four treatment and disposal ways in \textbf{a,} different ranks of
cities, \textbf{b,} Beijing and Shanghai, 2019. \textbf{c,} proportion
of dewatered sludge for four treatment and disposal ways in different
ranks of cities, 2009-2019.

\includegraphics[width=5.76806in,height=5.76806in]{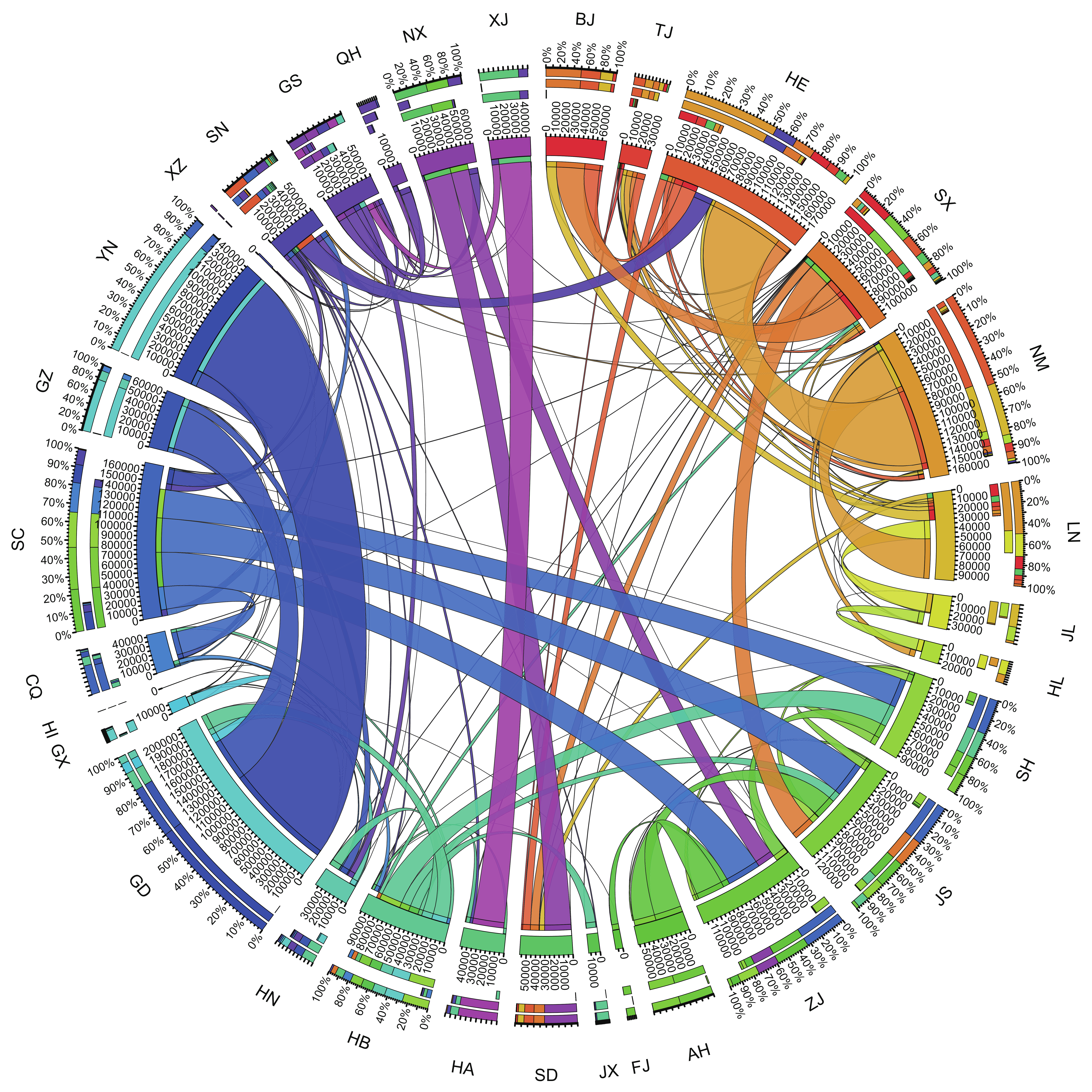}

\textbf{Supplementary Fig.} \textbf{8} Circos \textsuperscript{12} of
the interprovincial electricity transmission in 2017. Abbreviations of
the provinces are defined in Supplementary Table 15.

\includegraphics[width=2.56493in,height=2.04861in]{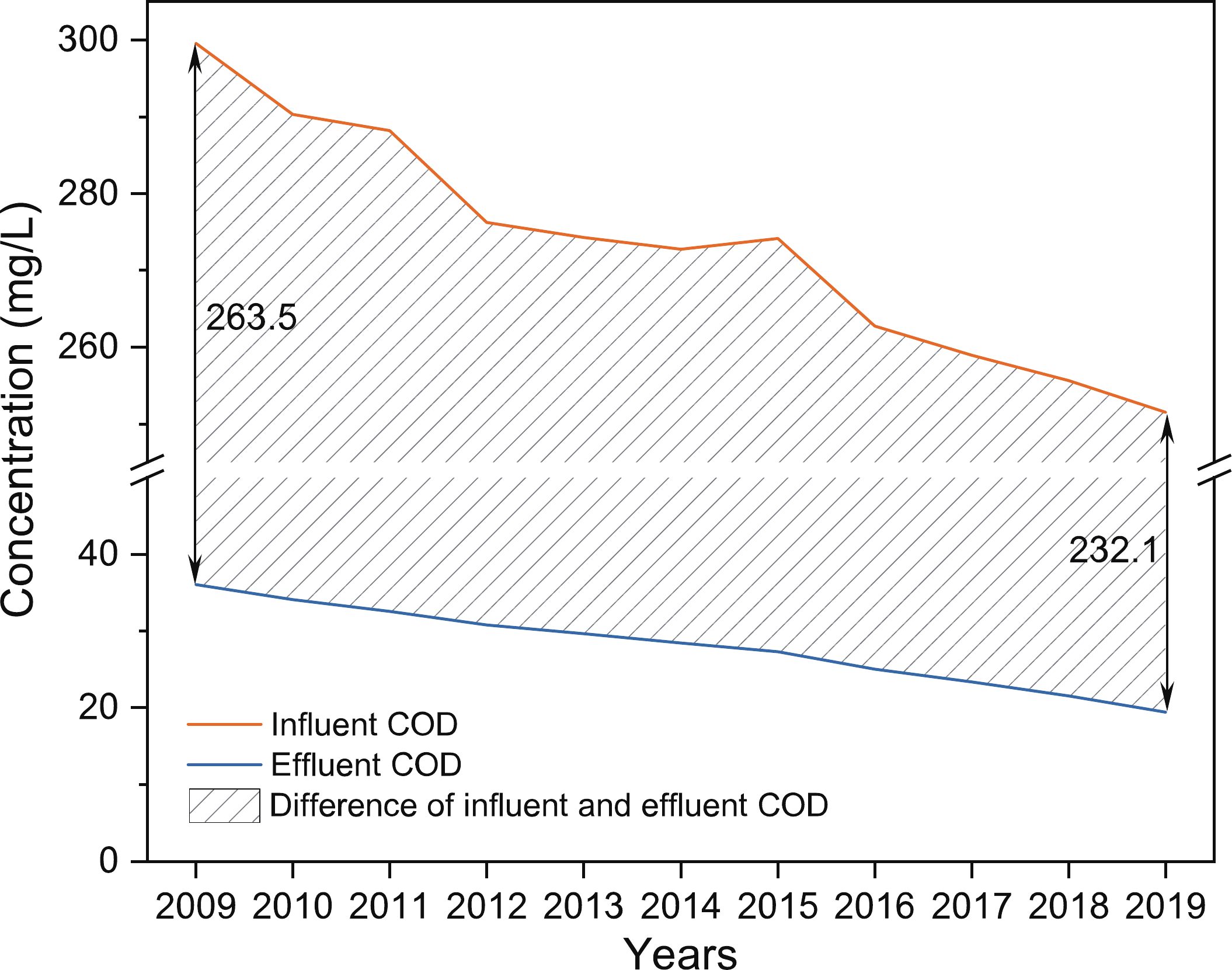}\includegraphics[width=2.5625in,height=2.09202in]{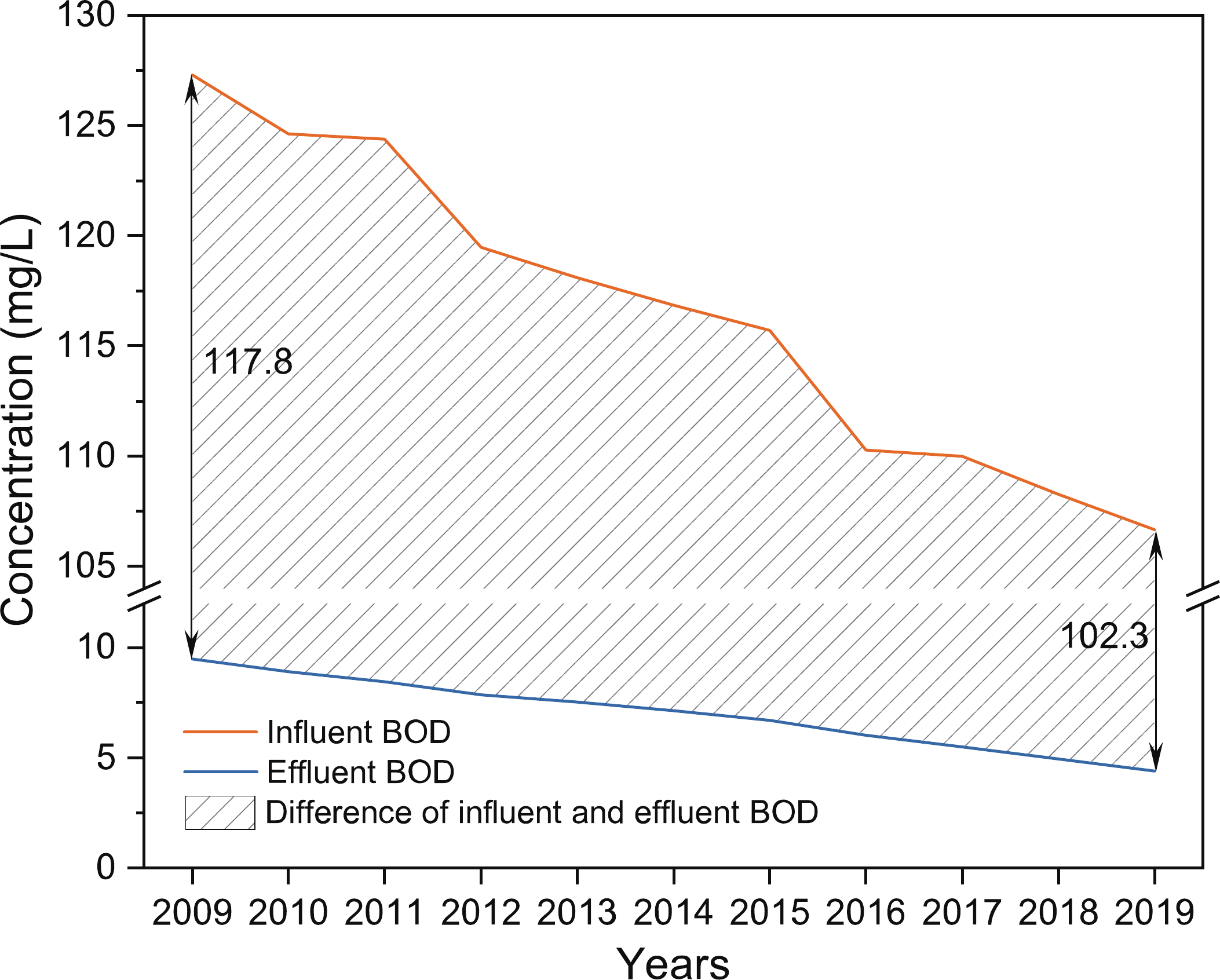}

\includegraphics[width=2.54861in,height=2.05982in]{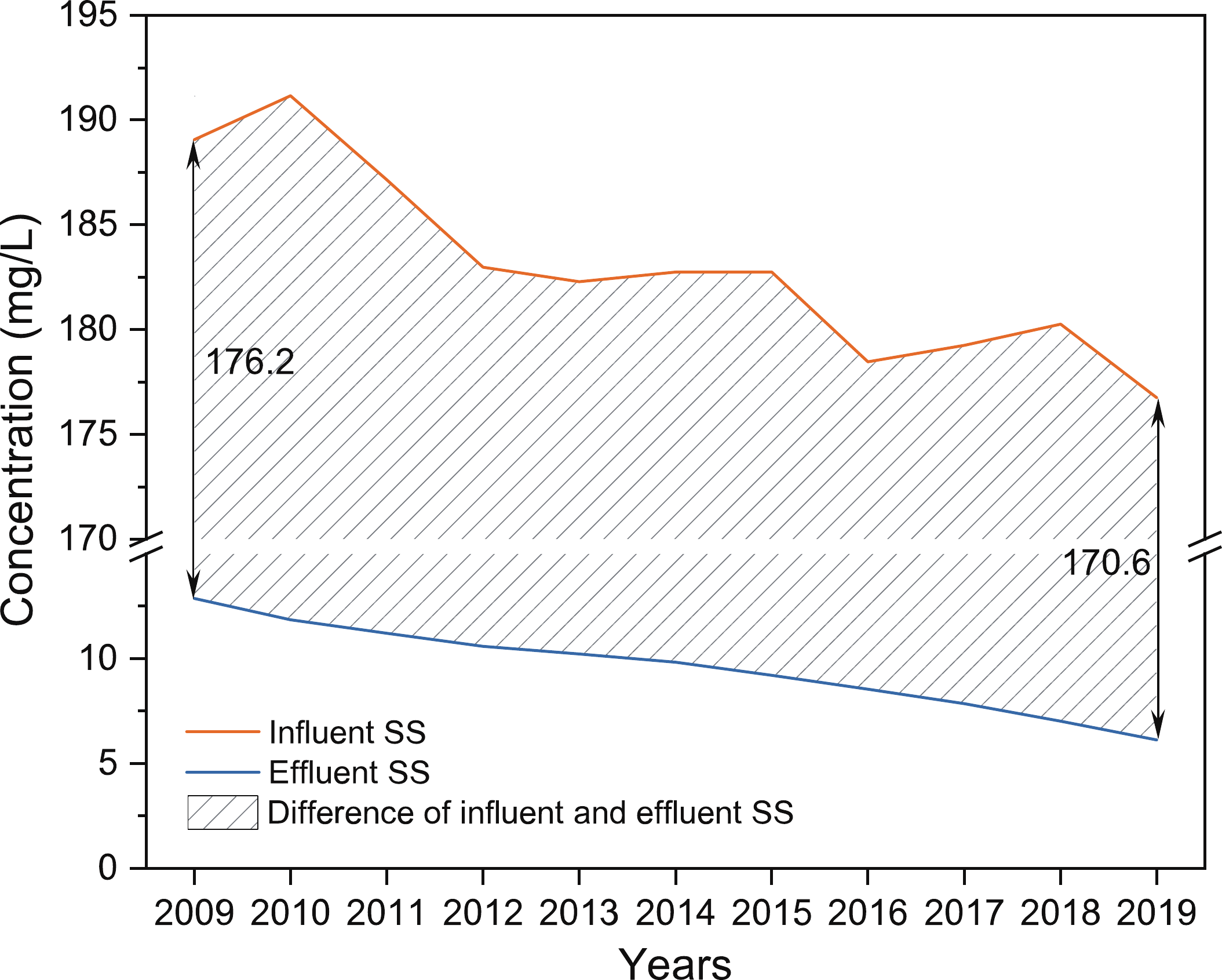}
\includegraphics[width=2.53472in,height=2.07148in]{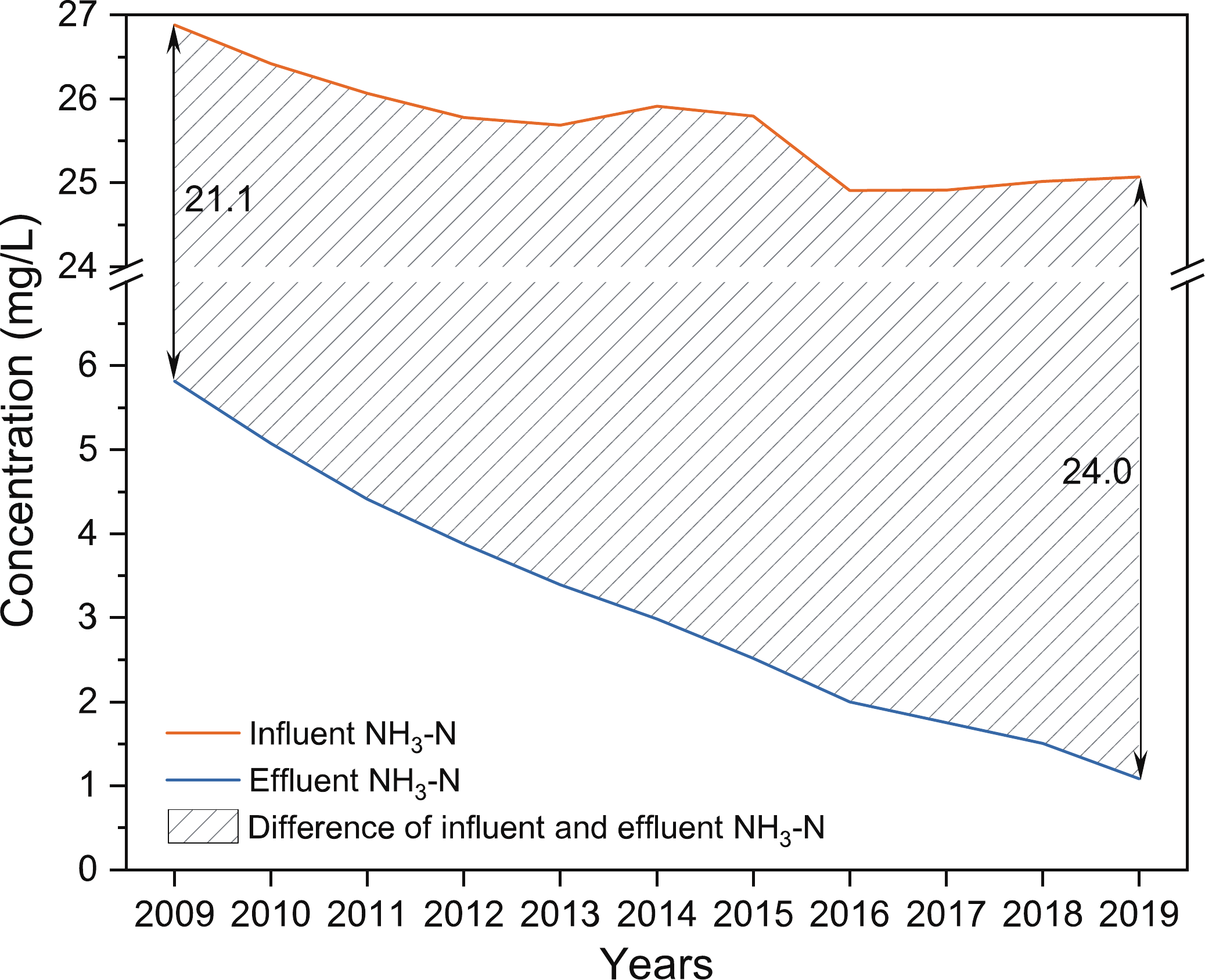}

\includegraphics[width=2.55292in,height=2.06944in]{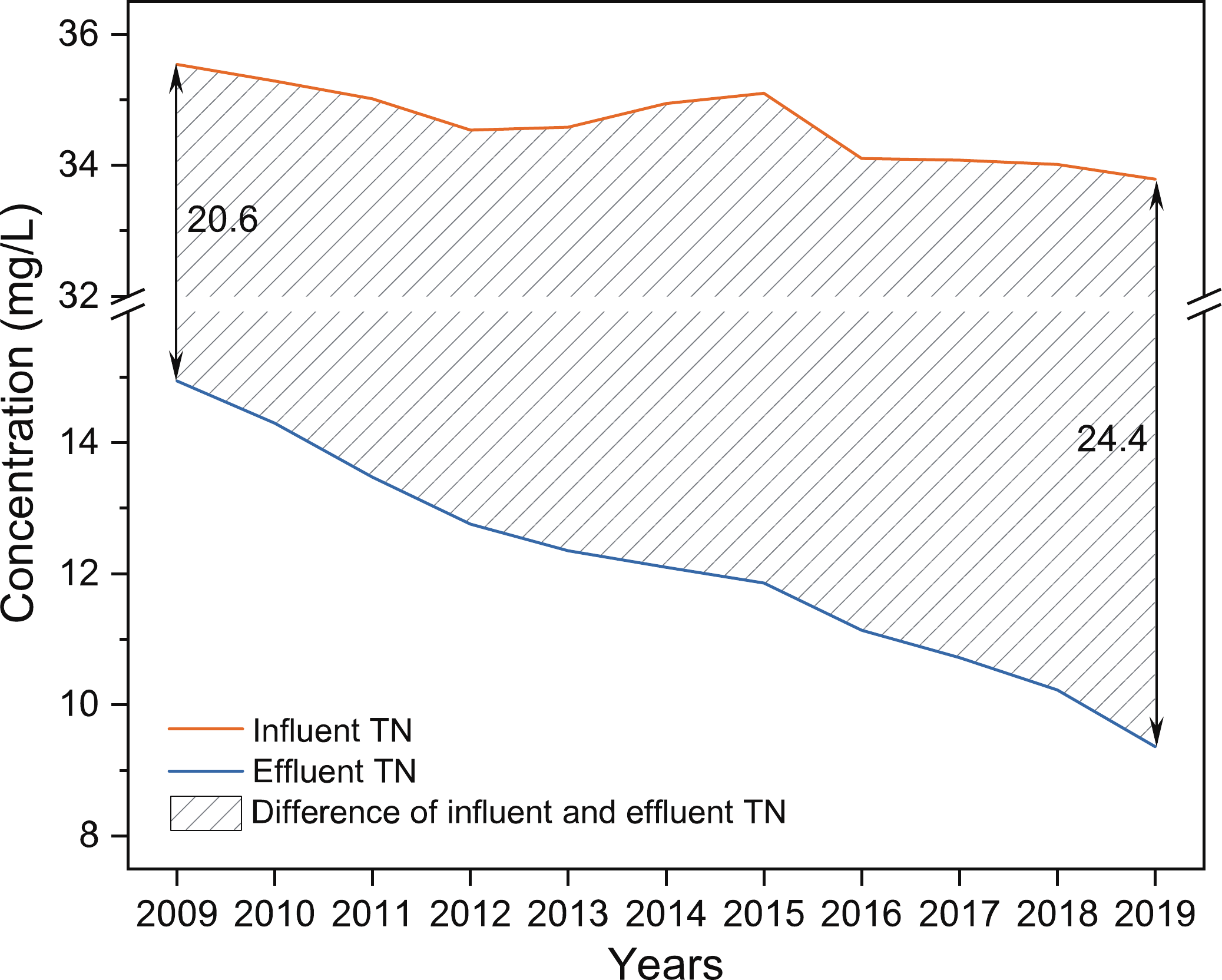}
\includegraphics[width=2.54167in,height=2.05854in]{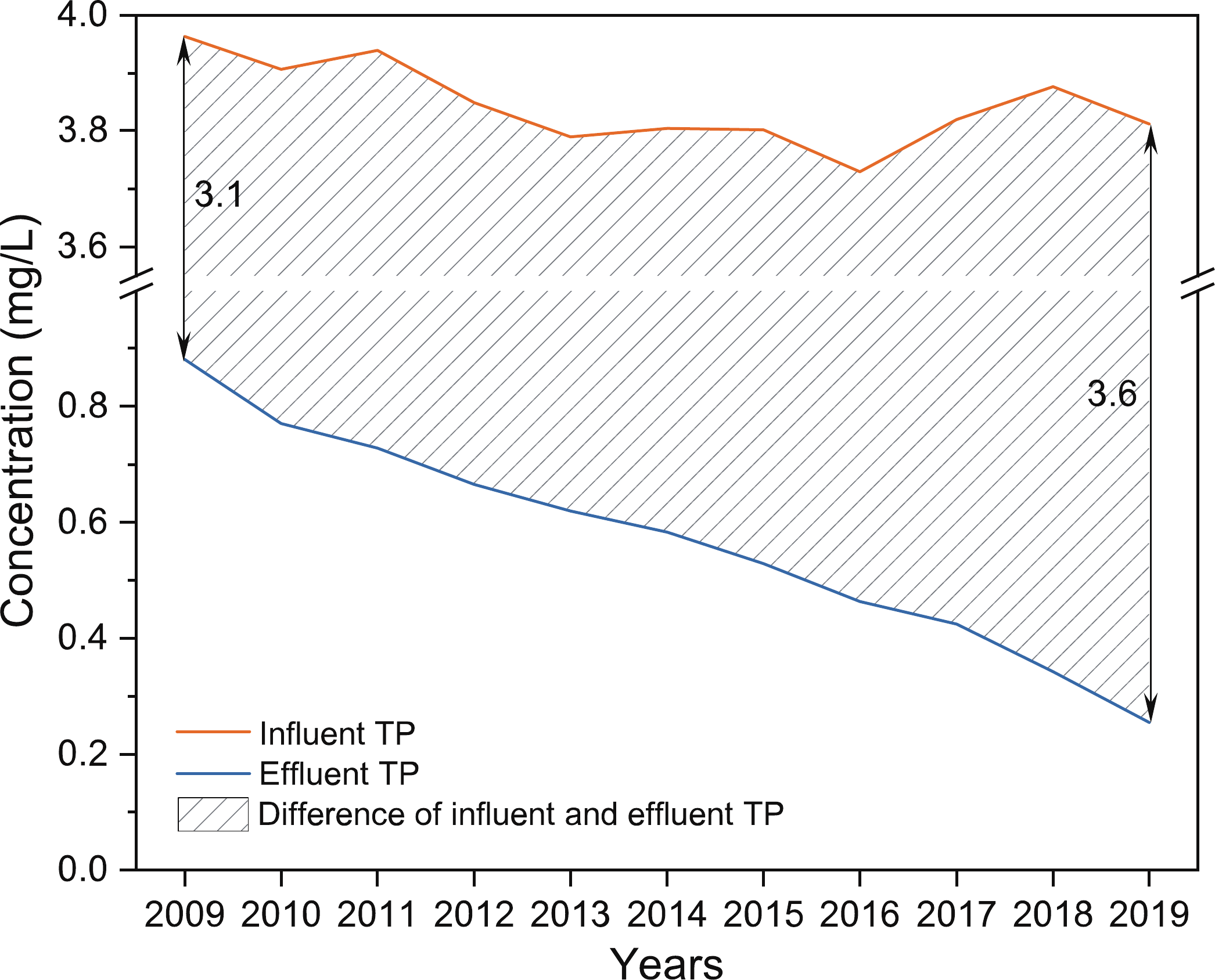}

\includegraphics[width=2.55669in,height=2.11806in]{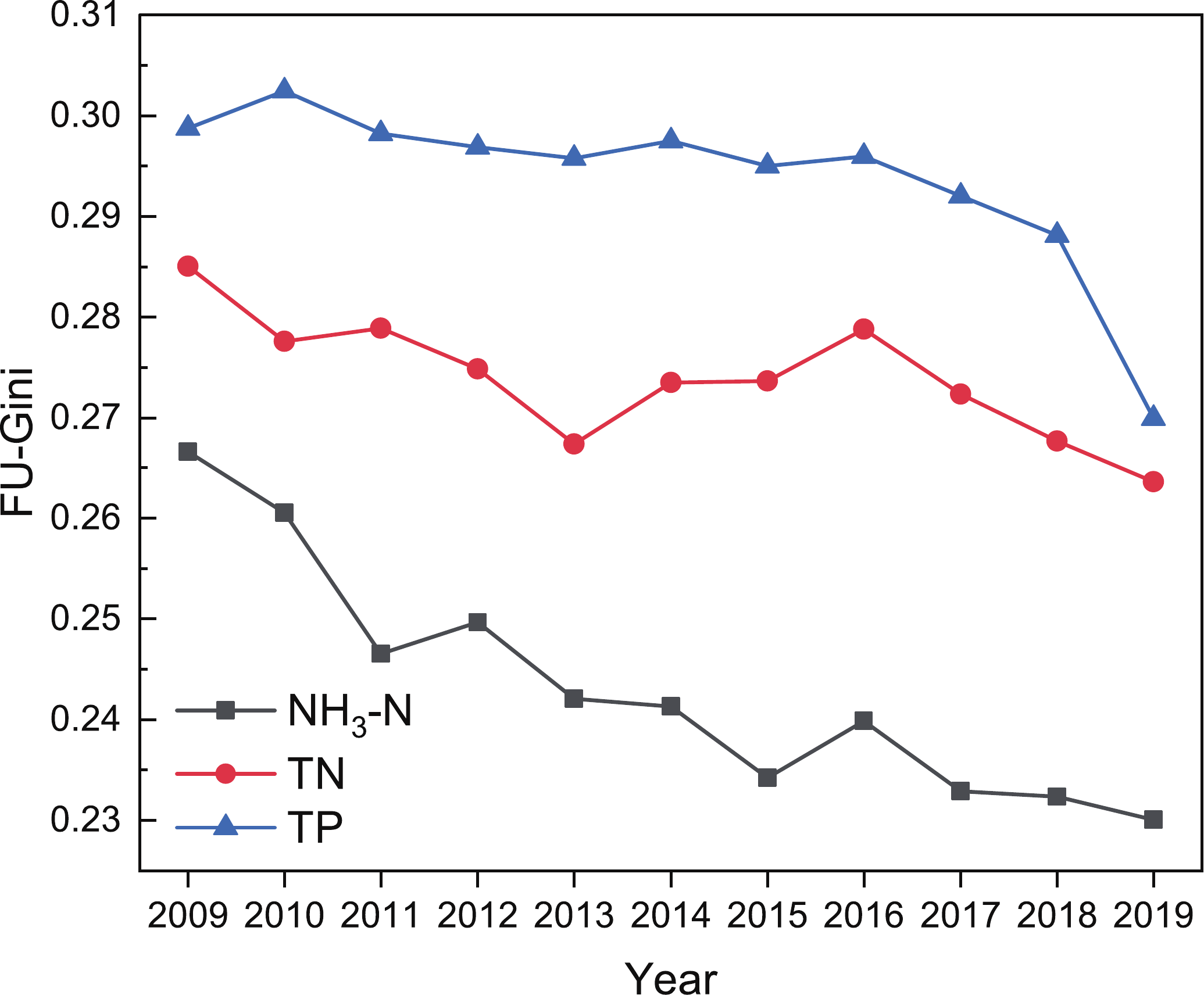}

\textbf{Supplementary Fig.} \textbf{9} Influent and effluent
concentration of \textbf{a,} COD, \textbf{b,} BOD, \textbf{c,} SS,
\textbf{d,} NH\textsubscript{3}-N, \textbf{e,} TN, \textbf{f,} TP, in
2009-2019. \textbf{g,} FU-Gini of reduced NH\textsubscript{3}-N, TN and
TP concentrations (volume of treated wastewater as functional unit).

\includegraphics[width=4.19792in,height=2.76731in]{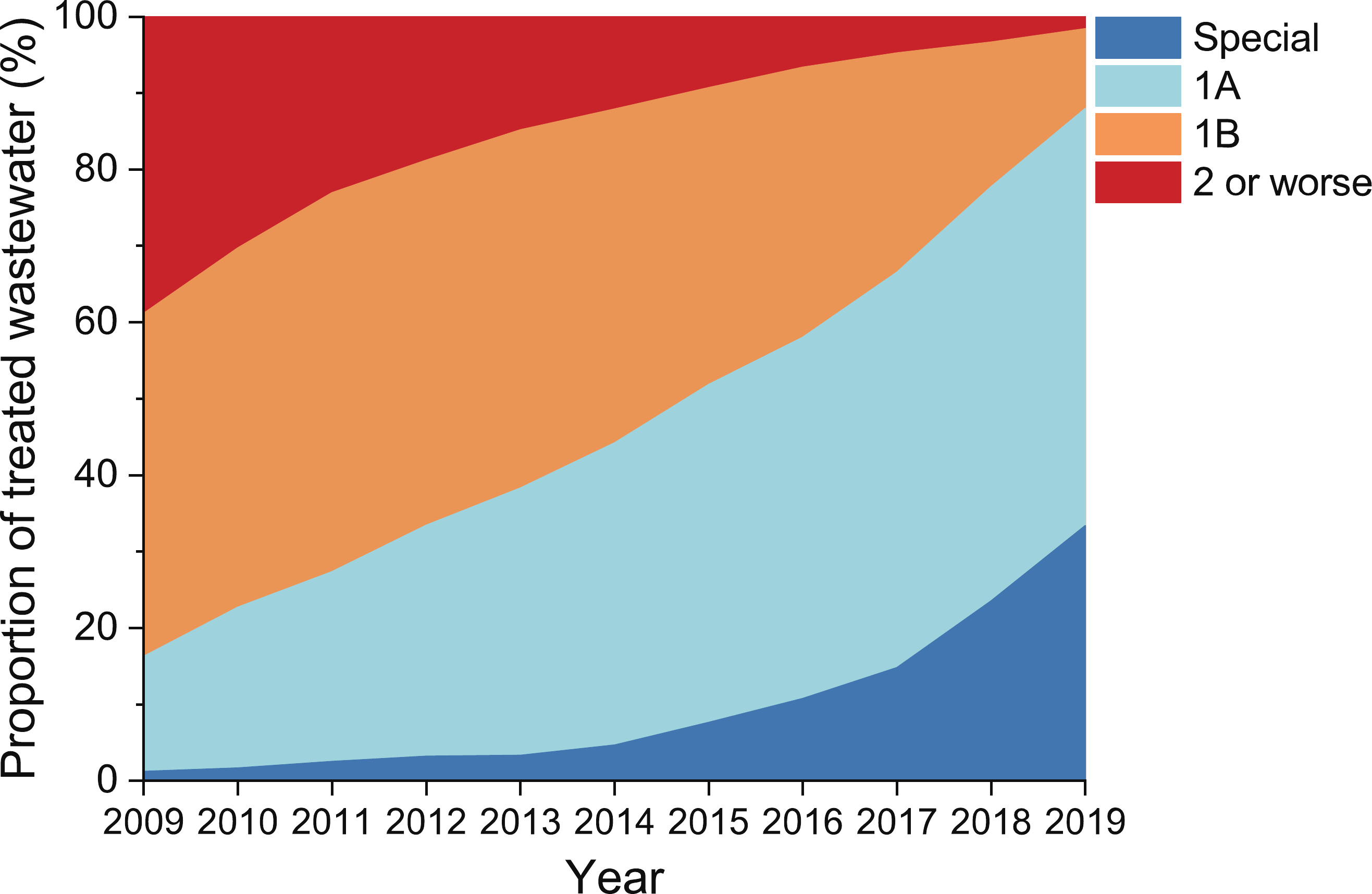}

\textbf{Supplementary Fig.} \textbf{10} In 2009-2019, Proportion of
treated wastewater reaching different standards, in China.

\includegraphics[width=3.24495in,height=1.99284in]{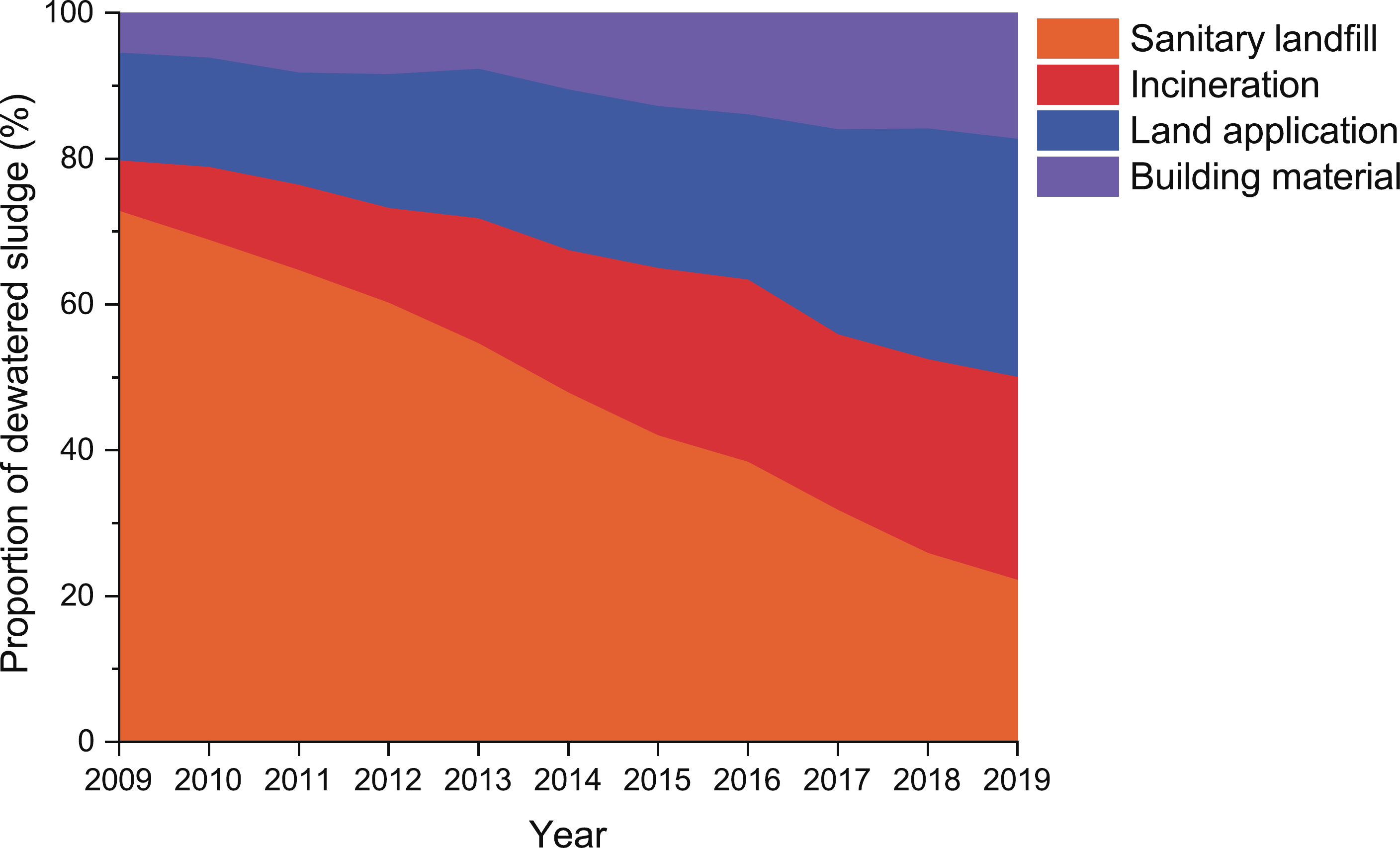}
\includegraphics[width=2.41772in,height=1.98314in]{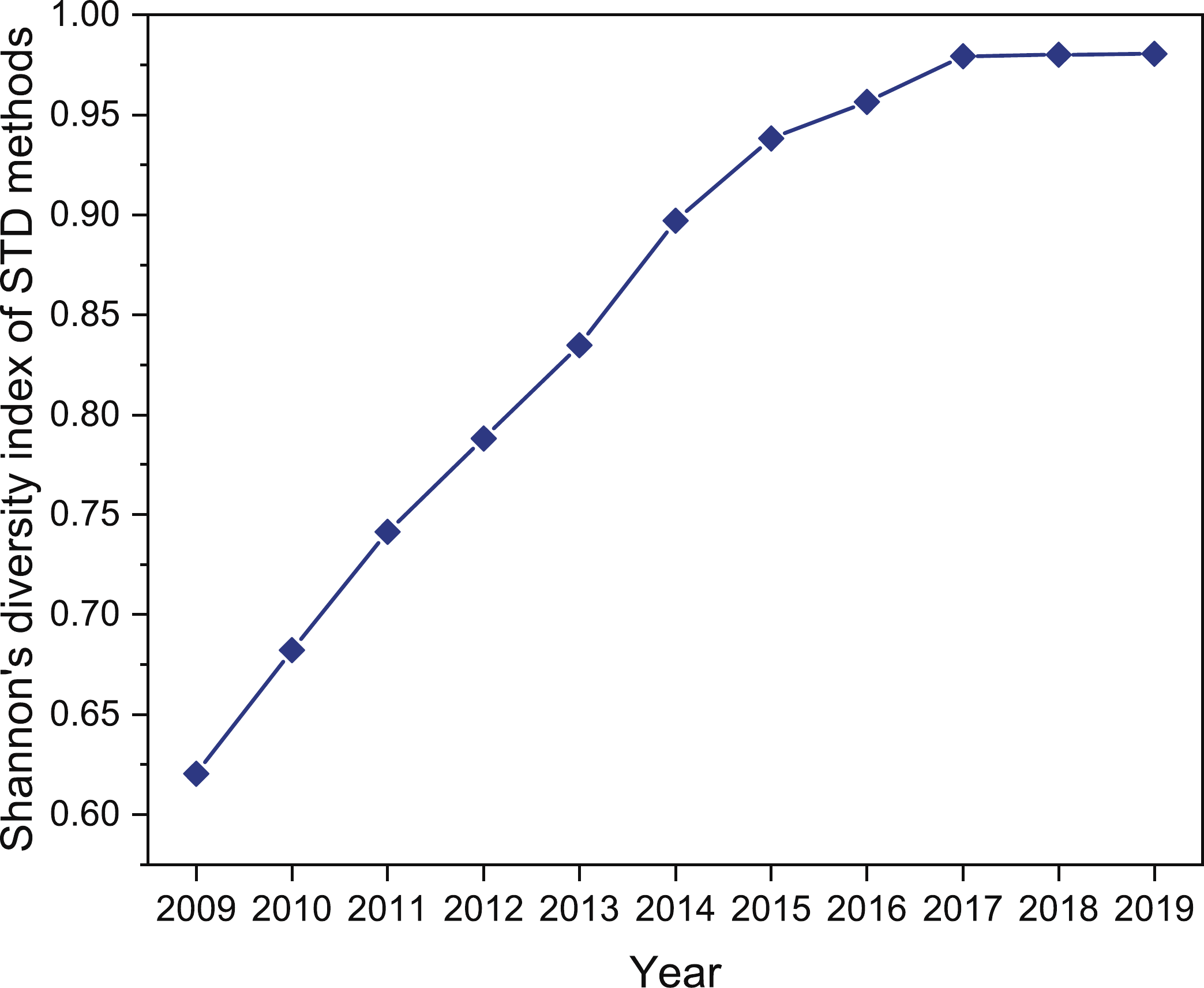}

\textbf{Supplementary Fig.} \textbf{11 a,} Proportion of dewatered
sludge having different ways of treatment and disposal in 2009-2019.
\textbf{b,} Growth of Shannon's diversity\textsuperscript{13} of STD
methods, indicating that the diversity of STD methods have been growing.

\includegraphics[width=2.7619in,height=1.80924in]{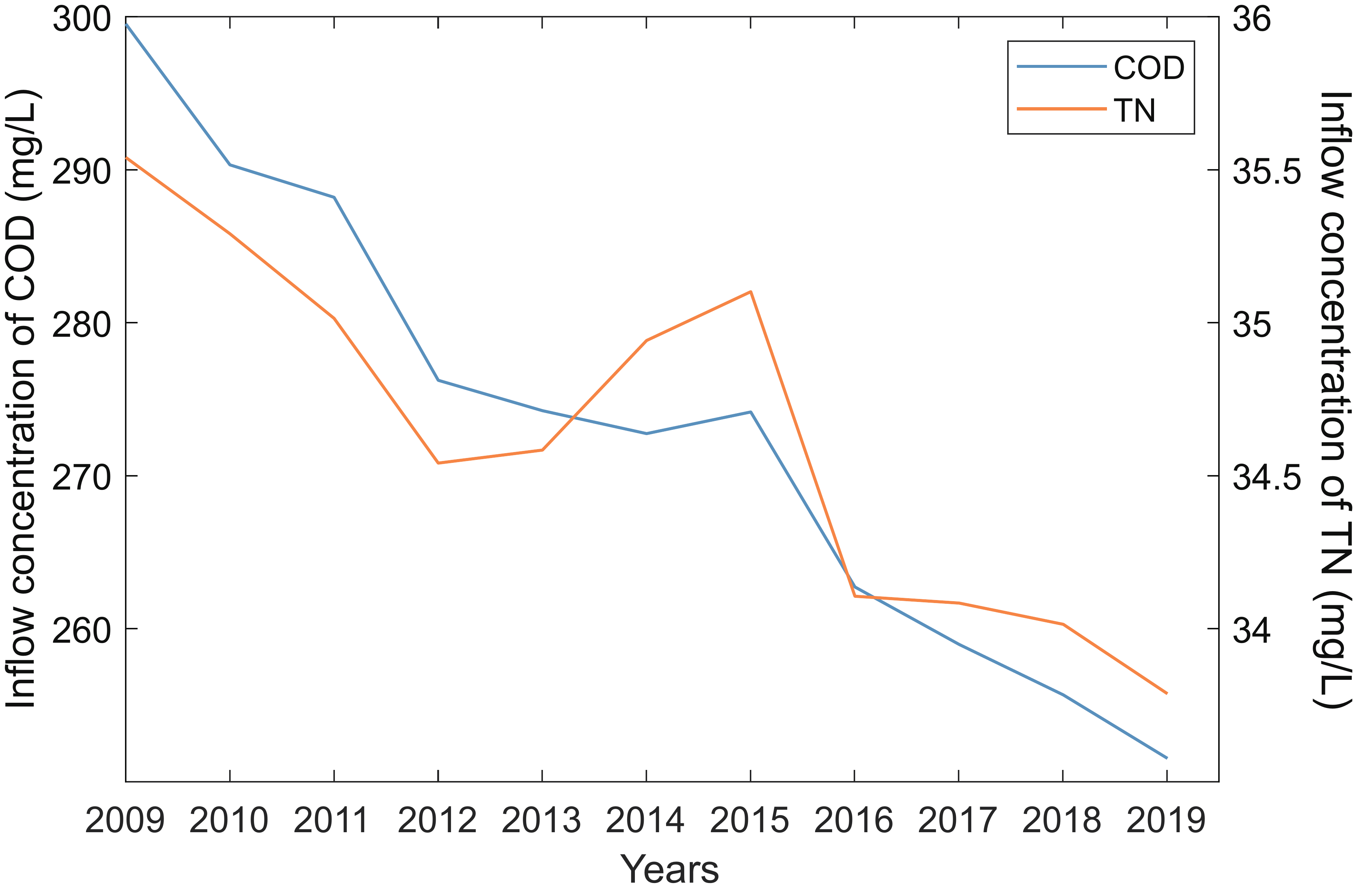}
\includegraphics[width=2.6255in,height=1.85714in]{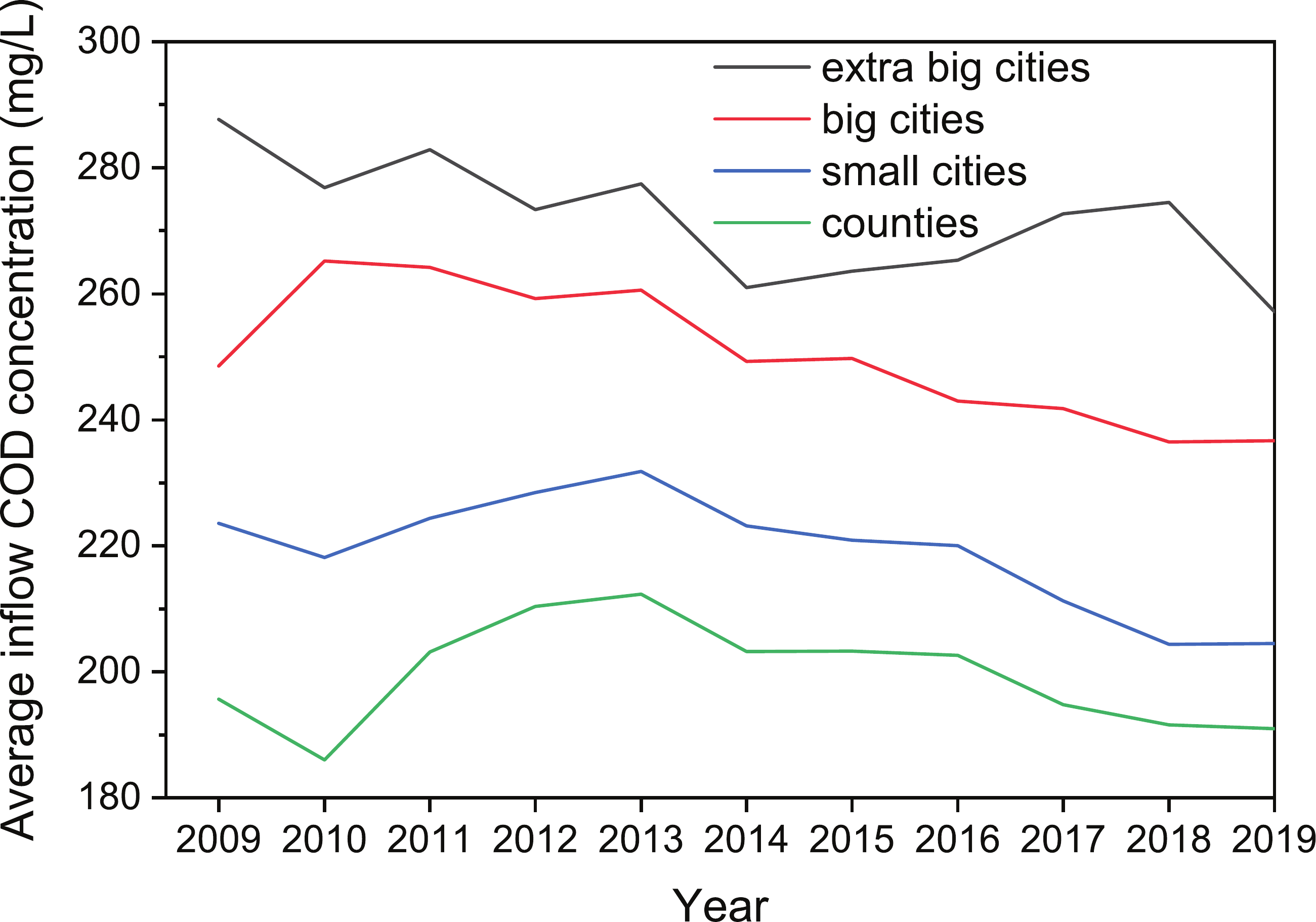}

\textbf{Supplementary Fig.} \textbf{12} Average COD concentration
(wastewater amount-weighted) in \textbf{a,} China (along with TN),
\textbf{b,} different scales of cities and counties, 2009-2019.

\includegraphics[width=5.72785in,height=3.03767in]{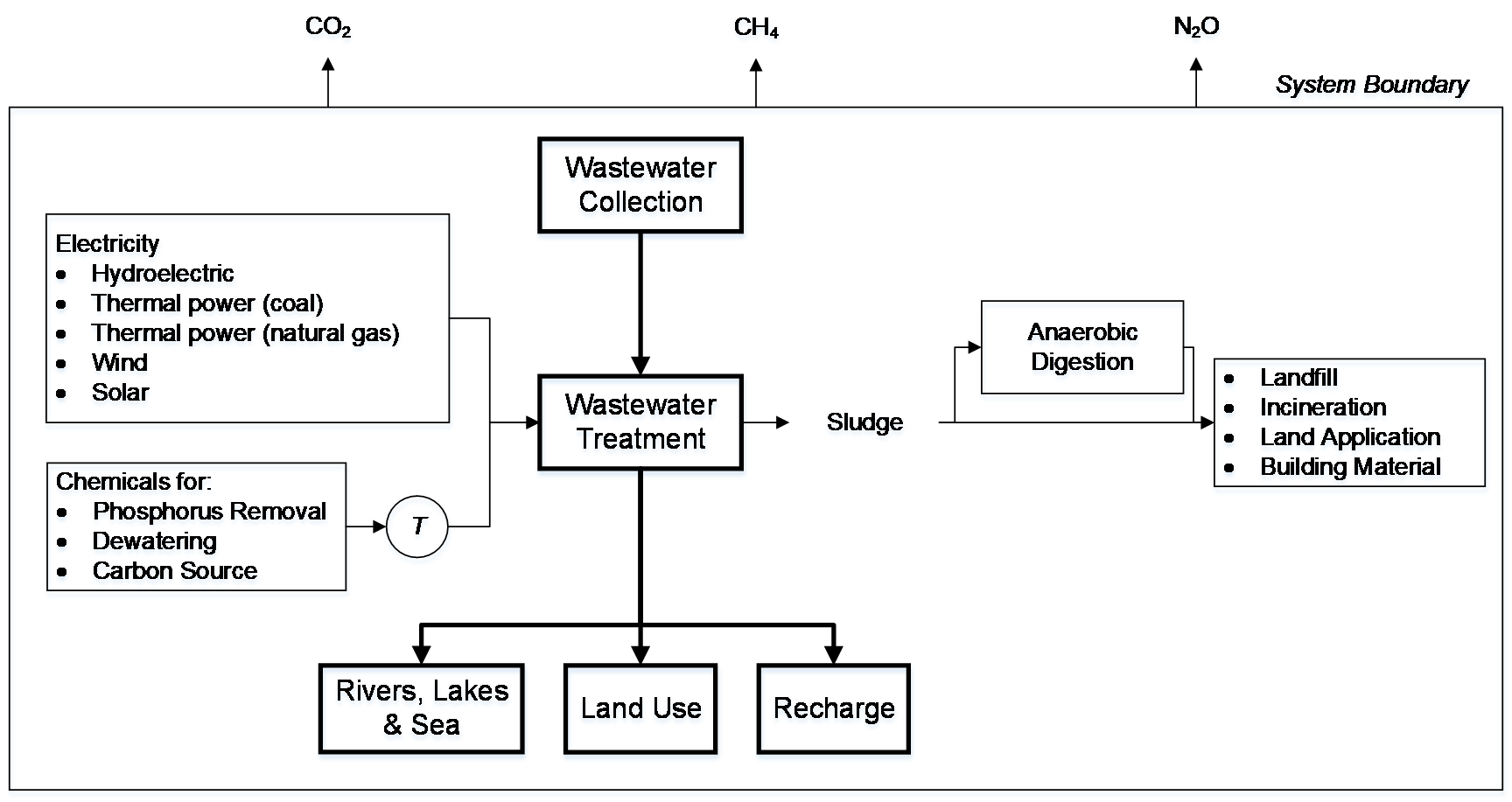}

\textbf{Supplementary Fig.} \textbf{13 Conceptual framework of
greenhouse gas (GHG) accounting for WWTP.} It contains different
situations of provinces. The main process is
collection-treatment-discharge (bold box). (T-transport).

\textbf{Supplementary Table} \textbf{1:} Comparison of China's WWTP
discharge standards (mg/L)

\begin{longtable}[]{@{}
  >{\raggedright\arraybackslash}p{(\columnwidth - 12\tabcolsep) * \real{0.3028}}
  >{\raggedright\arraybackslash}p{(\columnwidth - 12\tabcolsep) * \real{0.1033}}
  >{\raggedright\arraybackslash}p{(\columnwidth - 12\tabcolsep) * \real{0.1255}}
  >{\raggedright\arraybackslash}p{(\columnwidth - 12\tabcolsep) * \real{0.1255}}
  >{\raggedright\arraybackslash}p{(\columnwidth - 12\tabcolsep) * \real{0.1120}}
  >{\raggedright\arraybackslash}p{(\columnwidth - 12\tabcolsep) * \real{0.1320}}
  >{\raggedright\arraybackslash}p{(\columnwidth - 12\tabcolsep) * \real{0.0991}}@{}}
\toprule
\begin{minipage}[b]{\linewidth}\raggedright
\textbf{Parameter}
\end{minipage} & \begin{minipage}[b]{\linewidth}\raggedright
\textbf{Special limit}
\end{minipage} & \begin{minipage}[b]{\linewidth}\raggedright
\textbf{Class 1A}
\end{minipage} & \begin{minipage}[b]{\linewidth}\raggedright
\textbf{Class 1B}
\end{minipage} & \begin{minipage}[b]{\linewidth}\raggedright
\textbf{Class 2}
\end{minipage} & \begin{minipage}[b]{\linewidth}\raggedright
\textbf{Germany}
\end{minipage} & \begin{minipage}[b]{\linewidth}\raggedright
\textbf{United States}
\end{minipage} \\
\midrule
\endhead
Chemical oxygen demand (COD) & 30 & 50 & 60 & 80 & 75 & n/a \\
Biological oxygen demand (BOD\textsubscript{5}) & 6 & 10 & 20 & 30 & 15
& 30 \\
Suspended solids (SS) & 5 & 10 & 20 & 30 & n/a & 30 \\
Ammonium-nitrogen (NH\textsubscript{3}-N) & 3 (5) & 5 (8) & 8 (15) & 15
(20) & 10 & n/a \\
Total nitrogen (TN) & 15 & 15 & 20 & 25 & 13 & n/a \\
Total phosphorus (TP) & 0.3 & 0.5 & 1 & 1 & 1 & n/a \\
\bottomrule
\end{longtable}

Note: For NH\textsubscript{3}-N, figures in brackets are the indices for
conditions of water temperature higher than 12℃, and that out of
brackets are the indices for conditions of water temperature not higher
than 12℃. In this paper, the water temperature is assumed to be the same
as the air temperature at the representative weather station in
corresponding province.

\textbf{Supplementary Table} \textbf{2:} Top 3 cities of WWT GHG
emission and amount of treated wastewater

\begin{longtable}[]{@{}
  >{\raggedright\arraybackslash}p{(\columnwidth - 8\tabcolsep) * \real{0.1246}}
  >{\raggedright\arraybackslash}p{(\columnwidth - 8\tabcolsep) * \real{0.2071}}
  >{\raggedright\arraybackslash}p{(\columnwidth - 8\tabcolsep) * \real{0.2408}}
  >{\raggedright\arraybackslash}p{(\columnwidth - 8\tabcolsep) * \real{0.2154}}
  >{\raggedright\arraybackslash}p{(\columnwidth - 8\tabcolsep) * \real{0.2121}}@{}}
\toprule
\begin{minipage}[b]{\linewidth}\raggedright
\textbf{City}
\end{minipage} & \begin{minipage}[b]{\linewidth}\raggedright
\textbf{GHG emission (Mt)}
\end{minipage} & \begin{minipage}[b]{\linewidth}\raggedright
\textbf{Treated wastewater (10\textsuperscript{6} m\textsuperscript{3})}
\end{minipage} & \begin{minipage}[b]{\linewidth}\raggedright
\textbf{Removed COD (kt)}
\end{minipage} & \begin{minipage}[b]{\linewidth}\raggedright
\textbf{Removed BOD (kt)}
\end{minipage} \\
\midrule
\endhead
Beijing & 2.14 & 1980 & 757.3 & 338.3 \\
Shenzhen & 2.07 & 1811 & 482.0 & 193.6 \\
Shanghai & 1.65 & 2152 & 613.9 & 289.1 \\
\textbf{City} & \textbf{Removed SS (kt)} & \textbf{Removed
NH\textsubscript{3} (kt)} & \textbf{Removed TN (kt)} & \textbf{Removed
TP (kt)} \\
Beijing & 404.8 & 72.9 & 74.8 & 9.87 \\
Shenzhen & 479.9 & 42.3 & 43.6 & 8.60 \\
Shanghai & 349.8 & 50.0 & 50.3 & 8.07 \\
\bottomrule
\end{longtable}

\textbf{Supplementary Table} \textbf{3:} Length and proportion of
wastewater pipe in combined and separated sewer system in
2015\textsuperscript{1}

\begin{longtable}[]{@{}
  >{\raggedright\arraybackslash}p{(\columnwidth - 12\tabcolsep) * \real{0.1075}}
  >{\raggedright\arraybackslash}p{(\columnwidth - 12\tabcolsep) * \real{0.1488}}
  >{\raggedright\arraybackslash}p{(\columnwidth - 12\tabcolsep) * \real{0.1488}}
  >{\raggedright\arraybackslash}p{(\columnwidth - 12\tabcolsep) * \real{0.1488}}
  >{\raggedright\arraybackslash}p{(\columnwidth - 12\tabcolsep) * \real{0.1488}}
  >{\raggedright\arraybackslash}p{(\columnwidth - 12\tabcolsep) * \real{0.1488}}
  >{\raggedright\arraybackslash}p{(\columnwidth - 12\tabcolsep) * \real{0.1488}}@{}}
\toprule
\multirow{2}{*}{\begin{minipage}[b]{\linewidth}\raggedright
\textbf{City / County}
\end{minipage}} &
\multicolumn{2}{l}{\begin{minipage}[b]{\linewidth}\raggedright
\textbf{Pipe in}

\textbf{combined system}
\end{minipage}} &
\multicolumn{2}{l}{\begin{minipage}[b]{\linewidth}\raggedright
\textbf{Wastewater pipe in separated system}
\end{minipage}} &
\multicolumn{2}{l}{\begin{minipage}[b]{\linewidth}\raggedright
\textbf{Rainwater pipe in separated system}
\end{minipage}} \\
& \begin{minipage}[b]{\linewidth}\raggedright
\textbf{Length (10\textsuperscript{3} km)}
\end{minipage} & \begin{minipage}[b]{\linewidth}\raggedright
\textbf{Proportion (\%)}
\end{minipage} & \begin{minipage}[b]{\linewidth}\raggedright
\textbf{Length (10\textsuperscript{3} km)}
\end{minipage} & \begin{minipage}[b]{\linewidth}\raggedright
\textbf{Proportion (\%)}
\end{minipage} & \begin{minipage}[b]{\linewidth}\raggedright
\textbf{Length (10\textsuperscript{3} km)}
\end{minipage} & \begin{minipage}[b]{\linewidth}\raggedright
\textbf{Proportion (\%)}
\end{minipage} \\
\midrule
\endhead
City & 107.8 & 20.0 & 226.2 & 41.9 & 205.6 & 38.1 \\
County & 52.2 & 31.1 & 70.2 & 41.8 & 45.5 & 27.1 \\
\bottomrule
\end{longtable}

\textbf{Supplementary Table} \textbf{4:} Scale of
cities\textsuperscript{2}

\begin{longtable}[]{@{}
  >{\raggedright\arraybackslash}p{(\columnwidth - 4\tabcolsep) * \real{0.2387}}
  >{\raggedright\arraybackslash}p{(\columnwidth - 4\tabcolsep) * \real{0.4773}}
  >{\raggedright\arraybackslash}p{(\columnwidth - 4\tabcolsep) * \real{0.2839}}@{}}
\toprule
\begin{minipage}[b]{\linewidth}\raggedright
\textbf{Scale}
\end{minipage} & \begin{minipage}[b]{\linewidth}\raggedright
\textbf{Threshold of Population}
\end{minipage} & \begin{minipage}[b]{\linewidth}\raggedright
\textbf{Number of Cities}
\end{minipage} \\
\midrule
\endhead
Extra large & {[}5×10\textsuperscript{6}, $+\infty$) & 7 \\
Large & {[}1×10\textsuperscript{6}, 5×10\textsuperscript{6}) & 64 \\
Small & (0, 1×10\textsuperscript{6}) & 557 \\
\bottomrule
\end{longtable}

\textbf{Supplementary Table} \textbf{5:} Pearson and Spearman
correlation between GDP and other factors of 31 provinces

\begin{longtable}[]{@{}
  >{\raggedright\arraybackslash}p{(\columnwidth - 8\tabcolsep) * \real{0.3555}}
  >{\raggedright\arraybackslash}p{(\columnwidth - 8\tabcolsep) * \real{0.1610}}
  >{\raggedright\arraybackslash}p{(\columnwidth - 8\tabcolsep) * \real{0.1611}}
  >{\raggedright\arraybackslash}p{(\columnwidth - 8\tabcolsep) * \real{0.1611}}
  >{\raggedright\arraybackslash}p{(\columnwidth - 8\tabcolsep) * \real{0.1611}}@{}}
\toprule
\multirow{3}{*}{\begin{minipage}[b]{\linewidth}\raggedright
\textbf{Factors}
\end{minipage}} &
\multicolumn{4}{l}{\begin{minipage}[b]{\linewidth}\raggedright
\textbf{GDP}
\end{minipage}} \\
& \multicolumn{2}{l}{\begin{minipage}[b]{\linewidth}\raggedright
Pearson
\end{minipage}} &
\multicolumn{2}{l}{\begin{minipage}[b]{\linewidth}\raggedright
Spearman
\end{minipage}} \\
& \begin{minipage}[b]{\linewidth}\raggedright
Correlation coefficient
\end{minipage} & \begin{minipage}[b]{\linewidth}\raggedright
P-value
\end{minipage} & \begin{minipage}[b]{\linewidth}\raggedright
Correlation coefficient
\end{minipage} & \begin{minipage}[b]{\linewidth}\raggedright
P-value
\end{minipage} \\
\midrule
\endhead
GHG intensity & 0.562*** & 0.001 & 0.447** & 0.012 \\
Proportion of treated wastewater reaching 1A or stricter standard &
0.477*** & 0.007 & 0.488*** & 0.005 \\
Observed sludge yield & 0.429** & 0.016 & 0.468*** & 0.008 \\
\bottomrule
\end{longtable}

\textbf{Supplementary Table} \textbf{6:} One-way analysis of variance
(ANOVA) towards WWTPs in Beijing and Shanghai

\begin{longtable}[]{@{}
  >{\raggedright\arraybackslash}p{(\columnwidth - 12\tabcolsep) * \real{0.2048}}
  >{\raggedright\arraybackslash}p{(\columnwidth - 12\tabcolsep) * \real{0.2048}}
  >{\raggedright\arraybackslash}p{(\columnwidth - 12\tabcolsep) * \real{0.0854}}
  >{\raggedright\arraybackslash}p{(\columnwidth - 12\tabcolsep) * \real{0.1074}}
  >{\raggedright\arraybackslash}p{(\columnwidth - 12\tabcolsep) * \real{0.1999}}
  >{\raggedright\arraybackslash}p{(\columnwidth - 12\tabcolsep) * \real{0.0852}}
  >{\raggedright\arraybackslash}p{(\columnwidth - 12\tabcolsep) * \real{0.1126}}@{}}
\toprule
\multirow{2}{*}{\begin{minipage}[b]{\linewidth}\raggedright
City
\end{minipage}} &
\multicolumn{3}{l}{\begin{minipage}[b]{\linewidth}\raggedright
Sludge generating intensity
\end{minipage}} &
\multicolumn{3}{l}{\begin{minipage}[b]{\linewidth}\raggedright
Electricity intensity
\end{minipage}} \\
& \begin{minipage}[b]{\linewidth}\raggedright
Average

(kg/kg COD)
\end{minipage} & \begin{minipage}[b]{\linewidth}\raggedright
\emph{F}
\end{minipage} & \begin{minipage}[b]{\linewidth}\raggedright
\emph{P}
\end{minipage} & \begin{minipage}[b]{\linewidth}\raggedright
Average

(kWh/kg)
\end{minipage} & \begin{minipage}[b]{\linewidth}\raggedright
\emph{F}
\end{minipage} & \begin{minipage}[b]{\linewidth}\raggedright
\emph{P}
\end{minipage} \\
\midrule
\endhead
Beijing & 0.71 & \multirow{2}{*}{1.80} & \multirow{2}{*}{0.18} & 2.20 &
\multirow{2}{*}{3.27} & \multirow{2}{*}{0.074} \\
Shanghai & 0.65 & & & 1.83 \\
\bottomrule
\end{longtable}

Note: To meet the assumption of normal distribution and variance
homogeneity, data were log-transformed before analysis.

\textbf{Supplementary Table} \textbf{7:} FU-Gini coefficient of 5 GHG
sources in provinces of China in 2009.

\begin{longtable}[]{@{}
  >{\raggedright\arraybackslash}p{(\columnwidth - 12\tabcolsep) * \real{0.1695}}
  >{\raggedright\arraybackslash}p{(\columnwidth - 12\tabcolsep) * \real{0.1161}}
  >{\raggedright\arraybackslash}p{(\columnwidth - 12\tabcolsep) * \real{0.1429}}
  >{\raggedright\arraybackslash}p{(\columnwidth - 12\tabcolsep) * \real{0.1490}}
  >{\raggedright\arraybackslash}p{(\columnwidth - 12\tabcolsep) * \real{0.1405}}
  >{\raggedright\arraybackslash}p{(\columnwidth - 12\tabcolsep) * \real{0.1392}}
  >{\raggedright\arraybackslash}p{(\columnwidth - 12\tabcolsep) * \real{0.1429}}@{}}
\toprule
\begin{minipage}[b]{\linewidth}\raggedright
\textbf{Province}
\end{minipage} & \begin{minipage}[b]{\linewidth}\raggedright
\textbf{Total}
\end{minipage} & \begin{minipage}[b]{\linewidth}\raggedright
\textbf{Collection \& treatment}
\end{minipage} & \begin{minipage}[b]{\linewidth}\raggedright
\textbf{Discharged water}
\end{minipage} & \begin{minipage}[b]{\linewidth}\raggedright
\textbf{Sludge treatment \& disposal}
\end{minipage} & \begin{minipage}[b]{\linewidth}\raggedright
\textbf{Electricity use}
\end{minipage} & \begin{minipage}[b]{\linewidth}\raggedright
\textbf{Chemical use}
\end{minipage} \\
\midrule
\endhead
Beijing & 0.130 & 0.125 & 0.138 & 0.275 & 0.112 & \multirow{31}{*}{/} \\
Tianjin & 0.114 & 0.156 & 0.100 & 0.101 & 0.238 \\
Hebei & 0.181 & 0.288 & 0.268 & 0.411 & 0.281 \\
Shanxi & 0.204 & 0.247 & 0.224 & 0.395 & 0.276 \\
Neimenggu & 0.123 & 0.256 & 0.148 & 0.203 & 0.226 \\
Liaoning & 0.117 & 0.322 & 0.171 & 0.144 & 0.182 \\
Jilin & 0.110 & 0.123 & 0.122 & 0.185 & 0.112 \\
Heilongjiang & 0.107 & 0.057 & 0.063 & 0.187 & 0.166 \\
Shanghai & 0.131 & 0.161 & 0.147 & 0.213 & 0.127 \\
Jiangsu & 0.220 & 0.222 & 0.200 & 0.474 & 0.159 \\
Zhejiang & 0.209 & 0.149 & 0.194 & 0.363 & 0.324 \\
Anhui & 0.220 & 0.231 & 0.183 & 0.447 & 0.238 \\
Fujian & 0.134 & 0.152 & 0.162 & 0.315 & 0.143 \\
Jiangxi & 0.136 & 0.321 & 0.175 & 0.428 & 0.119 \\
Shandong & 0.259 & 0.226 & 0.219 & 0.622 & 0.234 \\
Henan & 0.156 & 0.166 & 0.156 & 0.445 & 0.169 \\
Hubei & 0.241 & 0.213 & 0.348 & 0.372 & 0.285 \\
Hunan & 0.126 & 0.206 & 0.137 & 0.286 & 0.151 \\
Guangdong & 0.214 & 0.201 & 0.245 & 0.371 & 0.237 \\
Guangxi & 0.176 & 0.207 & 0.173 & 0.290 & 0.181 \\
Hainan & 0.074 & 0.097 & 0.034 & 0.094 & 0.321 \\
Chongqing & 0.217 & 0.116 & 0.177 & 0.481 & 0.228 \\
Sichuan & 0.182 & 0.175 & 0.273 & 0.376 & 0.229 \\
Guizhou & 0.154 & 0.220 & 0.280 & 0.372 & 0.186 \\
Yunnan & 0.180 & 0.149 & 0.180 & 0.319 & 0.200 \\
Xizang & 0.000 & 0.000 & 0.000 & 0.000 & 0.000 \\
Shaanxi & 0.184 & 0.128 & 0.163 & 0.360 & 0.242 \\
Gansu & 0.301 & 0.332 & 0.326 & 0.456 & 0.273 \\
Qinghai & 0.000 & 0.000 & 0.000 & 0.000 & 0.000 \\
Ningxia & 0.232 & 0.189 & 0.162 & 0.379 & 0.303 \\
Xinjiang & 0.196 & 0.289 & 0.185 & 0.233 & 0.199 \\
\bottomrule
\end{longtable}

Note: For chemical use, only data since 2014 are available.

\textbf{Supplementary Table} \textbf{8:} FU-Gini coefficient of 5 GHG
sources in provinces of China in 2015.

\begin{longtable}[]{@{}
  >{\raggedright\arraybackslash}p{(\columnwidth - 12\tabcolsep) * \real{0.1686}}
  >{\raggedright\arraybackslash}p{(\columnwidth - 12\tabcolsep) * \real{0.1160}}
  >{\raggedright\arraybackslash}p{(\columnwidth - 12\tabcolsep) * \real{0.1440}}
  >{\raggedright\arraybackslash}p{(\columnwidth - 12\tabcolsep) * \real{0.1473}}
  >{\raggedright\arraybackslash}p{(\columnwidth - 12\tabcolsep) * \real{0.1422}}
  >{\raggedright\arraybackslash}p{(\columnwidth - 12\tabcolsep) * \real{0.1392}}
  >{\raggedright\arraybackslash}p{(\columnwidth - 12\tabcolsep) * \real{0.1429}}@{}}
\toprule
\begin{minipage}[b]{\linewidth}\raggedright
\textbf{Province}
\end{minipage} & \begin{minipage}[b]{\linewidth}\raggedright
\textbf{Total}
\end{minipage} & \begin{minipage}[b]{\linewidth}\raggedright
\textbf{Collection \& treatment}
\end{minipage} & \begin{minipage}[b]{\linewidth}\raggedright
\textbf{Discharged water}
\end{minipage} & \begin{minipage}[b]{\linewidth}\raggedright
\textbf{Sludge treatment \& disposal}
\end{minipage} & \begin{minipage}[b]{\linewidth}\raggedright
\textbf{Electricity use}
\end{minipage} & \begin{minipage}[b]{\linewidth}\raggedright
\textbf{Chemical use}
\end{minipage} \\
\midrule
\endhead
Beijing & 0.216 & 0.132 & 0.278 & 0.479 & 0.269 & 0.566 \\
Tianjin & 0.185 & 0.069 & 0.133 & 0.387 & 0.238 & 0.505 \\
Hebei & 0.191 & 0.231 & 0.168 & 0.430 & 0.227 & 0.730 \\
Shanxi & 0.195 & 0.132 & 0.256 & 0.470 & 0.231 & 0.530 \\
Neimenggu & 0.189 & 0.197 & 0.297 & 0.412 & 0.291 & 0.898 \\
Liaoning & 0.185 & 0.173 & 0.304 & 0.354 & 0.208 & 0.804 \\
Jilin & 0.235 & 0.211 & 0.170 & 0.485 & 0.209 & 0.891 \\
Heilongjiang & 0.198 & 0.168 & 0.180 & 0.362 & 0.262 & 0.696 \\
Shanghai & 0.168 & 0.099 & 0.158 & 0.359 & 0.148 & 0.899 \\
Jiangsu & 0.213 & 0.176 & 0.285 & 0.318 & 0.217 & 0.757 \\
Zhejiang & 0.215 & 0.171 & 0.206 & 0.329 & 0.226 & 0.797 \\
Anhui & 0.251 & 0.194 & 0.183 & 0.488 & 0.222 & 0.760 \\
Fujian & 0.225 & 0.168 & 0.189 & 0.400 & 0.243 & 0.675 \\
Jiangxi & 0.220 & 0.156 & 0.156 & 0.517 & 0.208 & 0.992 \\
Shandong & 0.260 & 0.181 & 0.236 & 0.541 & 0.227 & 0.656 \\
Henan & 0.172 & 0.177 & 0.198 & 0.440 & 0.210 & 0.725 \\
Hubei & 0.216 & 0.188 & 0.165 & 0.425 & 0.223 & 0.828 \\
Hunan & 0.229 & 0.207 & 0.213 & 0.418 & 0.230 & 0.938 \\
Guangdong & 0.240 & 0.172 & 0.244 & 0.410 & 0.226 & 0.762 \\
Guangxi & 0.278 & 0.222 & 0.183 & 0.576 & 0.256 & 0.922 \\
Hainan & 0.145 & 0.066 & 0.115 & 0.381 & 0.295 & 0.351 \\
Chongqing & 0.224 & 0.100 & 0.144 & 0.425 & 0.164 & 0.884 \\
Sichuan & 0.261 & 0.122 & 0.182 & 0.445 & 0.204 & 0.823 \\
Guizhou & 0.126 & 0.172 & 0.138 & 0.295 & 0.213 & 0.760 \\
Yunnan & 0.176 & 0.143 & 0.230 & 0.333 & 0.222 & 0.849 \\
Xizang & 0.000 & 0.000 & 0.000 & 0.000 & 0.000 & 0.000 \\
Shaanxi & 0.196 & 0.165 & 0.236 & 0.377 & 0.224 & 0.816 \\
Gansu & 0.249 & 0.167 & 0.248 & 0.439 & 0.314 & 0.956 \\
Qinghai & 0.169 & 0.150 & 0.169 & 0.208 & 0.300 & 0.943 \\
Ningxia & 0.186 & 0.163 & 0.204 & 0.539 & 0.239 & 0.895 \\
Xinjiang & 0.163 & 0.151 & 0.174 & 0.393 & 0.375 & 0.931 \\
\bottomrule
\end{longtable}

\textbf{Supplementary Table} \textbf{9:} FU-Gini coefficient of 5 GHG
sources in provinces of China in 2019.

\begin{longtable}[]{@{}
  >{\raggedright\arraybackslash}p{(\columnwidth - 12\tabcolsep) * \real{0.1695}}
  >{\raggedright\arraybackslash}p{(\columnwidth - 12\tabcolsep) * \real{0.1161}}
  >{\raggedright\arraybackslash}p{(\columnwidth - 12\tabcolsep) * \real{0.1429}}
  >{\raggedright\arraybackslash}p{(\columnwidth - 12\tabcolsep) * \real{0.1490}}
  >{\raggedright\arraybackslash}p{(\columnwidth - 12\tabcolsep) * \real{0.1405}}
  >{\raggedright\arraybackslash}p{(\columnwidth - 12\tabcolsep) * \real{0.1392}}
  >{\raggedright\arraybackslash}p{(\columnwidth - 12\tabcolsep) * \real{0.1429}}@{}}
\toprule
\begin{minipage}[b]{\linewidth}\raggedright
\textbf{Province}
\end{minipage} & \begin{minipage}[b]{\linewidth}\raggedright
\textbf{Total}
\end{minipage} & \begin{minipage}[b]{\linewidth}\raggedright
\textbf{Collection \& treatment}
\end{minipage} & \begin{minipage}[b]{\linewidth}\raggedright
\textbf{Discharged water}
\end{minipage} & \begin{minipage}[b]{\linewidth}\raggedright
\textbf{Sludge treatment \& disposal}
\end{minipage} & \begin{minipage}[b]{\linewidth}\raggedright
\textbf{Electricity use}
\end{minipage} & \begin{minipage}[b]{\linewidth}\raggedright
\textbf{Chemical use}
\end{minipage} \\
\midrule
\endhead
Beijing & 0.217 & 0.139 & 0.200 & 0.326 & 0.535 & 0.228 \\
Tianjin & 0.271 & 0.148 & 0.189 & 0.389 & 0.599 & 0.320 \\
Hebei & 0.242 & 0.227 & 0.221 & 0.611 & 0.486 & 0.241 \\
Shanxi & 0.216 & 0.141 & 0.184 & 0.547 & 0.425 & 0.213 \\
Neimenggu & 0.170 & 0.187 & 0.274 & 0.624 & 0.315 & 0.267 \\
Liaoning & 0.262 & 0.189 & 0.235 & 0.603 & 0.566 & 0.265 \\
Jilin & 0.239 & 0.178 & 0.178 & 0.917 & 0.597 & 0.204 \\
Heilongjiang & 0.195 & 0.166 & 0.234 & 0.746 & 0.509 & 0.262 \\
Shanghai & 0.149 & 0.092 & 0.170 & 0.745 & 0.238 & 0.170 \\
Jiangsu & 0.203 & 0.151 & 0.239 & 0.609 & 0.266 & 0.217 \\
Zhejiang & 0.220 & 0.157 & 0.193 & 0.589 & 0.311 & 0.212 \\
Anhui & 0.232 & 0.147 & 0.253 & 0.693 & 0.464 & 0.212 \\
Fujian & 0.288 & 0.112 & 0.207 & 0.520 & 0.517 & 0.201 \\
Jiangxi & 0.239 & 0.153 & 0.164 & 0.625 & 0.506 & 0.224 \\
Shandong & 0.301 & 0.168 & 0.228 & 0.536 & 0.525 & 0.252 \\
Henan & 0.229 & 0.169 & 0.223 & 0.664 & 0.466 & 0.233 \\
Hubei & 0.283 & 0.160 & 0.198 & 0.578 & 0.504 & 0.201 \\
Hunan & 0.257 & 0.199 & 0.242 & 0.740 & 0.410 & 0.271 \\
Guangdong & 0.272 & 0.145 & 0.259 & 0.676 & 0.410 & 0.250 \\
Guangxi & 0.255 & 0.169 & 0.216 & 0.724 & 0.547 & 0.261 \\
Hainan & 0.362 & 0.109 & 0.138 & 0.389 & 0.585 & 0.262 \\
Chongqing & 0.355 & 0.112 & 0.233 & 0.637 & 0.502 & 0.235 \\
Sichuan & 0.290 & 0.161 & 0.221 & 0.692 & 0.521 & 0.247 \\
Guizhou & 0.255 & 0.261 & 0.263 & 0.632 & 0.422 & 0.331 \\
Yunnan & 0.209 & 0.112 & 0.326 & 0.807 & 0.469 & 0.240 \\
Xizang & 0.136 & 0.200 & 0.136 & 0.685 & 0.264 & 0.234 \\
Shaanxi & 0.249 & 0.137 & 0.214 & 0.718 & 0.482 & 0.251 \\
Gansu & 0.223 & 0.180 & 0.258 & 0.770 & 0.423 & 0.267 \\
Qinghai & 0.285 & 0.251 & 0.160 & 0.353 & 0.455 & 0.329 \\
Ningxia & 0.266 & 0.119 & 0.139 & 0.392 & 0.594 & 0.232 \\
Xinjiang & 0.211 & 0.131 & 0.326 & 0.812 & 0.485 & 0.371 \\
\bottomrule
\end{longtable}

\textbf{Supplementary Table} \textbf{10:} FU-Gini coefficient of 5 GHG
sources in different scales of cities or counties in China.

\begin{longtable}[]{@{}
  >{\raggedright\arraybackslash}p{(\columnwidth - 14\tabcolsep) * \real{0.0848}}
  >{\raggedright\arraybackslash}p{(\columnwidth - 14\tabcolsep) * \real{0.1186}}
  >{\raggedright\arraybackslash}p{(\columnwidth - 14\tabcolsep) * \real{0.0848}}
  >{\raggedright\arraybackslash}p{(\columnwidth - 14\tabcolsep) * \real{0.1356}}
  >{\raggedright\arraybackslash}p{(\columnwidth - 14\tabcolsep) * \real{0.1491}}
  >{\raggedright\arraybackslash}p{(\columnwidth - 14\tabcolsep) * \real{0.1424}}
  >{\raggedright\arraybackslash}p{(\columnwidth - 14\tabcolsep) * \real{0.1424}}
  >{\raggedright\arraybackslash}p{(\columnwidth - 14\tabcolsep) * \real{0.1424}}@{}}
\toprule
\begin{minipage}[b]{\linewidth}\raggedright
\textbf{Year}
\end{minipage} & \begin{minipage}[b]{\linewidth}\raggedright
\textbf{Scale of cities or counties}
\end{minipage} & \begin{minipage}[b]{\linewidth}\raggedright
\textbf{Total}
\end{minipage} & \begin{minipage}[b]{\linewidth}\raggedright
\textbf{Collection \& treatment}
\end{minipage} & \begin{minipage}[b]{\linewidth}\raggedright
\textbf{Discharged water}
\end{minipage} & \begin{minipage}[b]{\linewidth}\raggedright
\textbf{Sludge treatment \& disposal}
\end{minipage} & \begin{minipage}[b]{\linewidth}\raggedright
\textbf{Electricity use}
\end{minipage} & \begin{minipage}[b]{\linewidth}\raggedright
\textbf{Chemical use}
\end{minipage} \\
\midrule
\endhead
\multirow{4}{*}{2009} & Extra big & 0.199 & 0.177 & 0.205 & 0.353 &
0.207 & / \\
& Big & 0.188 & 0.210 & 0.231 & 0.391 & 0.246 & / \\
& Small & 0.243 & 0.249 & 0.238 & 0.539 & 0.246 & / \\
& County & 0.204 & 0.254 & 0.230 & 0.486 & 0.266 & / \\
\multirow{4}{*}{2015} & Extra big & 0.254 & 0.134 & 0.254 & 0.474 &
0.270 & 0.734 \\
& Big & 0.272 & 0.176 & 0.269 & 0.506 & 0.272 & 0.766 \\
& Small & 0.262 & 0.202 & 0.228 & 0.509 & 0.297 & 0.857 \\
& County & 0.222 & 0.205 & 0.243 & 0.495 & 0.289 & 0.843 \\
\multirow{4}{*}{2019} & Extra big & 0.290 & 0.141 & 0.234 & 0.518 &
0.277 & 0.646 \\
& Big & 0.308 & 0.166 & 0.246 & 0.520 & 0.286 & 0.646 \\
& Small & 0.291 & 0.191 & 0.245 & 0.500 & 0.299 & 0.705 \\
& County & 0.279 & 0.202 & 0.260 & 0.517 & 0.306 & 0.694 \\
\bottomrule
\end{longtable}

\textbf{Supplementary Table} \textbf{11:} Operation data of WWTPs in
provinces

\begin{longtable}[]{@{}
  >{\raggedright\arraybackslash}p{(\columnwidth - 12\tabcolsep) * \real{0.1375}}
  >{\raggedright\arraybackslash}p{(\columnwidth - 12\tabcolsep) * \real{0.0383}}
  >{\raggedright\arraybackslash}p{(\columnwidth - 12\tabcolsep) * \real{0.2890}}
  >{\raggedright\arraybackslash}p{(\columnwidth - 12\tabcolsep) * \real{0.1756}}
  >{\raggedright\arraybackslash}p{(\columnwidth - 12\tabcolsep) * \real{0.1385}}
  >{\raggedright\arraybackslash}p{(\columnwidth - 12\tabcolsep) * \real{0.1369}}
  >{\raggedright\arraybackslash}p{(\columnwidth - 12\tabcolsep) * \real{0.0841}}@{}}
\toprule
\multicolumn{2}{l}{\begin{minipage}[b]{\linewidth}\raggedright
\textbf{Province}
\end{minipage}} & \begin{minipage}[b]{\linewidth}\raggedright
\textbf{Increased electricity intensity (2013-2015)}

\textbf{(kWh/m\textsuperscript{3})}
\end{minipage} &
\multicolumn{2}{l}{\begin{minipage}[b]{\linewidth}\raggedright
\textbf{Increased proportion of wastewater treated to Class 1A
(2013-2015) (\%)}
\end{minipage}} &
\multicolumn{2}{l}{\begin{minipage}[b]{\linewidth}\raggedright
\textbf{Emissions factors in 2015}

\textbf{(kg CO\textsubscript{2} eq/kWh)}
\end{minipage}} \\
\midrule
\endhead
\multicolumn{2}{l}{Beijing} & 0.155 & \multicolumn{2}{l}{0.4} &
\multicolumn{2}{l}{988.6} \\
\multicolumn{2}{l}{Tianjin} & 0.126 & \multicolumn{2}{l}{-1.2} &
\multicolumn{2}{l}{1003.9} \\
\multicolumn{2}{l}{Hebei} & 0.085 & \multicolumn{2}{l}{11.9} &
\multicolumn{2}{l}{955.3} \\
\multicolumn{2}{l}{Shanxi} & 0.005 & \multicolumn{2}{l}{11.6} &
\multicolumn{2}{l}{979.6} \\
\multicolumn{2}{l}{Neimenggu} & 0.270 & \multicolumn{2}{l}{-18.0} &
\multicolumn{2}{l}{921.4} \\
\multicolumn{2}{l}{Liaoning} & 0.199 & \multicolumn{2}{l}{15.2} &
\multicolumn{2}{l}{879.5} \\
\multicolumn{2}{l}{Jilin} & 0.176 & \multicolumn{2}{l}{3.2} &
\multicolumn{2}{l}{896.1} \\
\multicolumn{2}{l}{Heilongjiang} & -0.019 & \multicolumn{2}{l}{-4.3} &
\multicolumn{2}{l}{944.7} \\
\multicolumn{2}{l}{Shanghai} & 0.107 & \multicolumn{2}{l}{-1.2} &
\multicolumn{2}{l}{736.4} \\
\multicolumn{2}{l}{Jiangsu} & 0.114 & \multicolumn{2}{l}{11.7} &
\multicolumn{2}{l}{925.1} \\
\multicolumn{2}{l}{Zhejiang} & 0.052 & \multicolumn{2}{l}{20.5} &
\multicolumn{2}{l}{753.6} \\
\multicolumn{2}{l}{Anhui} & 0.016 & \multicolumn{2}{l}{15.1} &
\multicolumn{2}{l}{1015.1} \\
\multicolumn{2}{l}{Fujian} & 0.109 & \multicolumn{2}{l}{6.8} &
\multicolumn{2}{l}{623.2} \\
\multicolumn{2}{l}{Jiangxi} & 0.240 & \multicolumn{2}{l}{-0.1} &
\multicolumn{2}{l}{824.1} \\
\multicolumn{2}{l}{Shandong} & 0.077 & \multicolumn{2}{l}{16.3} &
\multicolumn{2}{l}{1014.0} \\
\multicolumn{2}{l}{Henan} & 0.176 & \multicolumn{2}{l}{16.7} &
\multicolumn{2}{l}{979.9} \\
\multicolumn{2}{l}{Hubei} & -0.018 & \multicolumn{2}{l}{4.5} &
\multicolumn{2}{l}{482.5} \\
\multicolumn{2}{l}{Hunan} & 0.000 & \multicolumn{2}{l}{1.7} &
\multicolumn{2}{l}{592.4} \\
\multicolumn{2}{l}{Guangdong} & -0.112 & \multicolumn{2}{l}{6.2} &
\multicolumn{2}{l}{631.6} \\
\multicolumn{2}{l}{Guangxi} & 0.051 & \multicolumn{2}{l}{-0.7} &
\multicolumn{2}{l}{436.3} \\
\multicolumn{2}{l}{Hainan} & 0.146 & \multicolumn{2}{l}{-2.9} &
\multicolumn{2}{l}{967.4} \\
\multicolumn{2}{l}{Chongqing} & -0.088 & \multicolumn{2}{l}{25.5} &
\multicolumn{2}{l}{544.4} \\
\multicolumn{2}{l}{Sichuan} & 0.099 & \multicolumn{2}{l}{22.3} &
\multicolumn{2}{l}{170.5} \\
\multicolumn{2}{l}{Guizhou} & -0.337 & \multicolumn{2}{l}{-2.8} &
\multicolumn{2}{l}{584.4} \\
\multicolumn{2}{l}{Yunnan} & -0.080 & \multicolumn{2}{l}{-4.2} &
\multicolumn{2}{l}{117.7} \\
\multicolumn{2}{l}{Xizang} & -0.699 & \multicolumn{2}{l}{-89.4} &
\multicolumn{2}{l}{42.4} \\
\multicolumn{2}{l}{Shaanxi} & 0.064 & \multicolumn{2}{l}{49.9} &
\multicolumn{2}{l}{907.0} \\
\multicolumn{2}{l}{Gansu} & 0.168 & \multicolumn{2}{l}{-0.4} &
\multicolumn{2}{l}{633.3} \\
\multicolumn{2}{l}{Qinghai} & -0.558 & \multicolumn{2}{l}{-6.3} &
\multicolumn{2}{l}{295.0} \\
\multicolumn{2}{l}{Ningxia} & -0.248 & \multicolumn{2}{l}{1.1} &
\multicolumn{2}{l}{896.6} \\
\multicolumn{2}{l}{Xinjiang} & 0.144 & \multicolumn{2}{l}{7.1} &
\multicolumn{2}{l}{879.9} \\
\multirow{4}{*}{Pearson correlation} & \multicolumn{3}{l}{} &
\multicolumn{2}{l}{Correlation coefficients} & \emph{P} \\
& \multicolumn{3}{l}{Increased electricity intensity (2013-2015) \&
Increased proportion of wastewater treated to Class 1A (2013-2015)} &
\multicolumn{2}{l}{0.545} & 0.002 \\
& \multicolumn{3}{l}{Increased electricity intensity (2013-2015) \&
Emissions factors in 2015} & \multicolumn{2}{l}{0.554} & 0.001 \\
& \multicolumn{3}{l}{Increased proportion of wastewater treated to Class
1A (2013-2015) \& Emissions factors in 2015} & \multicolumn{2}{l}{0.423}
& 0.018 \\
\bottomrule
\end{longtable}

\textbf{Supplementary Table} \textbf{12:} Effects of stricter standard
(Class 1A compared with Class 1B) on FU-Gini coefficient of sludge yield
with provincial fixed effect.

\begin{longtable}[]{@{}
  >{\raggedright\arraybackslash}p{(\columnwidth - 16\tabcolsep) * \real{0.1694}}
  >{\raggedright\arraybackslash}p{(\columnwidth - 16\tabcolsep) * \real{0.1185}}
  >{\raggedright\arraybackslash}p{(\columnwidth - 16\tabcolsep) * \real{0.1188}}
  >{\raggedright\arraybackslash}p{(\columnwidth - 16\tabcolsep) * \real{0.0904}}
  >{\raggedright\arraybackslash}p{(\columnwidth - 16\tabcolsep) * \real{0.0283}}
  >{\raggedright\arraybackslash}p{(\columnwidth - 16\tabcolsep) * \real{0.1186}}
  >{\raggedright\arraybackslash}p{(\columnwidth - 16\tabcolsep) * \real{0.1186}}
  >{\raggedright\arraybackslash}p{(\columnwidth - 16\tabcolsep) * \real{0.1186}}
  >{\raggedright\arraybackslash}p{(\columnwidth - 16\tabcolsep) * \real{0.1187}}@{}}
\toprule
\endhead
gini & Coef. & St.Err. & \multicolumn{2}{l}{t-value} & p-value &
\multicolumn{2}{l}{{[}95\% Conf Interval{]}} & Sig \\
: base 1A & 0 & . & \multicolumn{2}{l}{.} & . & . & . & \\
1B & 0.033 & 0.007 & \multicolumn{2}{l}{4.81} & 0 & 0.019 & 0.046 &
*** \\
: base Anhui & 0 & . & \multicolumn{2}{l}{.} & . & . & . & \\
Beijing & -0.080 & 0.019 & \multicolumn{2}{l}{-4.32} & 0 & -0.117 &
-0.044 & *** \\
Chongqing & -0.065 & 0.015 & \multicolumn{2}{l}{-4.25} & 0 & -0.096 &
-0.035 & *** \\
Fujian & -0.034 & 0.013 & \multicolumn{2}{l}{-2.58} & 0.010 & -0.060 &
-0.008 & ** \\
Gansu & -0.066 & 0.029 & \multicolumn{2}{l}{-2.29} & 0.022 & -0.122 &
-0.009 & ** \\
Guangdong & -0.008 & 0.011 & \multicolumn{2}{l}{-0.71} & 0.476 & -0.029
& 0.014 & \\
Guangxi & 0.063 & 0.017 & \multicolumn{2}{l}{3.62} & 0 & 0.029 & 0.097 &
*** \\
Guizhou & -0.013 & 0.023 & \multicolumn{2}{l}{-0.58} & 0.565 & -0.058 &
0.032 & \\
Hainan & -0.023 & 0.020 & \multicolumn{2}{l}{-1.15} & 0.250 & -0.063 &
0.017 & \\
Hebei & 0.020 & 0.013 & \multicolumn{2}{l}{1.58} & 0.116 & -0.005 &
0.045 & \\
Heilongjiang & 0.048 & 0.023 & \multicolumn{2}{l}{2.13} & 0.034 & 0.004
& 0.092 & ** \\
Henan & -0.020 & 0.016 & \multicolumn{2}{l}{-1.27} & 0.203 & -0.050 &
0.011 & \\
Hubei & 0.014 & 0.017 & \multicolumn{2}{l}{0.80} & 0.423 & -0.020 &
0.047 & \\
Hunan & 0.038 & 0.020 & \multicolumn{2}{l}{1.94} & 0.052 & 0 & 0.077 &
* \\
Jiangsu & 0.013 & 0.021 & \multicolumn{2}{l}{0.62} & 0.533 & -0.028 &
0.054 & \\
Jiangxi & -0.008 & 0.027 & \multicolumn{2}{l}{-0.30} & 0.762 & -0.061 &
0.045 & \\
Jilin & -0.048 & 0.024 & \multicolumn{2}{l}{-2.01} & 0.044 & -0.094 &
-0.001 & ** \\
Liaoning & 0.009 & 0.018 & \multicolumn{2}{l}{0.53} & 0.594 & -0.025 &
0.044 & \\
Neimenggu & 0.024 & 0.031 & \multicolumn{2}{l}{0.79} & 0.428 & -0.036 &
0.084 & \\
Ningxia & -0.083 & 0.039 & \multicolumn{2}{l}{-2.14} & 0.033 & -0.159 &
-0.007 & ** \\
Qinghai & -0.023 & 0.031 & \multicolumn{2}{l}{-0.74} & 0.457 & -0.084 &
0.038 & \\
Shaanxi & -0.040 & 0.021 & \multicolumn{2}{l}{-1.89} & 0.060 & -0.083 &
0.002 & * \\
Shandong & 0.011 & 0.012 & \multicolumn{2}{l}{0.88} & 0.378 & -0.013 &
0.035 & \\
Shanghai & -0.083 & 0.015 & \multicolumn{2}{l}{-5.50} & 0 & -0.113 &
-0.054 & *** \\
Shanxi & -0.034 & 0.014 & \multicolumn{2}{l}{-2.42} & 0.016 & -0.061 &
-0.006 & ** \\
Sichuan & -0.015 & 0.014 & \multicolumn{2}{l}{-1.13} & 0.259 & -0.042 &
0.011 & \\
Tianjin & -0.087 & 0.021 & \multicolumn{2}{l}{-4.12} & 0 & -0.129 &
-0.046 & *** \\
Xinjiang & -0.008 & 0.037 & \multicolumn{2}{l}{-0.23} & 0.818 & -0.081 &
0.064 & \\
Xizang & -0.172 & 0.044 & \multicolumn{2}{l}{-3.89} & 0 & -0.259 &
-0.085 & *** \\
Yunnan & -0.035 & 0.012 & \multicolumn{2}{l}{-2.97} & 0.003 & -0.059 &
-0.012 & *** \\
Zhejiang & -0.009 & 0.019 & \multicolumn{2}{l}{-0.50} & 0.616 & -0.046 &
0.027 & \\
Constant & 0.233 & 0.009 & \multicolumn{2}{l}{25.42} & 0 & 0.215 & 0.251
& *** \\
\multicolumn{2}{l}{Mean dependent var} & \multicolumn{2}{l}{0.230} &
\multicolumn{3}{l}{SD dependent var} & \multicolumn{2}{l}{0.095} \\
\multicolumn{2}{l}{R-squared} & \multicolumn{2}{l}{0.234} &
\multicolumn{3}{l}{Number of obs} & \multicolumn{2}{l}{641} \\
\multicolumn{2}{l}{F-test} & \multicolumn{2}{l}{7.033} &
\multicolumn{3}{l}{Prob \textgreater{} F} & \multicolumn{2}{l}{0.000} \\
\multicolumn{2}{l}{Akaike crit. (AIC)} & \multicolumn{2}{l}{-1308.622} &
\multicolumn{3}{l}{Bayesian crit. (BIC)} &
\multicolumn{2}{l}{-1165.805} \\
\multicolumn{9}{l}{\emph{*** p\textless0.01, ** p\textless0.05, *
p\textless0.1}} \\
\bottomrule
\end{longtable}

Note: A fixed effects model is used for the short panel data with a
structure of 11 years (2009-2019) and 31 provinces to estimate the
relationship between sludge yield inequality, measured by FU-Gini, and
stricter effluent standard (applying Class 1A rather than Class 1B) of
WWTPs, with clustered robust standard errors. Statistical evidence
reveals that FU-Gini would be lower with stricter effluent standard
(Class 1A). With provincial fixed effects, The estimated coefficient is
highly significant at the 1\% confidence level: 0.033 FU-Gini decrease
is associated with applying Class 1A but not Class 1B. This pattern is
similar to the Kuznets Curve theory. The Kuznets Curve theory
hypothesizes that growing economy may lead to a more equal income
distribution, which can be extended to the relationship between carbon
inequality and economic growth.

\textbf{Supplementary Table} \textbf{13:} Effects of stricter standard
(Class 1A compared with Class 1B) on FU-Gini coefficient of sludge yield
with year fixed effect.

\begin{longtable}[]{@{}
  >{\raggedright\arraybackslash}p{(\columnwidth - 20\tabcolsep) * \real{0.1526}}
  >{\raggedright\arraybackslash}p{(\columnwidth - 20\tabcolsep) * \real{0.1186}}
  >{\raggedright\arraybackslash}p{(\columnwidth - 20\tabcolsep) * \real{0.0024}}
  >{\raggedright\arraybackslash}p{(\columnwidth - 20\tabcolsep) * \real{0.1211}}
  >{\raggedright\arraybackslash}p{(\columnwidth - 20\tabcolsep) * \real{0.1138}}
  >{\raggedright\arraybackslash}p{(\columnwidth - 20\tabcolsep) * \real{0.0072}}
  >{\raggedright\arraybackslash}p{(\columnwidth - 20\tabcolsep) * \real{0.1211}}
  >{\raggedright\arraybackslash}p{(\columnwidth - 20\tabcolsep) * \real{0.1210}}
  >{\raggedright\arraybackslash}p{(\columnwidth - 20\tabcolsep) * \real{0.0049}}
  >{\raggedright\arraybackslash}p{(\columnwidth - 20\tabcolsep) * \real{0.1162}}
  >{\raggedright\arraybackslash}p{(\columnwidth - 20\tabcolsep) * \real{0.1211}}@{}}
\toprule
\endhead
gini & \multicolumn{2}{l}{Coef.} & St.Err. & \multicolumn{2}{l}{t-value}
& p-value & \multicolumn{3}{l}{{[}95\% Conf Interval{]}} & Sig \\
: base 1A & \multicolumn{2}{l}{0} & . & \multicolumn{2}{l}{.} & . & . &
\multicolumn{2}{l}{.} & \\
1B & \multicolumn{2}{l}{0.031} & 0.005 & \multicolumn{2}{l}{5.69} & 0 &
0.019 & \multicolumn{2}{l}{0.043} & *** \\
: base 2009 & \multicolumn{2}{l}{0} & . & \multicolumn{2}{l}{.} & . & .
& \multicolumn{2}{l}{.} & \\
2010 & \multicolumn{2}{l}{0.002} & 0.007 & \multicolumn{2}{l}{0.25} &
0.807 & -0.014 & \multicolumn{2}{l}{0.018} & \\
2011 & \multicolumn{2}{l}{-0.004} & 0.023 & \multicolumn{2}{l}{-0.17} &
0.871 & -0.055 & \multicolumn{2}{l}{0.048} & \\
2012 & \multicolumn{2}{l}{0} & 0.007 & \multicolumn{2}{l}{-0.04} & 0.965
& -0.017 & \multicolumn{2}{l}{0.016} & \\
2013 & \multicolumn{2}{l}{-0.017} & 0.008 & \multicolumn{2}{l}{-2.15} &
0.057 & -0.034 & \multicolumn{2}{l}{0.001} & * \\
2014 & \multicolumn{2}{l}{-0.012} & 0.013 & \multicolumn{2}{l}{-0.93} &
0.375 & -0.041 & \multicolumn{2}{l}{0.017} & \\
2015 & \multicolumn{2}{l}{-0.006} & 0.012 & \multicolumn{2}{l}{-0.46} &
0.652 & -0.033 & \multicolumn{2}{l}{0.021} & \\
2016 & \multicolumn{2}{l}{0.008} & 0.009 & \multicolumn{2}{l}{0.85} &
0.413 & -0.013 & \multicolumn{2}{l}{0.028} & \\
2017 & \multicolumn{2}{l}{0.006} & 0.008 & \multicolumn{2}{l}{0.77} &
0.460 & -0.012 & \multicolumn{2}{l}{0.025} & \\
2018 & \multicolumn{2}{l}{0.022} & 0.010 & \multicolumn{2}{l}{2.23} &
0.050 & 0 & \multicolumn{2}{l}{0.044} & * \\
2019 & \multicolumn{2}{l}{0.011} & 0.009 & \multicolumn{2}{l}{1.30} &
0.222 & -0.008 & \multicolumn{2}{l}{0.031} & \\
Constant & \multicolumn{2}{l}{0.261} & 0.008 & \multicolumn{2}{l}{34.04}
& 0 & 0.244 & \multicolumn{2}{l}{0.278} & *** \\
\multicolumn{2}{l}{Mean dependent var} & \multicolumn{3}{l}{0.277} &
\multicolumn{4}{l}{SD dependent var} & \multicolumn{2}{l}{0.021} \\
\multicolumn{2}{l}{R-squared} & \multicolumn{3}{l}{0.825} &
\multicolumn{4}{l}{Number of obs} & \multicolumn{2}{l}{22} \\
\multicolumn{2}{l}{F-test} & \multicolumn{3}{l}{8.712} &
\multicolumn{4}{l}{Prob \textgreater{} F} & \multicolumn{2}{l}{0.001} \\
\multicolumn{2}{l}{Akaike crit. (AIC)} & \multicolumn{3}{l}{-123.268} &
\multicolumn{4}{l}{Bayesian crit. (BIC)} &
\multicolumn{2}{l}{-110.176} \\
\multicolumn{11}{l}{\emph{*** p\textless0.01, ** p\textless0.05, *
p\textless0.1}} \\
\bottomrule
\end{longtable}

Note: A fixed effects model is used for the short panel data with a
structure of 11 years (2009-2019) to estimate the relationship between
sludge yield inequality, measured by FU-Gini, and stricter effluent
standard (applying Class 1A rather than Class 1B) of WWTPs, with
clustered robust standard errors. Statistical evidence reveals that
FU-Gini would be lower with stricter effluent standard (Class 1A). With
year fixed effects, The estimated coefficient is highly significant at
the 1\% confidence level: 0.031 FU-Gini decrease is associated with
applying Class 1A but not Class 1B. This pattern is similar to the
Kuznets Curve theory. The Kuznets Curve theory hypothesizes that growing
economy may lead to a more equal income distribution, which can be
extended to the relationship between carbon inequality and economic
growth.

\textbf{Supplementary Table} \textbf{14:} Effects of stricter standard
(Class 1A compared with Class 1B) on FU-Gini coefficient of electricity
intensity with provincial fixed effect.

\begin{longtable}[]{@{}
  >{\raggedright\arraybackslash}p{(\columnwidth - 16\tabcolsep) * \real{0.1692}}
  >{\raggedright\arraybackslash}p{(\columnwidth - 16\tabcolsep) * \real{0.1185}}
  >{\raggedright\arraybackslash}p{(\columnwidth - 16\tabcolsep) * \real{0.1186}}
  >{\raggedright\arraybackslash}p{(\columnwidth - 16\tabcolsep) * \real{0.0933}}
  >{\raggedright\arraybackslash}p{(\columnwidth - 16\tabcolsep) * \real{0.0253}}
  >{\raggedright\arraybackslash}p{(\columnwidth - 16\tabcolsep) * \real{0.1187}}
  >{\raggedright\arraybackslash}p{(\columnwidth - 16\tabcolsep) * \real{0.1188}}
  >{\raggedright\arraybackslash}p{(\columnwidth - 16\tabcolsep) * \real{0.1186}}
  >{\raggedright\arraybackslash}p{(\columnwidth - 16\tabcolsep) * \real{0.1190}}@{}}
\toprule
\endhead
gini & Coef. & St.Err. & \multicolumn{2}{l}{t-value} & p-value &
\multicolumn{2}{l}{{[}95\% Conf Interval{]}} & Sig \\
: base 1A & 0 & . & \multicolumn{2}{l}{.} & . & . & . & \\
1B & 0.029 & 0.005 & \multicolumn{2}{l}{5.69} & 0 & 0.019 & 0.039 &
*** \\
: base Anhui & 0 & . & \multicolumn{2}{l}{.} & . & . & . & \\
Beijing & -0.009 & 0.022 & \multicolumn{2}{l}{-0.39} & 0.693 & -0.052 &
0.034 & \\
Chongqing & -0.081 & 0.015 & \multicolumn{2}{l}{-5.35} & 0 & -0.110 &
-0.051 & *** \\
Fujian & -0.038 & 0.010 & \multicolumn{2}{l}{-3.72} & 0 & -0.058 &
-0.018 & *** \\
Gansu & -0.097 & 0.017 & \multicolumn{2}{l}{-5.57} & 0 & -0.131 & -0.063
& *** \\
Guangdong & -0.019 & 0.007 & \multicolumn{2}{l}{-2.57} & 0.01 & -0.034 &
-0.005 & ** \\
Guangxi & 0.003 & 0.010 & \multicolumn{2}{l}{0.29} & 0.769 & -0.016 &
0.022 & \\
Guizhou & -0.041 & 0.017 & \multicolumn{2}{l}{-2.46} & 0.014 & -0.074 &
-0.008 & ** \\
Hainan & -0.054 & 0.013 & \multicolumn{2}{l}{-4.31} & 0 & -0.079 &
-0.029 & *** \\
Hebei & -0.030 & 0.009 & \multicolumn{2}{l}{-3.27} & 0.001 & -0.048 &
-0.012 & *** \\
Heilongjiang & 0.019 & 0.012 & \multicolumn{2}{l}{1.62} & 0.106 & -0.004
& 0.043 & \\
Henan & -0.039 & 0.013 & \multicolumn{2}{l}{-3.05} & 0.002 & -0.064 &
-0.014 & *** \\
Hubei & -0.032 & 0.011 & \multicolumn{2}{l}{-2.81} & 0.005 & -0.054 &
-0.01 & *** \\
Hunan & -0.032 & 0.016 & \multicolumn{2}{l}{-2.02} & 0.044 & -0.063 &
-0.001 & ** \\
Jiangsu & -0.028 & 0.008 & \multicolumn{2}{l}{-3.55} & 0 & -0.044 &
-0.013 & *** \\
Jiangxi & -0.059 & 0.017 & \multicolumn{2}{l}{-3.42} & 0.001 & -0.093 &
-0.025 & *** \\
Jilin & -0.063 & 0.015 & \multicolumn{2}{l}{-4.10} & 0 & -0.093 & -0.033
& *** \\
Liaoning & -0.025 & 0.010 & \multicolumn{2}{l}{-2.54} & 0.011 & -0.045 &
-0.006 & ** \\
Neimenggu & -0.036 & 0.021 & \multicolumn{2}{l}{-1.69} & 0.091 & -0.077
& 0.006 & * \\
Ningxia & -0.102 & 0.028 & \multicolumn{2}{l}{-3.58} & 0 & -0.158 &
-0.046 & *** \\
Qinghai & -0.044 & 0.025 & \multicolumn{2}{l}{-1.78} & 0.076 & -0.093 &
0.005 & * \\
Shaanxi & -0.022 & 0.019 & \multicolumn{2}{l}{-1.11} & 0.266 & -0.060 &
0.017 & \\
Shandong & -0.047 & 0.010 & \multicolumn{2}{l}{-4.87} & 0 & -0.066 &
-0.028 & *** \\
Shanghai & -0.080 & 0.011 & \multicolumn{2}{l}{-7.34} & 0 & -0.102 &
-0.059 & *** \\
Shanxi & -0.014 & 0.012 & \multicolumn{2}{l}{-1.15} & 0.252 & -0.038 &
0.010 & \\
Sichuan & 0.004 & 0.013 & \multicolumn{2}{l}{0.33} & 0.742 & -0.021 &
0.030 & \\
Tianjin & -0.111 & 0.016 & \multicolumn{2}{l}{-6.84} & 0 & -0.143 &
-0.079 & *** \\
Xinjiang & -0.003 & 0.026 & \multicolumn{2}{l}{-0.13} & 0.895 & -0.055 &
0.048 & \\
Xizang & -0.197 & 0.014 & \multicolumn{2}{l}{-13.95} & 0 & -0.225 &
-0.169 & *** \\
Yunnan & -0.029 & 0.010 & \multicolumn{2}{l}{-2.84} & 0.005 & -0.050 &
-0.009 & *** \\
Zhejiang & -0.036 & 0.016 & \multicolumn{2}{l}{-2.23} & 0.026 & -0.069 &
-0.004 & ** \\
Constant & 0.209 & 0.006 & \multicolumn{2}{l}{34.30} & 0 & 0.197 & 0.221
& *** \\
\multicolumn{2}{l}{Mean dependent var} & \multicolumn{2}{l}{0.184} &
\multicolumn{3}{l}{SD dependent var} & \multicolumn{2}{l}{0.073} \\
\multicolumn{2}{l}{R-squared} & \multicolumn{2}{l}{0.289} &
\multicolumn{3}{l}{Number of obs} & \multicolumn{2}{l}{642} \\
\multicolumn{2}{l}{F-test} & \multicolumn{2}{l}{13.181} &
\multicolumn{3}{l}{Prob \textgreater{} F} & \multicolumn{2}{l}{0.000} \\
\multicolumn{2}{l}{Akaike crit. (AIC)} & \multicolumn{2}{l}{-1700.589} &
\multicolumn{3}{l}{Bayesian crit. (BIC)} &
\multicolumn{2}{l}{-1557.722} \\
\multicolumn{9}{l}{\emph{*** p\textless0.01, ** p\textless0.05, *
p\textless0.1}} \\
\bottomrule
\end{longtable}

Note: A fixed effects model is used for the short panel data with a
structure of 11 years (2009-2019) and 31 provinces to estimate the
relationship between electricity itnensity inequality, measured by
FU-Gini, and stricter effluent standard (applying Class 1A rather than
Class 1B) of WWTPs, with clustered robust standard errors. Statistical
evidence reveals that FU-Gini would be lower with stricter effluent
standard (Class 1A). With provincial fixed effects, The estimated
coefficient is highly significant at the 1\% confidence level: 0.029
FU-Gini decrease is associated with applying Class 1A but not Class 1B.
This pattern is similar to the Kuznets Curve theory. The Kuznets Curve
theory hypothesizes that growing economy may lead to a more equal income
distribution, which can be extended to the relationship between carbon
inequality and economic growth.

\textbf{Supplementary Table} \textbf{15:} Effects of stricter standard
(Class 1A compared with Class 1B) on FU-Gini coefficient of electricity
intensity with year fixed effect.

\begin{longtable}[]{@{}
  >{\raggedright\arraybackslash}p{(\columnwidth - 18\tabcolsep) * \real{0.1526}}
  >{\raggedright\arraybackslash}p{(\columnwidth - 18\tabcolsep) * \real{0.1186}}
  >{\raggedright\arraybackslash}p{(\columnwidth - 18\tabcolsep) * \real{0.1235}}
  >{\raggedright\arraybackslash}p{(\columnwidth - 18\tabcolsep) * \real{0.1025}}
  >{\raggedright\arraybackslash}p{(\columnwidth - 18\tabcolsep) * \real{0.0185}}
  >{\raggedright\arraybackslash}p{(\columnwidth - 18\tabcolsep) * \real{0.1211}}
  >{\raggedright\arraybackslash}p{(\columnwidth - 18\tabcolsep) * \real{0.1210}}
  >{\raggedright\arraybackslash}p{(\columnwidth - 18\tabcolsep) * \real{0.0049}}
  >{\raggedright\arraybackslash}p{(\columnwidth - 18\tabcolsep) * \real{0.1162}}
  >{\raggedright\arraybackslash}p{(\columnwidth - 18\tabcolsep) * \real{0.1211}}@{}}
\toprule
\endhead
gini & Coef. & St.Err. & \multicolumn{2}{l}{t-value} & p-value &
\multicolumn{3}{l}{{[}95\% Conf Interval{]}} & Sig \\
: base 1A & 0 & . & \multicolumn{2}{l}{.} & . & . &
\multicolumn{2}{l}{.} & \\
1B & 0.040 & 0.003 & \multicolumn{2}{l}{14.37} & 0 & 0.034 &
\multicolumn{2}{l}{0.046} & *** \\
: base 2009 & 0 & . & \multicolumn{2}{l}{.} & . & . &
\multicolumn{2}{l}{.} & \\
2010 & 0.002 & 0.004 & \multicolumn{2}{l}{0.45} & 0.663 & -0.007 &
\multicolumn{2}{l}{0.010} & \\
2011 & 0.010 & 0.003 & \multicolumn{2}{l}{3.16} & 0.010 & 0.003 &
\multicolumn{2}{l}{0.018} & ** \\
2012 & 0.018 & 0.006 & \multicolumn{2}{l}{3.00} & 0.013 & 0.005 &
\multicolumn{2}{l}{0.031} & ** \\
2013 & 0.013 & 0.006 & \multicolumn{2}{l}{2.07} & 0.065 & -0.001 &
\multicolumn{2}{l}{0.027} & * \\
2014 & 0.009 & 0.004 & \multicolumn{2}{l}{2.30} & 0.044 & 0 &
\multicolumn{2}{l}{0.017} & ** \\
2015 & 0.021 & 0.005 & \multicolumn{2}{l}{4.54} & 0.001 & 0.01 &
\multicolumn{2}{l}{0.031} & *** \\
2016 & 0.024 & 0.004 & \multicolumn{2}{l}{6.60} & 0 & 0.016 &
\multicolumn{2}{l}{0.032} & *** \\
2017 & 0.028 & 0.004 & \multicolumn{2}{l}{7.48} & 0 & 0.020 &
\multicolumn{2}{l}{0.037} & *** \\
2018 & 0.043 & 0.012 & \multicolumn{2}{l}{3.67} & 0.004 & 0.017 &
\multicolumn{2}{l}{0.068} & *** \\
2019 & 0.029 & 0.005 & \multicolumn{2}{l}{5.29} & 0 & 0.017 &
\multicolumn{2}{l}{0.041} & *** \\
Constant & 0.196 & 0.004 & \multicolumn{2}{l}{55.60} & 0 & 0.189 &
\multicolumn{2}{l}{0.204} & *** \\
\multicolumn{2}{l}{Mean dependent var} & \multicolumn{2}{l}{0.234} &
\multicolumn{4}{l}{SD dependent var} & \multicolumn{2}{l}{0.024} \\
\multicolumn{2}{l}{R-squared} & \multicolumn{2}{l}{0.966} &
\multicolumn{4}{l}{Number of obs} & \multicolumn{2}{l}{22} \\
\multicolumn{2}{l}{F-test} & \multicolumn{2}{l}{39.885} &
\multicolumn{4}{l}{Prob \textgreater{} F} & \multicolumn{2}{l}{0.000} \\
\multicolumn{2}{l}{Akaike crit. (AIC)} & \multicolumn{2}{l}{-152.457} &
\multicolumn{4}{l}{Bayesian crit. (BIC)} &
\multicolumn{2}{l}{-139.364} \\
\multicolumn{10}{l}{\emph{*** p\textless0.01, ** p\textless0.05, *
p\textless0.1}} \\
\bottomrule
\end{longtable}

Note: A fixed effects model is used for the short panel data with a
structure of 11 years (2009-2019) to estimate the relationship between
electricity intensity inequality, measured by FU-Gini, and stricter
effluent standard (applying Class 1A rather than Class 1B) of WWTPs,
with clustered robust standard errors. Statistical evidence reveals that
FU-Gini would be lower with stricter effluent standard (Class 1A). With
year fixed effects, The estimated coefficient is highly significant at
the 1\% confidence level: 0.040 FU-Gini decrease is associated with
applying Class 1A but not Class 1B. This pattern is similar to the
Kuznets Curve theory. The Kuznets Curve theory hypothesizes that growing
economy may lead to a more equal income distribution, which can be
extended to the relationship between carbon inequality and economic
growth.

\textbf{Supplementary Table} \textbf{16:} GWP of different
GHG\textsuperscript{3}

\begin{longtable}[]{@{}
  >{\raggedright\arraybackslash}p{(\columnwidth - 2\tabcolsep) * \real{0.5000}}
  >{\raggedright\arraybackslash}p{(\columnwidth - 2\tabcolsep) * \real{0.5000}}@{}}
\toprule
\begin{minipage}[b]{\linewidth}\raggedright
\textbf{Substance}
\end{minipage} & \begin{minipage}[b]{\linewidth}\raggedright
\textbf{GWP (kg CO\textsubscript{2} eq./kg GHG)}
\end{minipage} \\
\midrule
\endhead
Carbon Dioxide (CO\textsubscript{2}) (biogenic) & 0 \\
CO\textsubscript{2} (non-biogenic) & 1 \\
CO\textsubscript{2} eq. & 1 \\
Methane (CH\textsubscript{4}) & 30 \\
Nitrous Oxide (N\textsubscript{2}O) & 265 \\
\bottomrule
\end{longtable}

\textbf{Supplementary Table} \textbf{17:} CF of Wastewater Treatment,
and Sludge Treatment and Disposal

\begin{longtable}[]{@{}
  >{\raggedright\arraybackslash}p{(\columnwidth - 8\tabcolsep) * \real{0.1720}}
  >{\raggedright\arraybackslash}p{(\columnwidth - 8\tabcolsep) * \real{0.1207}}
  >{\raggedright\arraybackslash}p{(\columnwidth - 8\tabcolsep) * \real{0.1036}}
  >{\raggedright\arraybackslash}p{(\columnwidth - 8\tabcolsep) * \real{0.4330}}
  >{\raggedright\arraybackslash}p{(\columnwidth - 8\tabcolsep) * \real{0.1707}}@{}}
\toprule
\begin{minipage}[b]{\linewidth}\raggedright
\textbf{Process}
\end{minipage} & \begin{minipage}[b]{\linewidth}\raggedright
\textbf{Input Flow}
\end{minipage} & \begin{minipage}[b]{\linewidth}\raggedright
\textbf{GHG type}
\end{minipage} & \begin{minipage}[b]{\linewidth}\raggedright
\textbf{Technology}
\end{minipage} & \begin{minipage}[b]{\linewidth}\raggedright
\textbf{CF (kg GHG/kg Input Flow)}
\end{minipage} \\
\midrule
\endhead
\multirow{5}{*}{Wastewater Collection and Treatment\textsuperscript{4}}
& \multirow{2}{*}{BOD} & \multirow{2}{*}{CH\textsubscript{4}} &
Absorption Biodegradation (AB) & 0.180 \\
& & & Except AB & 0.0180 \\
& \multirow{3}{*}{TN} & \multirow{3}{*}{N\textsubscript{2}O} & AB & 0 \\
& & & Soil Treatment & 0.0157 \\
& & & Except AB and Soil Treatment & 0.0251 \\
\multirow{8}{*}{Sludge disposal} & \multirow{8}{*}{Dry Solids (DS) of
Sludge} & \multirow{8}{*}{CO\textsubscript{2} eq.} & Landfill (No
Anaerobic Digestion)\textsuperscript{5} & 1.881 \\
& & & Landfill (with Anaerobic Digestion)\textsuperscript{6} & 0.550 \\
& & & Incineration (No Anaerobic Digestion) \textsuperscript{6} &
5.810 \\
& & & Incineration (with Anaerobic Digestion)\textsuperscript{7} &
3.764 \\
& & & Land Application (No Anaerobic Digestion)\textsuperscript{5} &
0.263 \\
& & & Land Application (with Anaerobic Digestion)\textsuperscript{8} &
0.731 \\
& & & Building Material (No Anaerobic Digestion) & 5.800 \\
& & & Building Material (with Anaerobic Digestion)\textsuperscript{6} &
0.824 \\
\bottomrule
\end{longtable}

Note: For sludge disposal, GHG emissions from electricity use are
excluded according to corresponding article, assuming that sludge
disposal is finished in WWTPs and the electricity consumption is
included into the operational data.

\textbf{Supplementary Table} \textbf{18:} CF of Treated Water
Discharge\textsuperscript{4}

\begin{longtable}[]{@{}
  >{\raggedright\arraybackslash}p{(\columnwidth - 8\tabcolsep) * \real{0.1720}}
  >{\raggedright\arraybackslash}p{(\columnwidth - 8\tabcolsep) * \real{0.1207}}
  >{\raggedright\arraybackslash}p{(\columnwidth - 8\tabcolsep) * \real{0.1553}}
  >{\raggedright\arraybackslash}p{(\columnwidth - 8\tabcolsep) * \real{0.2933}}
  >{\raggedright\arraybackslash}p{(\columnwidth - 8\tabcolsep) * \real{0.2587}}@{}}
\toprule
\begin{minipage}[b]{\linewidth}\raggedright
\textbf{Process}
\end{minipage} & \begin{minipage}[b]{\linewidth}\raggedright
\textbf{Input Flow}
\end{minipage} & \begin{minipage}[b]{\linewidth}\raggedright
\textbf{GHG type}
\end{minipage} & \begin{minipage}[b]{\linewidth}\raggedright
\textbf{Receive destination type}
\end{minipage} & \begin{minipage}[b]{\linewidth}\raggedright
\textbf{CF (kg GHG/kg Input Flow)}
\end{minipage} \\
\midrule
\endhead
\multirow{5}{*}{Treated Water Discharge} & \multirow{3}{*}{BOD} &
\multirow{3}{*}{CH\textsubscript{4}} & Farmland Irrigated by Wastewater
& 0 \\
& & & Recharge to lakes & 0.114 \\
& & & Other & 0.0680 \\
& \multirow{2}{*}{TN} & \multirow{2}{*}{N\textsubscript{2}O} & Farmland
Irrigated by Wastewater & 0 \\
& & & Other & 0.00786 \\
\bottomrule
\end{longtable}

\textbf{Supplementary Table} \textbf{19:} CF of Power Generation

\begin{longtable}[]{@{}
  >{\raggedright\arraybackslash}p{(\columnwidth - 8\tabcolsep) * \real{0.1547}}
  >{\raggedright\arraybackslash}p{(\columnwidth - 8\tabcolsep) * \real{0.1380}}
  >{\raggedright\arraybackslash}p{(\columnwidth - 8\tabcolsep) * \real{0.1207}}
  >{\raggedright\arraybackslash}p{(\columnwidth - 8\tabcolsep) * \real{0.4211}}
  >{\raggedright\arraybackslash}p{(\columnwidth - 8\tabcolsep) * \real{0.1655}}@{}}
\toprule
\begin{minipage}[b]{\linewidth}\raggedright
\textbf{Process}
\end{minipage} & \begin{minipage}[b]{\linewidth}\raggedright
\textbf{Input Flow}
\end{minipage} & \begin{minipage}[b]{\linewidth}\raggedright
\textbf{GHG type}
\end{minipage} & \begin{minipage}[b]{\linewidth}\raggedright
\textbf{Technology}
\end{minipage} & \begin{minipage}[b]{\linewidth}\raggedright
\textbf{CF (kg CO\textsubscript{2} eq/MWh)}
\end{minipage} \\
\midrule
\endhead
\multirow{6}{*}{Power Generation for Electricity Use} &
\multirow{6}{*}{Electricity} & \multirow{6}{*}{CO\textsubscript{2} eq.}
& Hydroelectric\textsuperscript{9} & 18 \\
& & & Coal\textsuperscript{10} & 820 \\
& & & Natural Gas\textsuperscript{10} & 640 \\
& & & Nuclear\textsuperscript{10} & 17 \\
& & & Wind\textsuperscript{10} & 14 \\
& & & Solar\textsuperscript{10} & 76 \\
\bottomrule
\end{longtable}

Note: These data are life-cycle emission factors.

\textbf{Supplementary Table} \textbf{20:} Regional grids and
corresponding province

\begin{longtable}[]{@{}
  >{\raggedright\arraybackslash}p{(\columnwidth - 4\tabcolsep) * \real{0.3519}}
  >{\raggedright\arraybackslash}p{(\columnwidth - 4\tabcolsep) * \real{0.3373}}
  >{\raggedright\arraybackslash}p{(\columnwidth - 4\tabcolsep) * \real{0.3107}}@{}}
\toprule
\begin{minipage}[b]{\linewidth}\raggedright
\textbf{Regional Grid}
\end{minipage} & \begin{minipage}[b]{\linewidth}\raggedright
\textbf{Province/Municipality}
\end{minipage} & \begin{minipage}[b]{\linewidth}\raggedright
\textbf{Abbreviation}
\end{minipage} \\
\midrule
\endhead
North China Grid & Beijing, Tianjin, Hebei, Shandong, Shanxi, Inner
Mongolia & BJ, TJ, HE, SD, SX, NM \\
Central China Grid & Hubei, Henan, Hunan, Jiangxi, Sichuan, Chongqing &
HB, HA, HN, JX, SC, CQ \\
East China Grid & Anhui, Fujian, Shanghai, Jiangsu, Zhejiang & AH, FJ,
SH, JS, ZJ \\
Northeast Grid & Liaoning, Jilin, Heilongjiang & LN, JL, HL \\
Northwest Grid & Shaanxi, Gansu, Ningxia, Qinghai, Xinjiang, Tibet & SN,
GS, NX, QH, XJ, XZ \\
South Grid & Guangdong, Hainan, Guangxi, Yunnan, Guizhou & GD, HI, GX,
YN, GZ \\
\bottomrule
\end{longtable}

\textbf{References}

1. China Urban Water Association. \emph{Urban Sewage Yearbook}. (China
Urban Water Association, 2018).

2. State Council of China. Circular of the State Council on adjusting
the criteria for the division of urban scale. (2014).

3. Stocker, T. F. \emph{Climate Change 2013: The Physical Science
Basis}.

4. IPCC. 2019 Refinement to the 2006 IPCC Guidelines for National
Greenhouse Gas Inventories. \emph{Institute for Global Environmental
Strategies (IGES), Japan} (2019).

5. Liu, B., Wei, Q., Zhang, B. \& Bi, J. Life cycle GHG emissions of
sewage sludge treatment and disposal options in Tai Lake Watershed,
China. \emph{Science of The Total Environment} \textbf{447}, 361--369
(2013).

6. Lam, C.-M., Lee, P.-H. \& Hsu, S.-C. Eco-efficiency analysis of
sludge treatment scenarios in urban cities: the case of Hong Kong.
\emph{Journal of Cleaner Production} \textbf{112}, 3028--3039 (2016).

7. Xu, C., Chen, W. \& Hong, J. Life-cycle environmental and economic
assessment of sewage sludge treatment in China. \emph{Journal of Cleaner
Production} \textbf{67}, 79--87 (2014).

8. Fang, Y. R., Li, S., Zhang, Y. \& Xie, G. H. Spatio-temporal
distribution of sewage sludge, its methane production potential, and a
greenhouse gas emissions analysis. \emph{Journal of Cleaner Production}
\textbf{238}, 117895 (2019).

9. Weisser, D. A guide to life-cycle greenhouse gas (GHG) emissions from
electric supply technologies. \emph{Energy} \textbf{32}, 1543--1559
(2007).

10. Sharifzadeh, M., Hien, R. K. T. \& Shah, N. China's roadmap to
low-carbon electricity and water: Disentangling greenhouse gas (GHG)
emissions from electricity-water nexus via renewable wind and solar
power generation, and carbon capture and storage. \emph{Applied Energy}
\textbf{235}, 31--42 (2019).

11. MOHURD. \emph{Statistical yearbook of urban and rural construction}.
(2020).

12. Krzywinski, M. \emph{et al.} Circos: An information aesthetic for
comparative genomics. \emph{GENOME RESEARCH} vol. 19 1639--1645 (2009).

13. Shannon, C. E. A mathematical theory of communication. \emph{The
Bell System Technical Journal} \textbf{27}, 379--423 (1948).